%% file: main.tex
\documentclass[12pt]{article}

\usepackage{amssymb}
\usepackage{mathtools}
\usepackage{amsthm}
\input{math_command}

\usepackage[titletoc,toc,title]{appendix}
\usepackage[margin=1in]{geometry}     
\usepackage{multicol}       
\usepackage{setspace}   
\usepackage{url}                        
\usepackage[sort]{natbib}               
\bibpunct{(}{)}{;}{a}{,}{,}            
\usepackage{mathptmx} 
\DeclareMathAlphabet{\mathcal}{OMS}{cmsy}{m}{n} 
\usepackage{adjustbox}
\usepackage{tikz}
\usepackage{bm}
\usepackage[backref=page]{hyperref}
\usepackage{etoolbox}
\makeatletter
\patchcmd{\BR@backref}{\newblock}{\newblock(Cited in page~}{}{}
\patchcmd{\BR@backref}{\par}{)\par}{}{}
\makeatother
\usepackage{booktabs} 

\hypersetup{
	colorlinks = true,
	linkcolor = blue,
	citecolor = blue,
	runcolor = blue,
	urlcolor = blue
}

\allowbreak 

\title{Artificial Intelligence and Algorithmic Price Collusion in Two-sided Markets\thanks{This work was partially supported by NSF award DMS-2124913.}
}
\author{Cristian Chica, Yinglong Guo, and Gilad Lerman\thanks{School of Mathematics, University of Minnesota. Email addresses: chica013@umn.edu, guo00413@umn.edu, lerman@umn.edu.}}

\begin{document}
	
	\maketitle
	
	\begin{abstract}
		Algorithmic price collusion facilitated by artificial intelligence (AI) algorithms raises significant concerns. We examine how AI agents using $Q$-learning engage in tacit  collusion in two-sided markets. Our experiments reveal that AI-driven platforms achieve higher collusion levels compared to Bertrand competition. Increased network externalities significantly enhance collusion, suggesting AI algorithms exploit them to maximize profits. Higher user heterogeneity and greater utility from outside options generally reduce collusion, while higher discount rates increase it. Tacit collusion remains feasible even at low discount rates. To mitigate collusive behavior and inform potential regulatory measures, we propose incorporating a penalty term in the $Q$-learning algorithm.

		
	\end{abstract}

	\section{Introduction}
	\input{introduction}

	\section{Review of Our Economic Framework}
	\label{sect:platform_competition}
	\input{econmodel}

	\section{Simulation Framework}
	\label{sec:MARL}
	\input{reinforcement_learning}

	\section{Experimental Results}
	\label{sect:experiments}

\input{result}

	\section{Economic and Policy Discussion}
	
	We examine the economic implications of the numerical results presented in Section \ref{sect:experiments}. Specifically, we analyze the impact of network externalities on the collusive level and discuss the ramifications of these findings on real-life markets. We then review the Preventing Algorithmic Collusion Act of 2024, introduced by U.S. Senator Amy Klobuchar, and suggest an additional policy recommendation to ensure the safer use of $Q$-learning.
	
	\subsection{Economic Discussion}
	\label{sect:Discussion}
	\input{discussion}

	\subsection{Policy Discussion}
	\label{sect:policyDiscussion}
	\input{policy}

	\section{Concluding Remarks}

\input{concludingRemarks}

	\newpage
	\clearpage
	\bibliographystyle{apalike}
	\bibliography{ref}
	
	\newpage
	\setcounter{table}{0}
	\renewcommand{\thetable}{A\arabic{table}}
	\setcounter{figure}{0}
	\renewcommand{\thefigure}{A\arabic{figure}}
	\appendix
	\section{Appendix}\label{appendix}

\input{appendix}

\end{document}

%% file: math_command.tex
\usepackage{amsmath,amsfonts,bm}
\usepackage{xcolor}
\usepackage[normalem]{ulem}

\usepackage{graphicx}
\usepackage{caption}
\usepackage{subcaption}

\usepackage{algorithm,algorithmic}

\usepackage{amsthm}
\usepackage{footmisc}
\usepackage{float}

\theoremstyle{plain}

\theoremstyle{definition}

\theoremstyle{remark}

\def\eqref#1{(\ref{#1})}

\def\1{\bm{1}}

\def\va{{\bm{a}}}

\def\vp{{\bm{p}}}

\def\vs{{\bm{s}}}

\def\vx{{\bm{x}}}

\DeclareMathAlphabet{\mathsfit}{\encodingdefault}{\sfdefault}{m}{sl}
\SetMathAlphabet{\mathsfit}{bold}{\encodingdefault}{\sfdefault}{bx}{n}

\def\gA{{\mathcal{A}}}

\def\gP{{\mathcal{P}}}

\def\gS{{\mathcal{S}}}
\def\gT{{\mathcal{T}}}

\def\sE{{\mathbb{E}}}

\def\sN{{\mathbb{N}}}

\def\sR{{\mathbb{R}}}

\DeclareMathOperator*{\argmax}{arg\,max}

%% file: introduction.tex
Algorithmic price collusion occurs when economic agents set prices using artificial intelligence (AI) algorithms. Through repeated interactions, these agents learn that tacit collusion is optimal, as noted by \cite{calvano2020protecting}.\footnote{Tacit collusion happens when  firms coordinate behavior without explicit communication. \cite{du2023does,bertomeu2021tacit} have documented evidence of this in the U.S.~multihospital system and automotive industry.} 
Economists and antitrust authorities have expressed significant concerns about this form of collusion. The  \cite{OECD2017} raised concerns about pricing algorithms learning to collude via  tacit coordination. \cite{clark2023algorithmic} suggested that algorithmic pricing in Germany's retail gasoline market increased price margins by approximately 15\%. U.S.~Senator Amy Klobuchar introduced the S.3686 - Preventing Algorithmic Collusion Act of 2024 to curb anticompetitive behavior through algorithmic pricing using nonpublic competitor data. 

Recent experiments (\cite{calvano2020artificial}, \cite{klein2021autonomous}) have demonstrated that collusion can be achieved in Bertrand and Stackelberg competition models by simulated economic agents using $Q$-learning, a benchmark reinforcement learning algorithm. 
Building on these findings, this study experimentally investigates AI-driven platforms using $Q$-learning in a repeated two-sided platform competition game. 
We illustrate how these AI agents facilitate collusion compared to Bertrand competition. Our focus is particularly on the impact of network externalities on collusion.

In our model of repeated two-sided platform competition, 
multiple horizontally differentiated platforms compete to serve buyers and sellers, collectively referred to as users. These platforms repeatedly interact and independently choose prices using $Q$-learning, with the last period price as the state variable. This implies that platforms have bounded memory and employ one-memory strategies \citep{barlo2009repeated}. 
In each repetition, users can choose to join one of the platforms or opt for the outside option. Buyers who join a platform receive network externality benefits proportional to the number of buyers (within-side externalities) and sellers (cross-side externalities) on the same platform. Sellers who join the market receive both types of externalities as well. 

Our experiments show that even with zero network externalities, AI-driven platforms achieve higher collusion levels 
compared to those reported by \cite{calvano2020artificial} for Bertrand competition. This is likely due to the larger action space, which allows more information exchange. 
Furthermore, increased network externalities lead to significantly high collusion levels, suggesting AI-driven platforms can leverage these externalities to boost profits. 
In particular, algorithmic pricing can increase collusion in markets with significant positive within-side externalities (e.g., online/cloud gaming) and positive cross-side externalities (e.g., video streaming, social media).

Our findings indicate that higher user heterogeneity or greater utility from the outside option generally decrease collusion levels, except in certain local regions. In contrast, collusion levels typically rise with higher discount rates, especially in the presence of significant network externalities. 
Additionally, 
tacit collusion remains feasible even at very low discount factors. This contrasts with traditional literature on firms' collusion without AI agents, which suggests that collusion is feasible only at high discount factors \citep{tirole1988theory,obara2017collusion}. 

Finally, we propose reducing collusion through a penalty term in the $Q$-learning algorithm. 

\textbf{Related Literature.} There is a growing literature on algorithmic price collusion, with a particular emphasis on using numerical simulations to show that $Q$-learning results in tacit collusion. 
\cite{waltman2008q} showed that firms using $Q$-learning in repeated Cournot oligopoly games produce lower quantities than the competitive Nash equilibrium. 
 \cite{calvano2020artificial} showed that $Q$-learning firms choose high prices in repeated Bertrand games and learn strategies consistent with tacit collusion.  
Similar work was done by \cite{klein2021autonomous} for repeated Stackelberg games. \cite{clark2023algorithmic} is the first work that uses real-life data to show that firms may increase price margins with the adoption of algorithmic pricing. Our work extends the numerical understanding of algorithmic pricing, particularly in two-sided markets with network externalities.

Studies by \cite{johnson2023platform} and \cite{brero2022learning} on single platforms with AI-driven sellers show how platform-designed rules can promote competition and reduce collusion. 
Nevertheless, this setting does not apply to ours, where multiple platforms apply AI algorithms.  

Our model of repeated two-sided platform competition uses the model in \cite{cristian2023competition}, which, in turn, builds upon previous models by \cite{white2016insulated,tan2021effects,chica2021exclusive}. These models analyze network externality effects on equilibrium outputs. Our simulations use their insights to study the impact of these  externalities on the collusive levels. 

\cite{ruhmer2010platform} finds that higher cross-side 
externalities make collusion harder to sustain, when following the model of \citet{armstrong2006competition} without AI agents. This is consistent with our numerical results, even though we consider algorithmic pricing and follow the model of  \citet{cristian2023competition}.

Theoretical work in economics on algorithmic price collusion includes \citet{brown2023competition}, which demonstrates that simple pricing algorithms can elevate price levels. Additionally, \citet{arslantas2024strategizing} illustrates how a sophisticated agent can exploit another agent using a naive version of $Q$-learning, provided the former agent knows the algorithm being used.

%% file: econmodel.tex
We introduce the economics framework  used in our experiments.  Section \ref{sec:model1} presents the baseline platform competition game. Section \ref{sec:model2} extends the latter model to an infinite repeated game.

\subsection{The Baseline Platform Competition Game}
\label{sec:model1}

The baseline platform competition game consists of two stages. In stage I, a set of horizontally differentiated platforms strategically choose prices to maximize profits. In stage II, given the prices determined by the platforms, users on each of the two sides of the market choose whether to participate or not and if they participate they also choose which platform to join. The solution concept for the baseline game is backward induction. More specifically, $N$ platforms provide service options for users on two sides of a market, buyers and sellers. Users in these two sides of a market are denoted with $k\in \{b,s\}$, where $b$ and $s$ represent buyers and sellers, respectively. These users 
have $N+1$ choices, where $N\geq 2$. They can either opt out of the market by choosing the outside option, or join one of the $N$ horizontally differentiated platforms, each one denoted with $i\in [N] := \{1,\dots, N\}$. The users on side $k$ opting out of the market receive a deterministic outside option utility $u^{(0)}_k\in \mathbb{R}$. The users on side $k$ joining platform $i\in [N]$ receive a deterministic utility 
$$u^{(i)}_k :=  \phi_k(x^{(i)}_b,x^{(i)}_s)-p^{(i)}_k,$$ where $p^{(i)}_k$ is the price paid by the user to access services provided by the platform $i$; $x^{(i)}_k$ is the total mass of users on side $k$ joining platform $i$; and $$\phi_k(x^{(i)}_b,x^{(i)}_s):=\phi_{kb}x^{(i)}_b+\phi_{ks}x^{(i)}_s, \ \text{ with } \ \phi_{kb}, \phi_{ks}\in \mathbb{R},$$ is the network externality function that captures the network benefits users enjoy by joining platform $i$. The network externalities are captured by the following linear transformation
\begin{equation}
    (\phi_b(x_b^{(i)}, x_s^{(i)}), \phi_s(x_b^{(i)}, x_s^{(i)}))^T =\Phi \vx^{(i)}, \ \text{ where }\ \Phi = \left[\begin{array}{cc}
        \phi_{bb} & \phi_{bs} \\
        \phi_{sb} & \phi_{ss}
    \end{array}\right].
    \label{eqn:intro_phi}
\end{equation}
To save space, we write $\Phi = [\phi_{bb}, \phi_{bs}; \phi_{sb}, \phi_{ss}]$ when specifying choices of $\Phi$. The endogenous mass of users on each side of the market subscribed to platform $i$ is denoted by $\vx^{(i)}:=(x_b^{(i)},x_s^{(i)})\in [0,1]^2$ and the mass of users not participating in the market is denoted by $\vx^{(0)}:=(x_b^{(0)},x_s^{(0)}) \in [0,1]^2$. Assuming all users have Gumbel-distributed idiosyncratic preferences with parameters $(\mu_k,\beta_k)$, $\mu_k\in\mathbb{R}$ and $\beta_k>0$,\footnote{The Gumbel-distribution parameter $\beta_k$ measures the standard deviation of  $\epsilon_k^i$ and it captures the degree of heterogeneity in users' tastes. Unlike $\beta_k$, $\mu_k$ does not affect the equilibrium output of the model (see \eqref{xki}).\label{Footnote:betak_interpretation}} the quantities $x_k^{(i)}$ are determined through a maximization process conducted by users on side $k$ who solve the following equations (see \cite{cristian2023competition}):\footnote{We  
note that the model presented here is equivalent to a model in which users have Logistic-distributed preferences for the platforms, and the outside option utility is deterministic.} 
\begin{equation}
x_k^{(i)} = 1 - \left({1 + \exp\left(u^{(i)}_k/\beta_k - \ln\left(\sum\limits_{j=0,1,\cdots,N, j\neq i} e^{u^{(j)}_k / \beta_k}\right)\right)}\right)^{-1} \ i \in [N]\cup \{0\}, \ k\in \{b, s\}.\label{xki}
\end{equation}
The platforms incorporate \eqref{xki} into their profit  maximization problem as follows, where $\pi^{(i)}$ denotes the profit of platform $i$ and $\Pi_{\text{tot}}$ denotes the total profits of $N$ colluding platforms: 

(i) when competing, they solve
\begin{equation}\label{pi}
\max_{\{p^{(i)}_b,p_s^{(i)}\}}\text{  } \pi^{(i)} (p_b^{(i)},p^{(i)}_s) , \text{ where }  \pi^{(i)} (p_b^{(i)},p^{(i)}_s ) :=  x^{(i)}_b p^{(i)}_b+x^{(i)}_s p^{(i)}_s;\end{equation}

(ii) when colluding, they solve
\begin{equation}
    \max_{\substack{p_b,p_s}} \ 
    \Pi_{\text{tot}}(p_b,p_s), \text{ where }
    \Pi_{\text{tot}}(p_b,p_s) 
 := \sum_{i=1}^N\left(x^{(i)}_b p_b+x^{(i)}_s p_s\right).\label{pim}
\end{equation}
The maximizer of \eqref{pi} is called the \textit{competitive Nash Equilibrium} (CNE) and the corresponding equilibrium quantities are denoted by $p^{(i),*}_k$ and $x^{(i),*}_k$ for $i\in[N]\cup \{0\}$ and $k\in \{b,s\}$. 
The maximizer of \eqref{pim} is called the \textit{collusive equilibrium} (CE) and the corresponding equilibrium quantities are denoted by  $p^{(i),\text{C}}_k$ and $x^{(i),\text{C}}_k$,  $i\in[N]\cup \{0\}$, $k\in \{b,s\}$. 
In the symmetric equilibrium, $p_k^{(i),*} = p_k^*$, $p_k^{(i),\text{C}} = p_k^\text{C}$, $x_k^{(i),*} = x_k^*$ and $x_k^{(i),\text{C}} = x_k^\text{C}$ for all $i\in [N]$. Propositions 3.2 and 3.4 in \cite{cristian2023competition} provide first-order conditions for solving $p_k^\ast$ and $p_k^\text{C}$. Similarly, Propositions 3.3 and 3.5 in the same work provides sufficient conditions for the existence and uniqueness of symmetric CNE and CE equilibria. The symmetric CNE and CE individual platform profits are respectively defined by
\begin{equation}
    \label{one_stage_eq_profits}
        \pi^* := \pi^{(i)}(p_b^*, p_s^*) \  \textnormal{ and }\
        \pi^\text{C} := \frac{\Pi_{tot}(p_b^\text{C},p_s^\text{C})}{N}.
\end{equation}

\subsection{The Infinite Repeated Game}
\label{sec:model2}
The infinite repeated game consists of a sequence of games, where at time $t\in \sN\cup\{0\}$, platforms and users interact following the rules of the baseline platform competition game, introduced in Section \ref{sec:model1}, and additional ones. At each time step $t$, we use the same notations as above, but with a subscript $t$.
We assume that users on all sides are \textit{myopic}, i.e., they make decisions to maximize the utility at current time $t$ by solving \eqref{xki} which depends solely on the current prices observed in the market. 
We further assume that platforms compete and act strategically and determine the charged prices to maximize the total discounted future rewards at every step $t$ based on the past market states, which we clarify next after introducing some notation and definitions. 
Given a discounting rate $\delta\in(0, 1)$, we define the total discounted future rewards at time $t$ for platform $i$ by
\begin{equation}
    r_t^{(i)} := \sum_{\tau=0}^\infty \delta^{\tau} \sE [\pi_{t+\tau}^{(i)}], \  \textnormal{ where } \ \pi_{t+\tau}^{(i)} = \sum_{k\in\{b,s\}}x_{t+\tau,k}^{(i)}p_{t+\tau,k}^{(i)}
    \label{eqn:append:def_reward} 
\end{equation}
and $x_{t+\tau,k}^{(i)}$ is the mass of users on side $k$ joining platform $i$ at time $t+\tau$, and $p_{t+\tau,k}^{(i)}$ is the price that platform $i$ charges on side $k$ at time $t+\tau$. 
Note that from \eqref{xki}, $x_{t+\tau,k}^{(i)}$ is a function of all platforms prices at time $t+\tau$. Furthermore, this observation and \eqref{pi} imply that  $\pi_{t+\tau}^{(i)}$ can be written as a function of all platforms prices at time $t+\tau$, that is,
\begin{equation}
        \pi_{t+\tau}^{(i)} = \pi^{(i)}(\vp_{t+\tau}^{(1)},\vp_{t+\tau}^{(2)},\cdots, \vp_{t+\tau}^{(N)}).\label{eqn:reward_on_price}
\end{equation}
From the viewpoint of platform $i$, the policies of all other platforms are unknown, so their present and future prices are considered random variables.\footnote{For $\tau = 0$, the 
market share for platform $i$, $x_{t,k}^{(i)}$, is a random variable since it depends on all the prices charged by all other platforms (see \eqref{xki}).
For $\tau > 0$, the future policies of platforms are random variables. Since the future prices depend on the future states, which depend on the platform policies, they are also random variables.}

Each platform needs to strategically charge prices in order to maximize the expected total discounted future rewards \eqref{eqn:append:def_reward}. 
A common method to optimize the expectation of the total discounted future reward is Q-learning, which we introduce in \S~\ref{subsec:reinforce}. 

\begin{figure}[H]
    \centering
    \input{d1}
    \caption{Demonstration of the infinite repeated game framework}
    \label{fig:gameDiagram}
\end{figure}
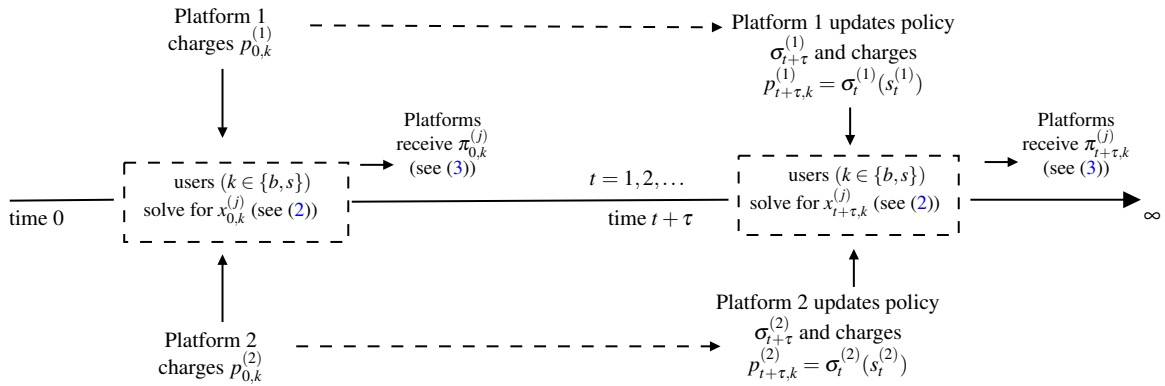

For $t\in \sN\cup\{0\}$, denote by  $\vp_t := (\vp_{t}^{(1)},\dots,\vp_{t}^{(N)})$ the vector of prices chosen by the $N$ platforms at time $t$, where $\vp_{t}^{(i)}:=(p_{t,b}^{(i)}, p_{t,s}^{(i)})$, $i\in[N]$. For $L\geq 1$, denoting previous time steps, and $t \geq L$, let 
$$\vs_{t,L}:=(\vp_{t-L}, \vp_{t-L+1}, \dots, \vp_{t-1}) \ \text{ and } \
H_{t,L} = \{\vs_{t,L}\in \mathbb{R}^{2LN} \},$$
where one typically constrains $H_{t,L}$ to be a discrete set (see Section \ref{subsec:problem}). 
The problem for each platform is to identify a policy $$\sigma_t^{(i)}: H_{t, L} \longrightarrow \sR^{2}$$ that inputs the current observed state $\vs_{t, L}$ and outputs the charged price $\vp_t^{(i)} \equiv (p_{t,b}^{(i)}, p_{t,s}^{(i)})$. 
During this infinitely repeated game, at each time step $t$, each platform $i$ updates the policy $\sigma_t^{(i)}$ based on the observed data (the states and rewards) to refine this policy that aims to maximize the expected total discounted future reward. Moreover, at each time step $t$, each platform $i$ uses the policy $\sigma_t^{(i)}$ to determine the charged prices $\vp^{(i)}_t$. A particular framework for doing this is discussed in Section \ref{sec:MARL}. 
Figure~\ref{fig:gameDiagram} demonstrates the different stages of the infinite repeated game. 

%% file: d1.tex
\tikzset{every picture/.style={line width=0.75pt}} 

\begin{tikzpicture}[x=0.75pt,y=0.75pt,yscale=-1,xscale=1]

\draw [line width=0.75]    (8.18,140.49) -- (60.67,140) ;
\draw [line width=0.75]    (180.67,141) -- (370.67,140) ;
\draw [line width=0.75]    (492.67,140) -- (575.67,140) ;
\draw [shift={(578.67,140)}, rotate = 180] [fill={rgb, 255:red, 0; green, 0; blue, 0 }  ][line width=0.08]  [draw opacity=0] (8.93,-4.29) -- (0,0) -- (8.93,4.29) -- cycle    ;
\draw    (115.48,73.96) -- (115.48,105.84) ;
\draw [shift={(115.48,108.84)}, rotate = 270] [fill={rgb, 255:red, 0; green, 0; blue, 0 }  ][line width=0.08]  [draw opacity=0] (5.36,-2.57) -- (0,0) -- (5.36,2.57) -- cycle    ;
\draw    (115.48,201.7) -- (115.48,170.52) ;
\draw [shift={(115.48,167.52)}, rotate = 90] [fill={rgb, 255:red, 0; green, 0; blue, 0 }  ][line width=0.08]  [draw opacity=0] (5.36,-2.57) -- (0,0) -- (5.36,2.57) -- cycle    ;
\draw    (184.91,122.39) -- (198.88,122.39) ;
\draw [shift={(201.88,122.39)}, rotate = 180] [fill={rgb, 255:red, 0; green, 0; blue, 0 }  ][line width=0.08]  [draw opacity=0] (5.36,-2.57) -- (0,0) -- (5.36,2.57) -- cycle    ;
\draw    (499.85,120.99) -- (513.82,120.99) ;
\draw [shift={(516.82,120.99)}, rotate = 180] [fill={rgb, 255:red, 0; green, 0; blue, 0 }  ][line width=0.08]  [draw opacity=0] (5.36,-2.57) -- (0,0) -- (5.36,2.57) -- cycle    ;
\draw    (432.02,93.44) -- (432.02,109.67) ;
\draw [shift={(432.02,112.67)}, rotate = 270] [fill={rgb, 255:red, 0; green, 0; blue, 0 }  ][line width=0.08]  [draw opacity=0] (5.36,-2.57) -- (0,0) -- (5.36,2.57) -- cycle    ;
\draw    (433.89,184.02) -- (433.89,167.8) ;
\draw [shift={(433.89,164.8)}, rotate = 90] [fill={rgb, 255:red, 0; green, 0; blue, 0 }  ][line width=0.08]  [draw opacity=0] (5.36,-2.57) -- (0,0) -- (5.36,2.57) -- cycle    ;
\draw  [dash pattern={on 4.5pt off 4.5pt}]  (159.67,52) -- (364.67,52.99) ;
\draw [shift={(367.67,53)}, rotate = 180.28] [fill={rgb, 255:red, 0; green, 0; blue, 0 }  ][line width=0.08]  [draw opacity=0] (5.36,-2.57) -- (0,0) -- (5.36,2.57) -- cycle    ;
\draw  [dash pattern={on 4.5pt off 4.5pt}]  (152.67,214) -- (363.67,214) ;
\draw [shift={(366.67,214)}, rotate = 180] [fill={rgb, 255:red, 0; green, 0; blue, 0 }  ][line width=0.08]  [draw opacity=0] (5.36,-2.57) -- (0,0) -- (5.36,2.57) -- cycle    ;

\draw (87,42) node [anchor=north west][inner sep=0.75pt]  [font=\small,xscale=0.75,yscale=0.75] [align=left] {\begin{minipage}[lt]{52.91pt}\setlength\topsep{0pt}
\begin{center}
Platform 1 charges $\displaystyle p_{0,k}^{( 1)}$
\end{center}

\end{minipage}};
\draw (82,205.95) node [anchor=north west][inner sep=0.75pt]  [font=\small,xscale=0.75,yscale=0.75] [align=left] {\begin{minipage}[lt]{52.91pt}\setlength\topsep{0pt}
\begin{center}
Platform 2 charges $\displaystyle p_{0,k}^{( 2)}$
\end{center}

\end{minipage}};
\draw (6.69,144.3) node [anchor=north west][inner sep=0.75pt]  [font=\small,xscale=0.75,yscale=0.75] [align=left] {{\small time $\displaystyle 0$}};
\draw  [dash pattern={on 4.5pt off 4.5pt}]  (65.91,121.31) -- (178.91,121.31) -- (178.91,161.31) -- (65.91,161.31) -- cycle  ;
\draw (68.91,125.31) node [anchor=north west][inner sep=0.75pt]  [font=\footnotesize,xscale=0.75,yscale=0.75] [align=left] {\begin{minipage}[lt]{99pt}\setlength\topsep{0pt}
\begin{center}
 $\quad$ users $\displaystyle ( k\in \{b,s\})$  solve for $\displaystyle x_{0,k}^{( j)}$ (see (\ref{xki}))
\end{center}
\end{minipage}};
\draw (198,95) node [anchor=north west][inner sep=0.75pt]  [font=\footnotesize,xscale=0.75,yscale=0.75] [align=left] {\begin{minipage}[lt]{55pt}\setlength\topsep{0pt}
\begin{center}
Platforms receive $\displaystyle \pi _{0,k}^{(j)}$ (see (\ref{pi}))
\end{center}

\end{minipage}};
\draw (299.75,124.32) node [anchor=north west][inner sep=0.75pt]  [font=\small,xscale=0.75,yscale=0.75] [align=left] {$\displaystyle t=1,2,\dotsc $};
\draw (308.8,144.26) node [anchor=north west][inner sep=0.75pt]  [font=\small,xscale=0.75,yscale=0.75] [align=left] {time $\displaystyle t+\tau $};
\draw  [dash pattern={on 4.5pt off 4.5pt}]  (375.78,117.8) -- (486.78,117.8) -- (486.78,157.8) -- (375.78,157.8) -- cycle  ;
\draw (378.78,121.8) node [anchor=north west][inner sep=0.75pt]  [font=\footnotesize,xscale=0.75,yscale=0.75] [align=left] {\begin{minipage}[lt]{99pt}\setlength\topsep{0pt}
\begin{center}
$\quad$ users $\displaystyle ( k\in \{b,s\})$ solve for $\displaystyle x_{t+\tau,k}^{( j)}$ (see (\ref{xki}))
\end{center}

\end{minipage}};
\draw (370,45) node [anchor=north west][inner sep=0.75pt]  [font=\small,xscale=0.75,yscale=0.75] [align=left] {\begin{minipage}[lt]{115pt}\setlength\topsep{0pt}
\begin{center}
Platform 1 updates policy $\displaystyle \sigma_{t+\tau}^{(1)}$ and charges $\displaystyle p_{t+\tau,k}^{( 1)} =\sigma_{t}^{(1)}(s_{t}^{(1)})$
\end{center}

\end{minipage}};
\draw (362.15,186.62) node [anchor=north west][inner sep=0.75pt]  [font=\small,xscale=0.75,yscale=0.75] [align=left] {\begin{minipage}[lt]{115pt}\setlength\topsep{0pt}
\begin{center}
Platform 2 updates policy $\displaystyle \sigma_{t+\tau}^{(2)}$ and charges $\displaystyle p_{t+\tau,k}^{( 2)} =\sigma_{t}^{(2)}(s_{t}^{( 2)})$
\end{center}

\end{minipage}};
\draw (515,95) node [anchor=north west][inner sep=0.75pt]  [font=\footnotesize,xscale=0.75,yscale=0.75] [align=left] {\begin{minipage}[lt]{60pt}\setlength\topsep{0pt}
\begin{center}
Platforms receive  $\displaystyle \pi _{t+\tau,k}^{(j)}$ (see (\ref{pi}))
\end{center}

\end{minipage}};
\draw (578.97,145.48) node [anchor=north west][inner sep=0.75pt]  [xscale=0.75,yscale=0.75] [align=left] {$\displaystyle \infty $};

\end{tikzpicture}

%% file: reinforcement_learning.tex
We first review the framework of multi-agent reinforcement learning in  Section~\ref{subsec:reinforce}. We then detail our simulation setting in Section~\ref{subsec:problem}, building upon the framework developed in Section~\ref{subsec:reinforce}. 

\subsection{Preliminaries: Multi-agent Reinforcement Learning}\label{subsec:reinforce}
Multi-agent reinforcement learning considers $N$ agents interacting in a dynamic environment. At each time $t\in \sN$, each agent $i\in[N]$ observes a state $s^{(i)}_t \in \gS$ and takes an action $a_t^{(i)}\in \gA$, based on this observed state and following a policy $\sigma^{(i)}_t: \gS \longrightarrow \gA$, which could be either deterministic or stochastic. Here, $\gS$ denotes the state space and $\gA$ denotes the action space.
Let
$\sigma_t = (\sigma^{(1)}_t, \sigma^{(2)}_t, \cdots, \sigma^{(N)}_t)$ and $A_t := (a_t^{(1)}, \cdots a_t^{(N)}) = (\sigma_t^{(1)}(s_t^{(1)}), \cdots \sigma_t^{(N)}(s_t^{(N)}))$ denote 
all policies and actions, respectively, at time $t$. 
We denote by $a_t^{(-i)}$,  $\vp_t^{(-i)}$, 
and $\sigma^{(-i)}_t$ 
the respective vectors of all actions $a_t^{(j)}$,  prices $\vp_t^{(j)}$,  
and policies $\sigma^{(j)}_t$with $j\neq i$.  The agent collects a reward $\pi_t^{(i)}$, which is a random variable conditioned on the state $s_t^{(i)}$ and actions $A_t$.  The state in the next time, $s_{t+1}^{(i)}$, is a random variable conditioned on the state $s_t^{(i)}$ and the actions $A_t$ taken by all the agents in the current time $t$. 
Given a discounting rate $\delta\in(0,1)$, at each time, each agent aims to find a policy in order to maximize the following expectation of the total discounted future reward given all observed states at time $t$:
\begin{equation}
    \sum_{\tau=0}^\infty\sE_{\pi, s, \sigma} \left[ \delta^\tau\pi_{t+\tau}^{(i)} (s_{t+\tau}^{(i)}, A_{t+\tau})\right].
    \label{eqn:future_reward}
\end{equation}
The expectation is needed due to the randomness in the rewards, the future states, and the future actions of all the agents. 

Q-learning is a classic method for finding the policy that maximizes \eqref{eqn:future_reward}. It uses the $Q$-function of agent $i$ at state $s$ given an action $a$, which is defined by
\begin{equation}\label{eqn:Q_future_reward}
    \begin{split}
         &Q^{(i)}(s, a, \sigma^{(i)}; \sigma^{(-i)}) :=\\
         & \sum_{\tau=0}^\infty \sE_{\pi, s, \sigma} \left[\delta^\tau\pi_{t+\tau}^{(i)} \big| s_t^{(i)}=\vs, a_t^{(i)} = a, a_{t+u}^{(i)} = \sigma^{(i)}(s_{t+u}^{(i)}), \ u \geq 1, \ a_{t+v}^{(-i)} = \sigma^{(-i)}(s_{t+v}), \ v \geq 0\right].
    \end{split}
\end{equation}
Note that \eqref{eqn:Q_future_reward} differs from \eqref{eqn:future_reward} by having agent $i$ follow the given action $a$ at time $t$ instead of the policy $\sigma^{(i)}$, whereas in both formulations all other agents at times $t, t+1, \ldots,$ and agent $i$ at times $t+1,t+2,\ldots,$ follow their policies. 

We denote an optimal policy for agent $i$ by $\sigma^{(i) \ast }$, which is hard to find.
Q-learning overcomes this difficulty by 
carefully estimating the solution $Q^{(i)\ast}(s_t, a_t; \sigma^{(-i)})$ to the  
following Bellman equation
\begin{equation}
    Q^{(i)\ast}(s_t, a_t; \sigma^{(-i)}) = \sE_\pi \left[\pi(s_t, A_t)\right] + \delta  \max_{a'} \sE_{s_{t+1}} \left[Q^{(i)\ast}(s_{t+1}, a'; \sigma^{(-i)})|A_t\right].\label{eqn:bellman}
\end{equation}
It then estimates $\sigma^{(i)\ast}$ using the following relationship between $\sigma^{(i)\ast}$ and  $Q^{(i)\ast} (x, a; \sigma^{(-i)})$:
\begin{equation}
\sigma^{(i) \ast}(s) = \argmax_a Q^{(i) \ast} (s, a; \sigma^{(-i)}).
\label{eqn:pi_ast_Q_ast}
\end{equation}
We detail the methods for estimating the $Q^\ast$-function in \eqref{eqn:bellman} in the following section.

\subsection{The Simulation Setup}
\label{subsec:problem}

We consider a market with two platforms, that is, we set $N=2$.\footnote{Since our model assumes  two sides of the market, each of the $N$ platforms must choose two prices. At each stage, our simulation estimates $2N$ different prices, and there are $M^{2N}$ possibilities for the vector of prices, where $M$ is the size of the set of price choices available to each platform. To make our simulations feasible, we choose $N=2$.} At time $t$, each platform $i\in\{1, 2\}$ observes the following state $\vs_t^{(i)}:=\vp_{t-1}$, which contains prices at the previous step.  Platform $i$ determines its prices $\vp^{(i)}_t$ based on the observed state $\vs_t^{(i)}$. 
At each time $t$, after all platforms have chosen prices $\vp_t^{(i)}$, they receive the reward $\pi_t^{(i)} = \sum_{k\in\{b,s\}} x_{t, k}^{(i)}p_{t, k}^{(i)}$, where $x_{t, k}^{(i)}$ is solved using \eqref{xki}. 

To simplify the computation, we allow platforms to choose from a discrete set of $M$ prices. 
While it is common to expect that $p_k^\ast < p_k^\text{C}$,\footnote{Note that Proposition 4.11 in \cite{cristian2023competition} provides sufficient conditions to guarantee that  $p^\ast_k < p^\text{C}_k$.} our model also allows the case $p^\text{C}_k < p^*_k$. 
We further introduce the parameter $\epsilon=0.1$ so the lowest price is slightly lower than $\min(p_k^*,p^\text{C}_k)$ and the highest one is slightly higher than $\max(p_k^*,p^\text{C}_k)$. 
For each $k\in \{b,s\}$, our set of prices is \begin{align}\label{def:Pkset}
\gP_k := \Big\{p_k^* - \epsilon (p_k^\text{C} - p_k^*) + \frac{j}{M-1}(1+2\epsilon)(p_k^\text{C} - p_k^*) \Big| j=0,\cdots, M-1 \Big\}.
\end{align}
Note that the cardinality of the price space $|\gP_k|$ is $M$ for both $k=b$ and $k=s$. The overall state space (for both platforms) and the action space for each platform are respectively defined by 
\begin{equation}
        \gS := (\gP_b\times \gP_s) \times (\gP_b\times \gP_s) \ \textnormal{ and} \
        \gA := \gP_b\times \gP_s.
        \label{eqn:def:gA_gS}
\end{equation}
We note that the size of the state space is $|\gS| = M^{4}$ and the size of the action space is $|\gA| = M^2$.

\textbf{Platform policy:} 
We denote the estimation at time $t$ of $Q^{(i)\ast}(\vs, \va;\sigma^{(-i)})$ by $Q^{(i)}_t(\vs, \va)$, where $ \vs\in\gS$, $\va\in\gA$ and $i \in \{1,2\}$ indexes the platform. Q-learning alternately estimates $Q_t^{(i)}$ and the stochastic policies at time $t$. We first assume that $Q_t^{(i)}$ is known and show how the platforms determine the stochastic policy at time $t$. We then explain the Q-learning estimation of $Q_t^{(i)}$. 
Instead of directly computing the policy as the maximum value in \eqref{eqn:pi_ast_Q_ast}, Q-learning computes a softmax value using a temperature parameter $\gT_t$. 
For this purpose, at time $t$ and given a state $\vs_t^{(i)}\in \gS$ and the $Q^\ast$-function estimate, $Q^{(i)}_t$, the policy of platform $i$ is the Boltzmann probability distribution: 
\begin{equation}
    P(\va_t^{(i)} = \va|\vs_t^{(i)})={\exp\left(Q_t^{(i)}(\vs_t^{(i)}, \va)/\gT_t\right)}/{\sum\limits_{\va'\in\gA}\exp\left(Q_t^{(i)}(\vs_t^{(i)}, \va')/\gT_t\right)}.\label{eqn:policy}
\end{equation}
We remark that all platforms independently determine their prices  based on \eqref{eqn:policy}.\footnote{As $\gT_t$ decreases, \eqref{eqn:policy} increasingly focuses on the optimal action based on $Q_t^{(i)}$. When $\gT_t \to 0$, the policy randomly selects between the actions that yield the maximal reward $Q_t^{(t)}$ with uniform probabilities. In the simulation, we set $\gT_0 = 1000$ to encourage exploration of possible actions, gradually decreasing it towards $0$ to exploit optimal actions.} 

\textbf{Q-learning estimation:} 
At each time step, after determining the price following \eqref{eqn:policy}, platform $i$ collects the reward $\pi_t^{(i)}$ defined by \eqref{eqn:reward_on_price}.
Next, platform $i$ updates the estimated values of the $Q^*$-function at the given state $\vs_t^{(i)}$ and the selected action $\va_t^{(i)}$ with a learning rate $\alpha$ as follows:
\begin{equation}
    Q^{(i)}_{t+1}(\vs_t^{(i)}, \va_t^{(i)}) := (1-\alpha) Q^{(i)}_t(\vs_t^{(i)}, \va_t^{(i)}) + \alpha \left(\pi_t^{(i)} + \delta \max_{\va} Q^{(i)}_t(\vs_t^{(i)}, \va)\right).\label{eqn:bellman_discrete}
\end{equation}
We remark that \eqref{eqn:bellman_discrete} is an approximation of \eqref{eqn:bellman} (see \cite{watkins1992q}).

We initialize the $Q^\ast$-function at $\vs\in\gS$ and $\va\in\gA$ assuming that in all future states platform $i$ charges $\va$ and all other platforms charge the prices in $\vs$. Therefore, for platform $i$, state $\vs$, a given action $\va$ and the price vector for platform $j\neq i$, which we denote by $\vp^{(j)}$ and it is part of the state $\vs$, 
the $Q^\ast$-function for platform $i$ is initialized by
\begin{equation}
    Q^{(i)}_0(\vs, \va) = \sum_{\tau=0}^\infty \delta^\tau \pi^{(i)}(\va,\vp^{(j)}) = \frac{\pi^{(i)}(\va,\vp^{(j)})}{1-\delta}.
    \label{eqn:init}
\end{equation}

\textbf{Parameter setup:} We choose exponentially decaying temperature parameter  $\gT_t = \gT_0 \lambda^t$ with $\gT_0:=1000/(1-\delta)$ and $\lambda=1-10^{-7}$. This choice encourages exploration in the early stages and exploits optimality in the later stages.
We choose both the idiosyncratic preference parameters and the outside option utilities to be the same on both sides of the market. Therefore we denote $\beta_k = \beta_b \equiv\beta_s$ and $u_k^{(0)} = u_b^{(0)}\equiv u_s^{(0)}$. 
We set the learning rate  $\alpha=0.15$, discount rate $\delta=0.05$, idiosyncratic preferences $\beta_k=1$, and  outside option utility $u_k^{(0)}=-2$ for each $k\in\{b,s\}$. We choose a small value for $\delta$, compared to the choice of the same parameter in \cite{calvano2020artificial,klein2021autonomous}, to emphasize that in our setting collusion is already present with a very small discount rate. 

\textbf{Reporting metric:}
We define the collusive level of platform $i$ at time $t$ as
\begin{equation}
    \Delta^{(i)}_t := \frac{\pi_t^{(i)} - \pi^*}{\pi^\text{C} -\pi^*},\label{eqn:def_Delta}
\end{equation}
where we recall (see   \eqref{one_stage_eq_profits})  $\pi^*$ and $\pi^\text{C}$ respectively denote the CNE and CE equilibrium profits of the baseline platform competition game.  When $\Delta_t^{(i)}=0$, platform $i$'s reward at time $t$ equals the CNE level, $\pi^*$; whereas when $\Delta_t^{(i)}=1$, it equals the CE level, $\pi^\text{C}$. Each simulation runs $T=5\times 10^8$ iterations and we report the overall collusive level in the last $K=1,000$ steps as follows:
\begin{equation}\label{eqn:delta_metric}
    \tilde{\Delta} := \frac{1}{K N}\sum_{s=0}^{K-1} \sum_{i=1}^N\Delta^{(i)}_{T-s}.
\end{equation}

%% file: result.tex
We report extensive numerical experiments using the setup of Section \ref{subsec:problem}.
Section \ref{subsection:phi_relationship_Delta}   investigates the general dependence of $\tilde{\Delta}$, defined in \eqref{eqn:delta_metric}, on the externality matrix $\Phi$.  
Section \ref{subsect:collusion_networkExternalities} 
further explores the latter dependence for concrete and useful choices of $\Phi$. 
Section \ref{subsect:colllusive_idiosyncratic} studies the dependence of $\tilde{\Delta}$ on the degree of heterogeneity in users' tastes, the outside option utility, and the discount rate.  
Lastly, Section~\ref{subsec:assymetric_collusion} explores two interesting scenarios: 1) long-run asymmetric equilibria outperform the symmetric equilibrium, and 2) competition prices are larger than collusion prices for one side of the market. 
A supplementary analysis in Appendix~\ref{appd:sensitive}  implies that our numerical results are consistent with platforms learning tacit collusion and equilibrium strategies.

\subsection{Dependence of the Collusive Level on the Network Externalities}
\label{subsection:phi_relationship_Delta}
We applied an additive model to infer the dependence of the collusive level, $\tilde{\Delta}$, on the externality matrix, $\Phi$. We ran 2,500 simulations according to the setting described in Section \ref{subsec:problem}. For each simulation, we randomly sampled the elements of the externality matrix $\Phi$ from independent normalized Gaussians (that is, $\phi_{kl}\sim N(0, 1)$ for $k, l\in\{b, s\}$), and recorded the final collusive level, $\tilde{\Delta}$. In order to infer the dependence of $\tilde{\Delta}$ on $\Phi$, we assume the following additive model:
\begin{equation}
\begin{split}
    \tilde{\Delta}(\Phi) &= \Delta_0 + f_{bb}(\phi_{bb}) + f_{ss}(\phi_{ss}) + f_{bs}(\phi_{bs}) + f_{sb}(\phi_{sb}) + f_{bb,ss}(\phi_{bb}, \phi_{ss}) + f_{sb,bs}(\phi_{sb}, \phi_{bs})  \\
    & + f_{bb,bs}(\phi_{bb}, \phi_{bs}) + f_{bb,sb}(\phi_{bb}, \phi_{sb})+ f_{ss,bs}(\phi_{ss}, \phi_{bs}) + f_{ss,sb}(\phi_{ss}, \phi_{sb}) + \epsilon,
\end{split}\label{eqn:fitting_Delta}
\end{equation}
where $\Delta_0$ is the sample mean of $\tilde{\Delta}$, the next 4  functions ($f_{bb}$, $f_{ss}$, $f_{bs}$, $f_{sb}$) represent the  univariate effects of the elements of $\Phi$ on $\tilde{\Delta}$, the last 6 functions ($f_{bb,ss}$, $f_{sb,bs}$, $f_{bb,bs}$, $f_{bb,sb}$, $f_{ss,bs}$,  $f_{ss,sb}$) represent the bivariate effects of the elements of $\Phi$ on $\tilde{\Delta}$ and $\epsilon$ is an error term, encompassing higher-order multivariate effects. 
Since the equilibrium values $\pi^*$ and $\pi^C$ depend on $\Phi$ nonlinearly (see Section \ref{sec:model1}), the 10 functions,  $f_{bb}, \cdots ,f_{ss, sb}$, are  nonlinear. 
We thus sequentially fit these functions using XGBoost  \citep{chen2016xgboost}, which is a popular non-parametric, nonlinear fitting method. To reduce the bias of the fitted functions, we alter the order of both the first four functions and the next six functions, during the sequential fitting procedure, and average the collusive level over the different orders. 
Appendix~\ref{appendix:fitting_method} contains more details of implementing XGBoost. 

We refer to $\Delta_0$ as the baseline collusive level, whereas $\tilde{\Delta}$ is the collusive level.  
Our simulations show that $\Delta_0$ is approximately $0.3$. 
Next, we report our estimates for the univariate and bivariate effects of the elements of $\Phi$ on $\tilde{\Delta}$. 

\begin{figure}[H]
    \centering
    \includegraphics[width=0.8\linewidth]{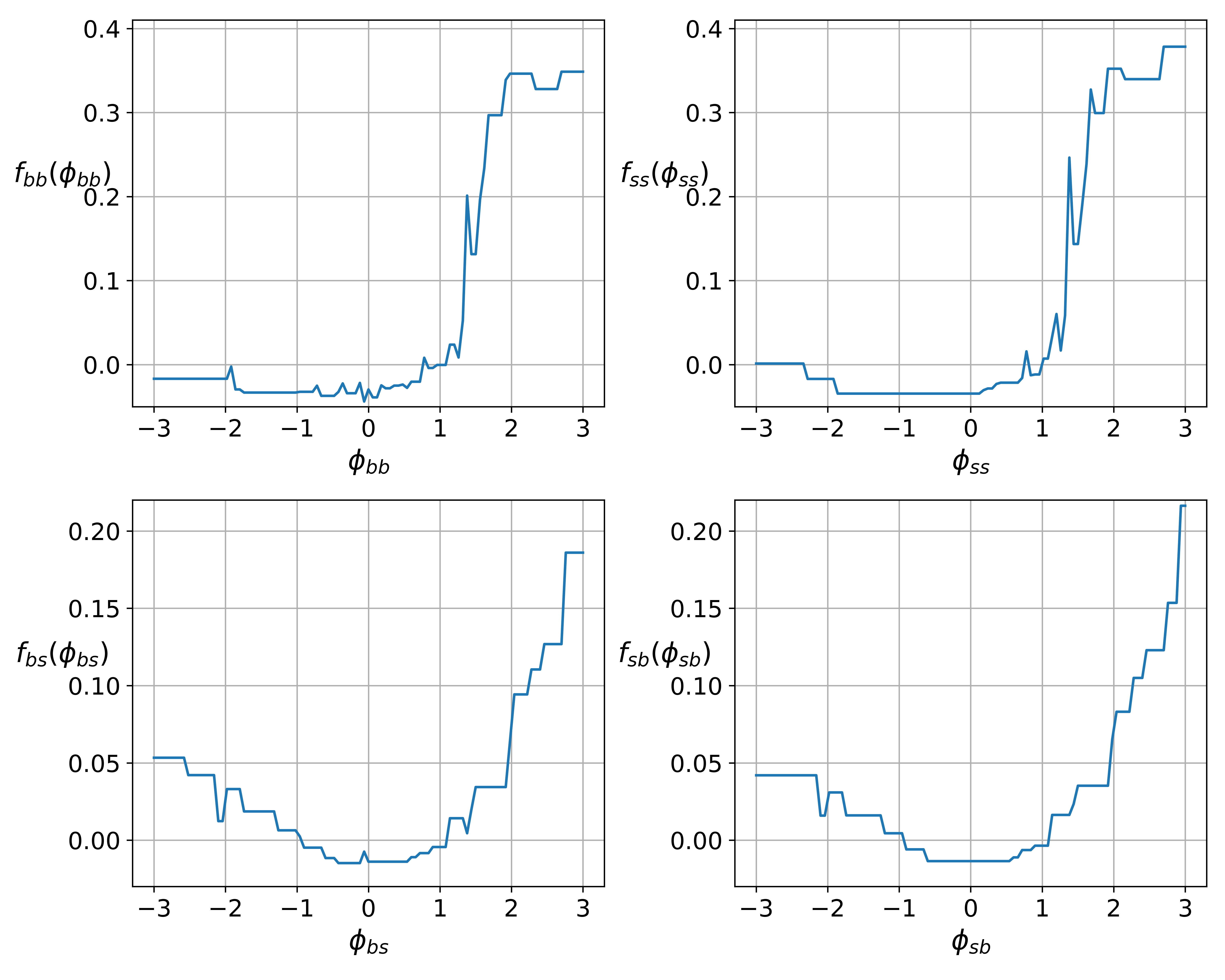}
    \caption{Demonstration of the dependence of the four fitted univariate functions on the externalities. Top left: $f_{bb}(\phi_{bb})$; Top right: $f_{ss}(\phi_{ss})$; Bottom left: $f_{bs}(\phi_{bs})$; Bottom right: $f_{sb}(\phi_{sb})$.}
    \label{fig:non_linear_main_effects}
\end{figure}

Figure~\ref{fig:non_linear_main_effects} illustrates the fitted functions $f_{bb}$, $f_{ss}$, $f_{bs}$ and $f_{sb}$, which capture the univariate effect of each entry in the externality matrix. 
The top two subfigures demonstrate the univariate effect of the within-side externalities, ($\phi_{bb}$ and $\phi_{ss}$). In this case, the  collusive level is close to zero when these externalities are less than 1, then increases sharply when these externalities increase from $1$ to $2$, and it is approximately flat when these externalities are above $2$ with a possible increase of the collusive level when the absolute values of the negative externalities increase. We remark that we cannot confidently conclude the latter increase from the current experimental results, but latter experiments in Section \ref{subsect:collusion_networkExternalities} support such an increase, especially when considering lower values of $\phi_{bb}$ and $\phi_{ss}$.    
The bottom two subfigures demonstrate the univariate effect of the cross-side externalities ($\phi_{bs}$ and $\phi_{sb}$). In this case, the dependence of the collusive level on the externalities is depicted by a J-shape function with a minimum when the externality is around zero. 
We thus note that in order to minimize the level of the algorithmic collusion, we would need to bound the values of the within-side externalities and the absolute values of the cross-side externalities. In our particular experimental setting, the desired bound is 1. 
In general, we expect 
there can be two different upper bounds for the within-side and cross-side externalities and they depend on the chosen parameters, in particular, $\{\delta,\beta_k,u_k^{(0)}\}$.

\begin{figure}[H]
    \centering
    \includegraphics[width=0.49\linewidth]{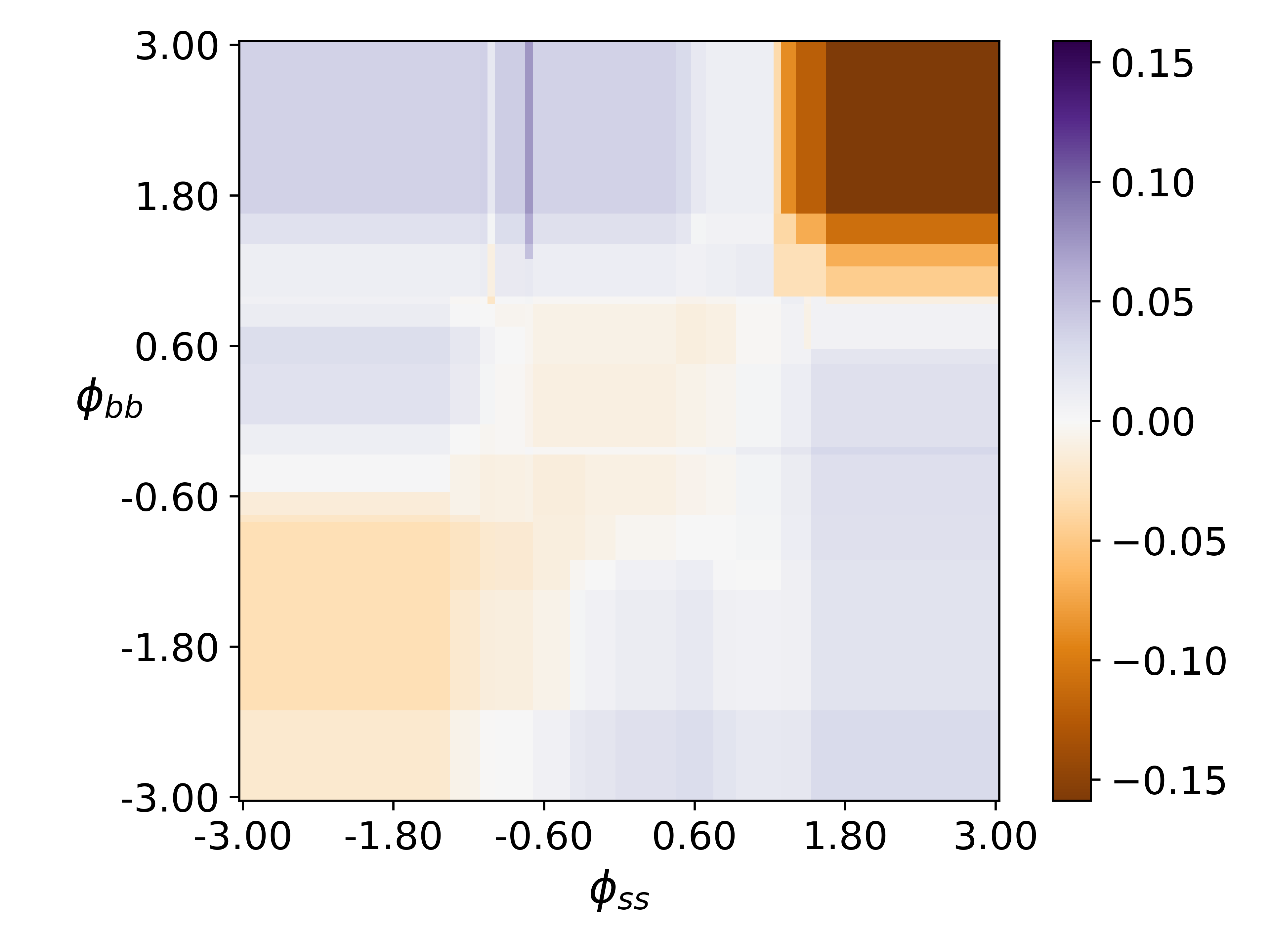}
    \includegraphics[width=0.49\linewidth]{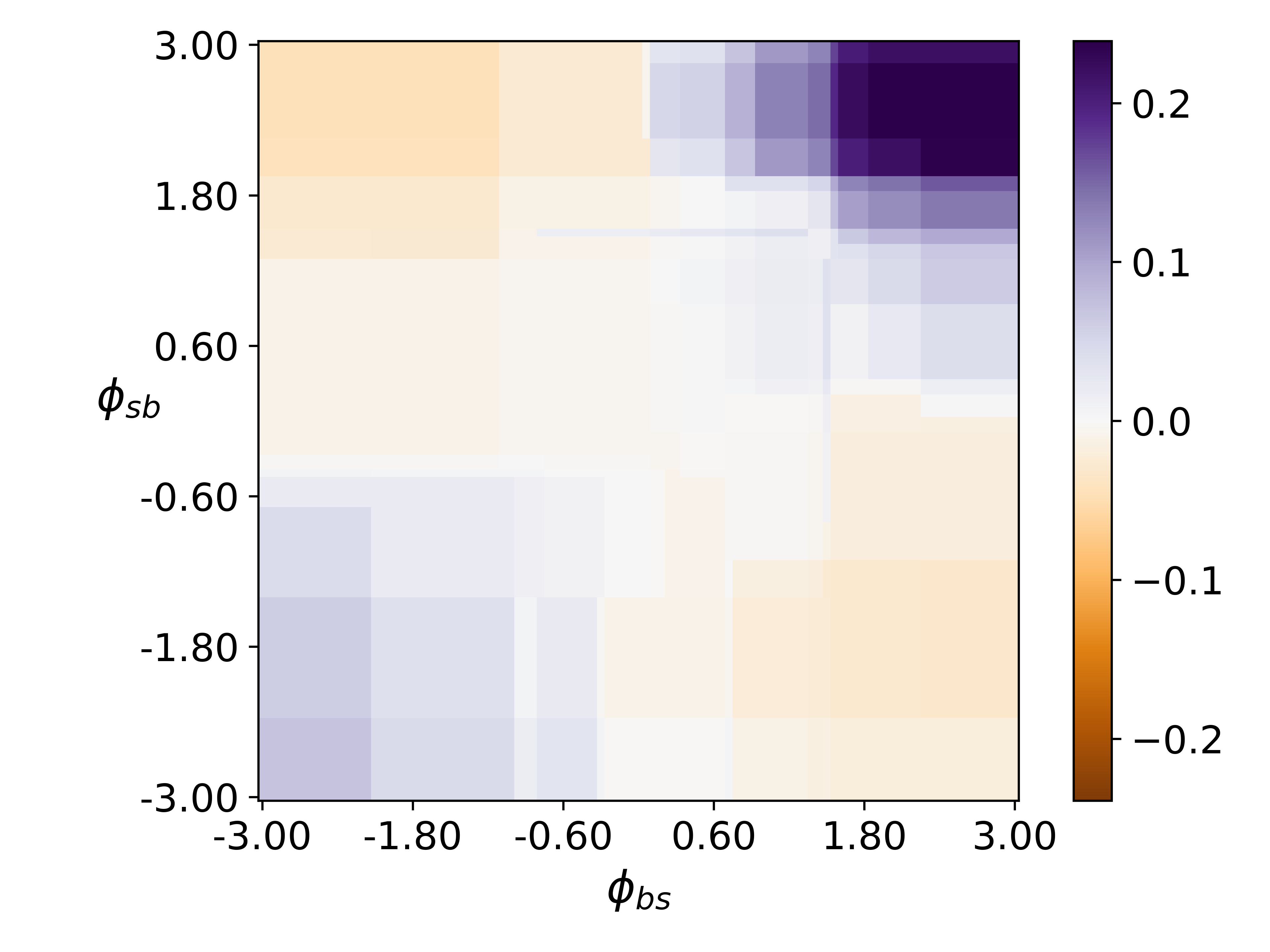}
    \caption{Heatmaps of the fitted functions $f_{bb,ss}(\phi_{bb}, \phi_{ss})$ (left) and $f_{sb,bs}(\phi_{bs}, \phi_{sb})$ (right), which capture the bivariate effect between $\phi_{bb}$ and $\phi_{ss}$, and between $\phi_{bs}$ and $\phi_{sb}$, respectively.}
    \label{fig:non_linear_bbss}
\end{figure}

Figure~\ref{fig:non_linear_bbss} demonstrates the fitted functions $f_{bb,ss}$ and $f_{
sb,bs}$, which capture the bivariate effects on collusion of the main diagonal and off-diagonal elements in the externality matrix. We present the images of these functions as heatmaps over their planar domains. For example, in the left-hand subfigure the domain is described by the within-side externality variables $\phi_{bb}$ and $\phi_{ss}$ and the collusive level is depicted by a heatmap, changing from purple (highly positive) to orange (highly negative). 
The left-hand subfigure implies that when the within-side externalities, $\phi_{bb}$ and $\phi_{ss}$, are both large, they result in the minimal value of the bivariate effect, which is negative. 
In this regime, the collusive level, resulting from both the univariate and bivariate effects, remains positive (recall that the univariate effect is demonstrated in the top subfigures of Figure~\ref{fig:non_linear_main_effects}).  
Similarly, when $\phi_{bb}$ and $\phi_{ss}$ are both sufficiently negative, their bivariate effect reduces the collusive level, albeit by a small amount. We remark that both the univariate and bivariate contributions in this case are rather small and it is hard to predict their combined effect from this experimental result, but another experiment in Section~\ref{subsect:collusion_networkExternalities} indicates that they cancel each other. 
The right-hand side subfigure indicates that the bivariate component of the cross-side externalities reduces the collusive level when these externalities are large in absolute values and have opposite signs. On the other hand, it increases the collusive level 
when the cross-side externalities have the same sign and have sufficiently large absolute values. The rate of increase is larger when they are both positive.

\begin{figure}[H]
    \centering
    \includegraphics[width=0.49\linewidth]{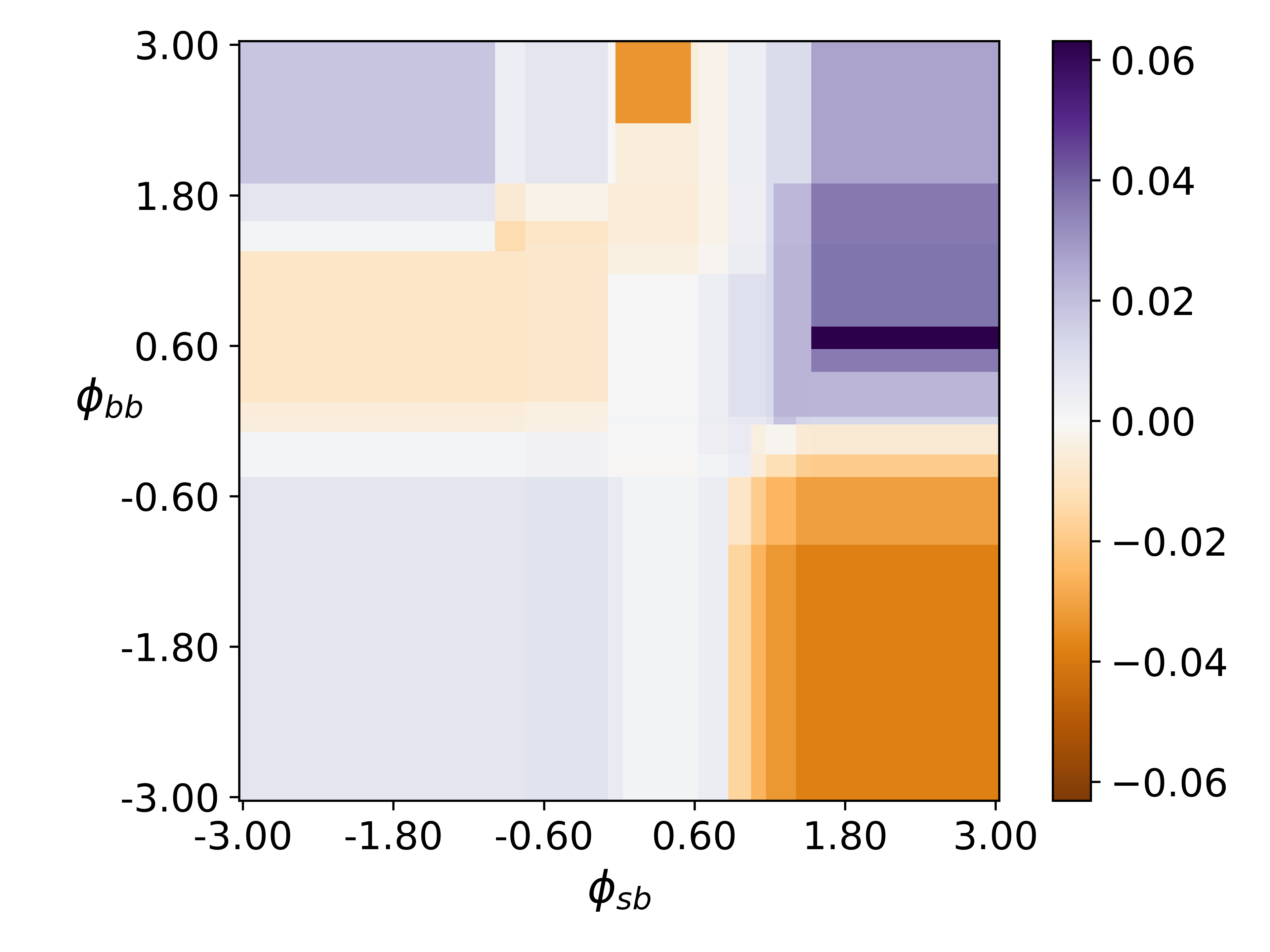}
    \includegraphics[width=0.49\linewidth]{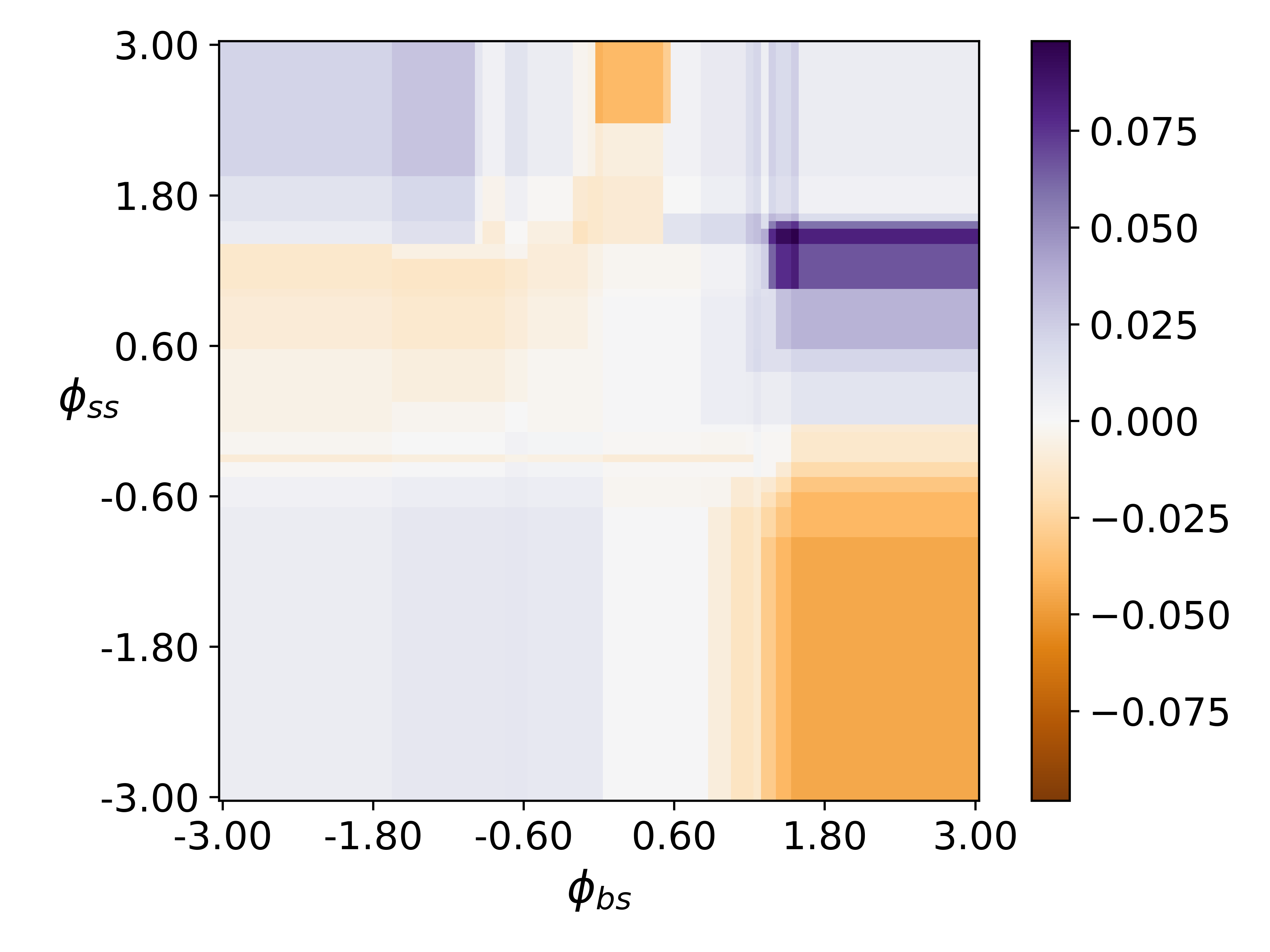}
    \caption{Heatmaps of the fitted functions $f_{bb,sb}(\phi_{bb}, \phi_{sb})$ (left) and $f_{ss,bs}(\phi_{ss}, \phi_{bs})$ (right), which capture the bivariate effect between $\phi_{bb}$ and $\phi_{sb}$, and between $\phi_{ss}$ and $\phi_{bs}$, respectively.}
    \label{fig:non_linear_bbsb_ssbs}
\end{figure}

Figure~\ref{fig:non_linear_bbsb_ssbs} demonstrates the bivariate effect on collusion when both buyers and sellers benefit from population joining the market on either side $b$ or $s$ (but not both at the same time). That is, it demonstrates the bivariate effect for $\phi_{bb}$ and $\phi_{sb}$ when considering side $b$ (left) and the bivariate effect  for $\phi_{ss}$ and $\phi_{bs}$ when considering side $s$ (right). The two subfigures are very similar and we thus only discuss the left one, with the variables $\phi_{bb}$ and $\phi_{sb}$.  
The bottom-right corner of this subfigure implies that if $\phi_{sb}$ sufficiently large and $\phi_{bb}$ is sufficiently negative, the bivariate effect on the collusive level is negative. 
In this regime, the collusive level, resulting from both the univariate and bivariate effects, remains positive (recall that the univariate effect is demonstrated in the top-left and bottom-right subfigures of Figure~\ref{fig:non_linear_main_effects}).  
On the other hand, the top-right corner in Figure~\ref{fig:non_linear_bbsb_ssbs} shows that when $\phi_{bb}$ and $\phi_{sb}$ are both large the bivariate effect on the collusive level is positive.

\begin{figure}[H]
    \centering
    \includegraphics[width=0.49\linewidth]{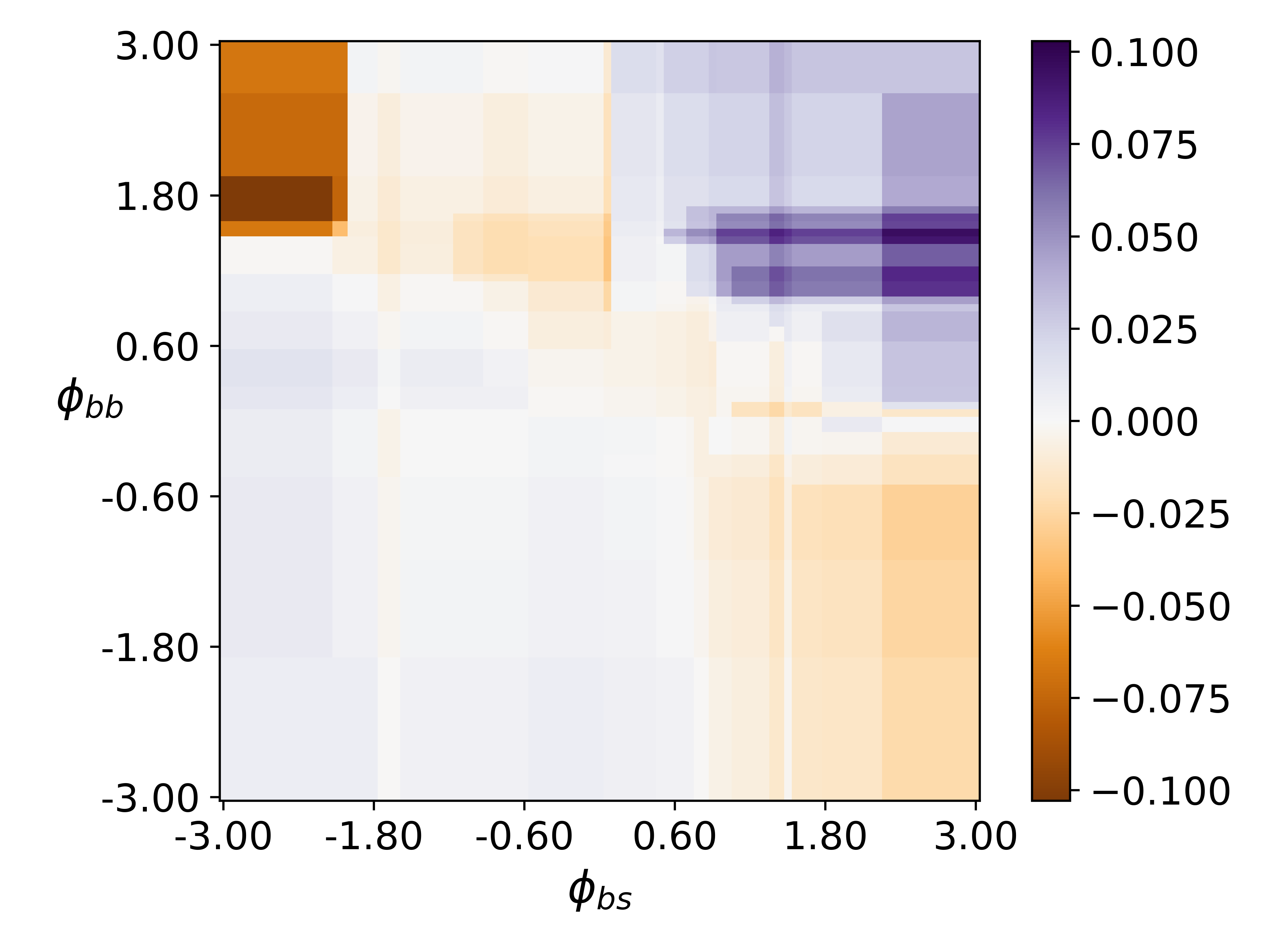}
    \includegraphics[width=0.49\linewidth]{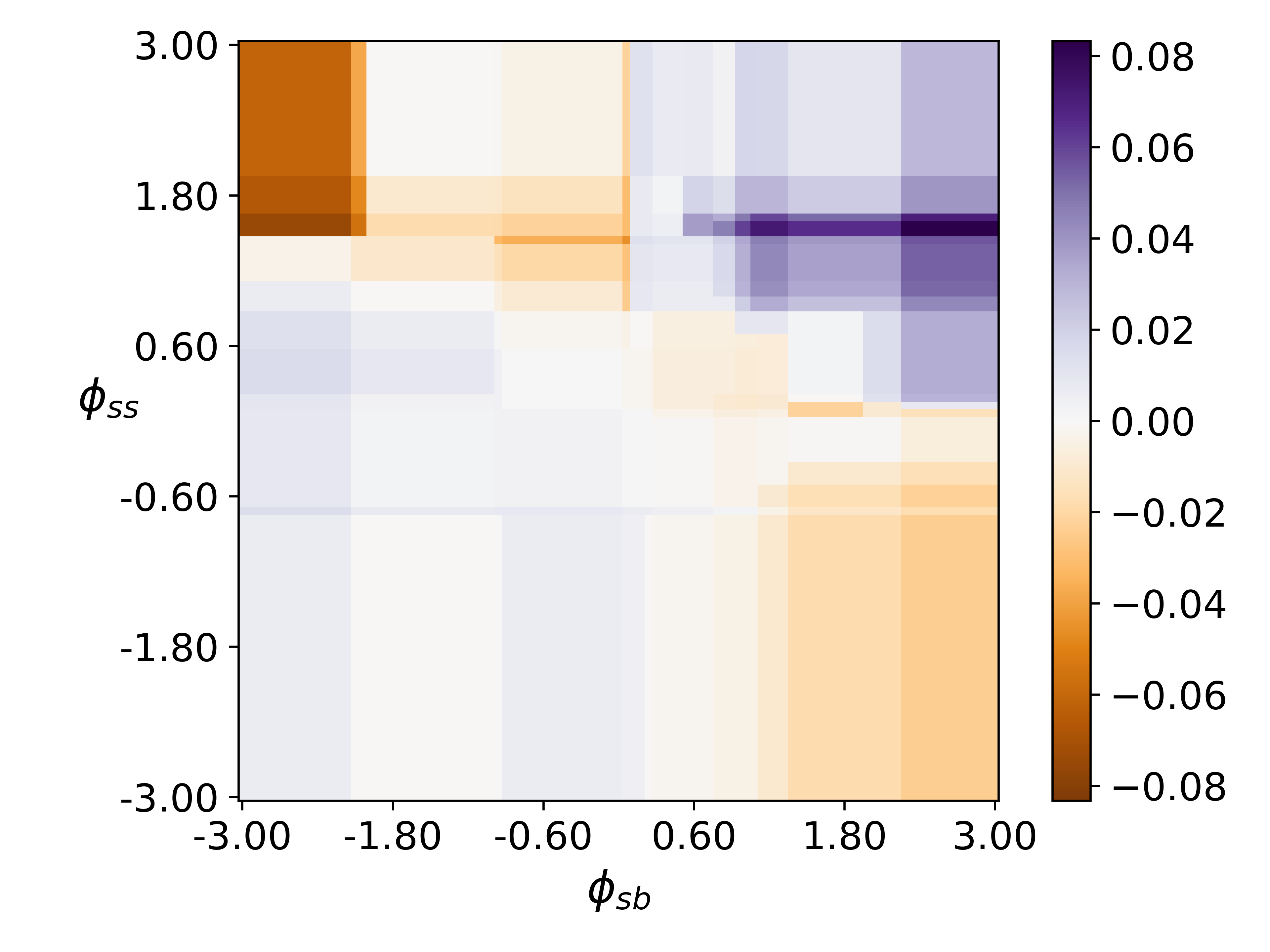}
    \caption{Heatmaps of the fitted functions $f_{bb,bs}(\phi_{bb}, \phi_{bs})$ (left) and $f_{ss,sb}(\phi_{ss}, \phi_{sb})$ (right), which capture the bivariate effect between $\phi_{bb}$ and $\phi_{bs}$, and between $\phi_{ss}$ and $\phi_{sb}$, respectively}
    \label{fig:non_linear_bbbs_sssb}
\end{figure}

Figure~\ref{fig:non_linear_bbbs_sssb} illustrates the bivariate effect on collusion when either buyers or sellers (but not both at the same time) benefit from population joining the market on sides $b$ or $s$. That is, it demonstrates the bivariate effect for $\phi_{bb}$ and $\phi_{bs}$ when considering only buyers (left subfigure) and the bivariate effect for $\phi_{ss}$ and $\phi_{sb}$ when considering only sellers (right subfigure). The two subfigures are very similar and we thus only discuss the left one, with the variables $\phi_{bb}$ and $\phi_{bs}$. We notice that when the $\phi_{bb}$ is sufficiently large and $\phi_{bs}$ is sufficiently negative, the bivariate effect on the collusion is negative. In this regime, the collusive level, resulting from both the univariate and bivariate effects, remains positive (the univariate effect is demonstrated in the left subfigures of Figure~\ref{fig:non_linear_main_effects}).
We further notice that when $\phi_{bs}$ is sufficiently large and $\phi_{bb}$ is sufficiently negative the bivariate effect on the collusion is also negative, but smaller than the latter one. Similarly, the collusive level, resulting from both the univariate and bivariate effects, remains positive. 
On the other hand, when both $\phi_{bb}$ and $\phi_{bs}$ are sufficiently large, the bivariate effect on the collusive level is positive. 

\subsection{A Study of the Collusive Level under Special Network Externalities }\label{subsect:collusion_networkExternalities}

We assume special parameterizations of the network externality matrices, $\Phi$, and explore the dependence of $\tilde{\Delta}$ on any such $\Phi$. This allows us to track more carefully the dependence of $\tilde{\Delta}$ on $\Phi$ in some special settings. For each specific $\Phi$, we ran 100 simulations. Our figures present the dependence of the overall collusive level on the elements of $\Phi$, where their main curves represent the average of the collusive levels from the 100 runs and their shaded areas represent the uncertainty level, which was computed using bootstrapping with a 99\% confidence interval.

\begin{figure}[H]
    \centering
    \includegraphics[width=0.45\linewidth]{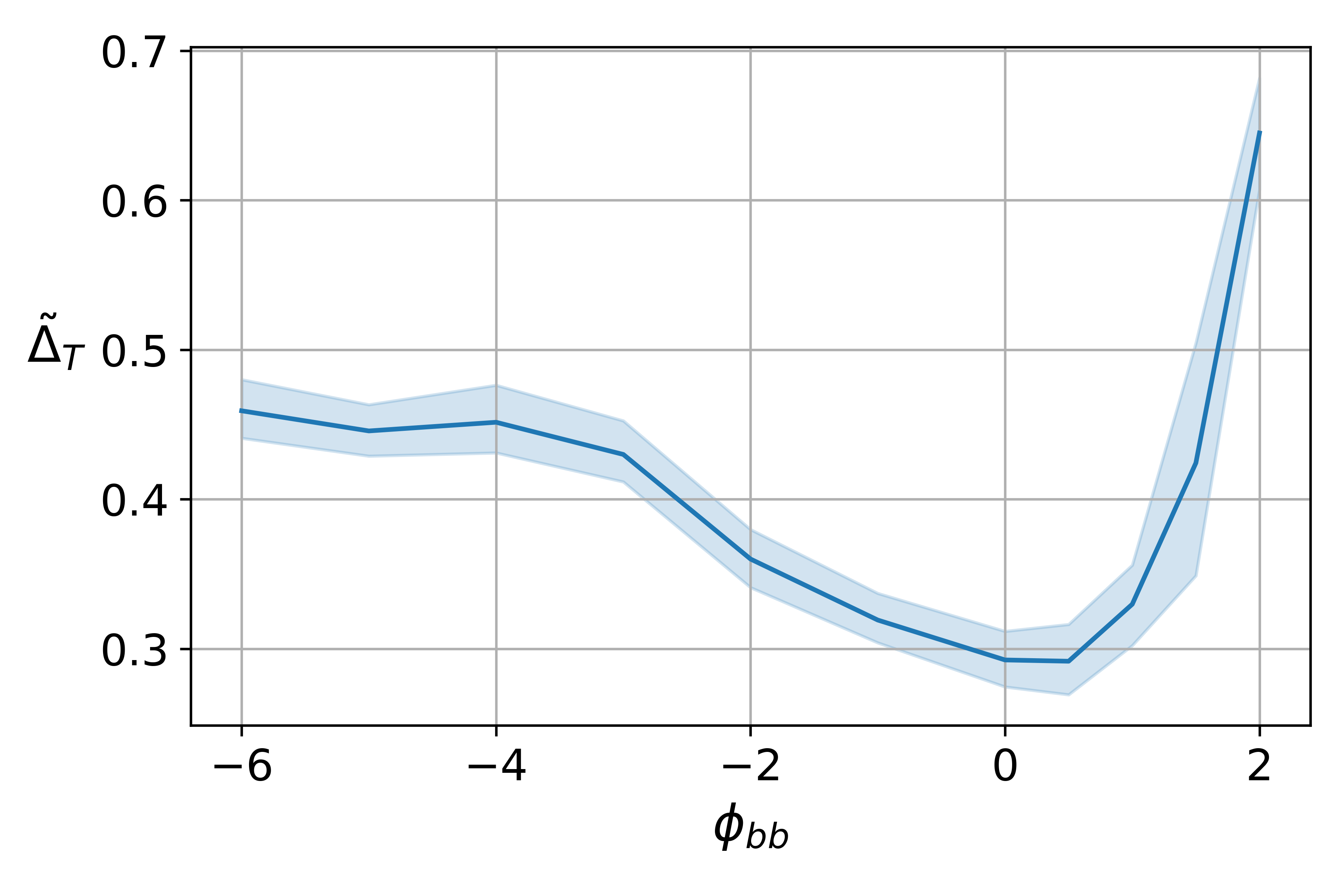}
    \includegraphics[width=0.45\linewidth]{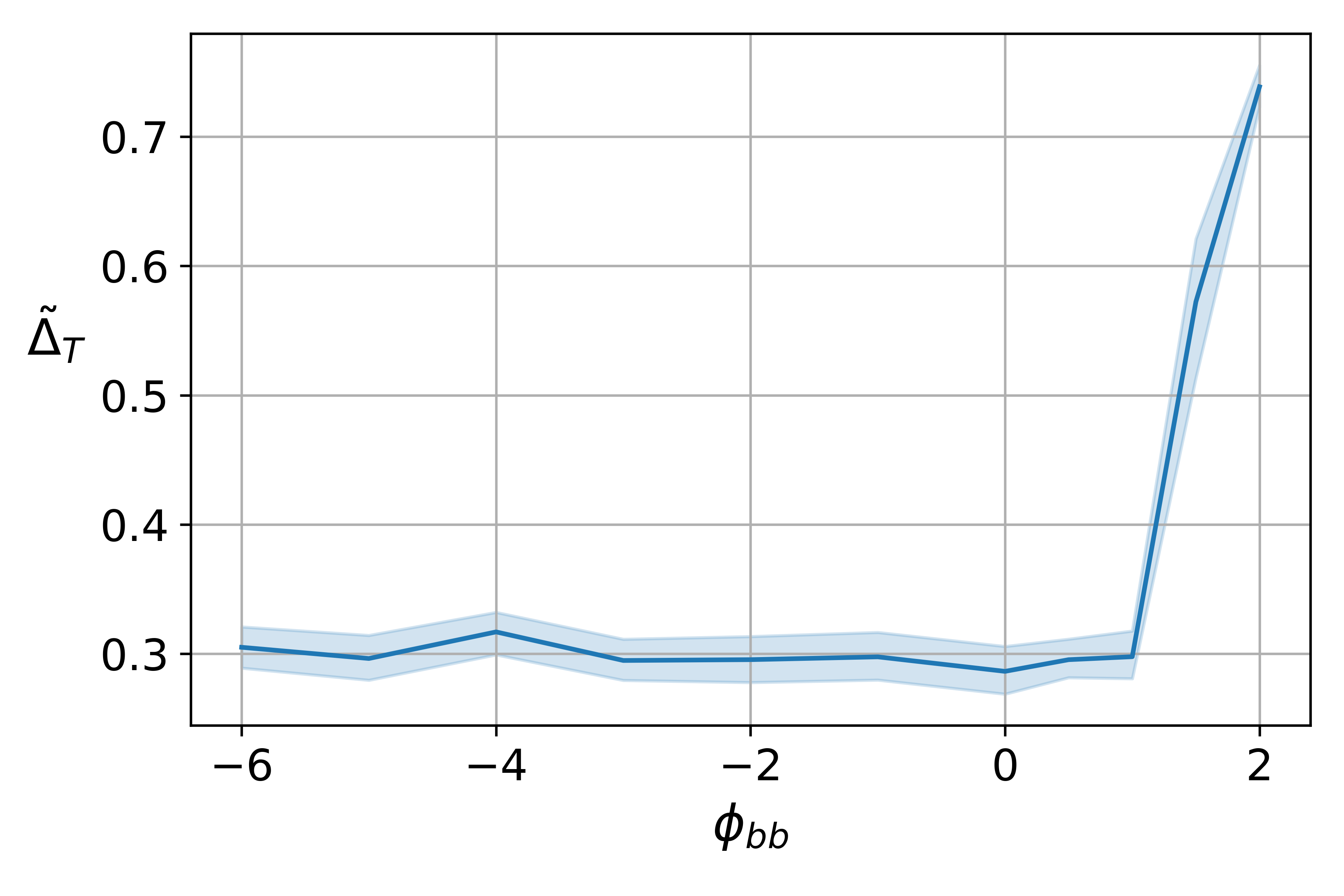}
    \caption{Collusive level with varying $\phi_{bb}$: $\Phi = [\phi_{bb}, 0; 0, 0]$ (left) and $\Phi = [\phi_{bb}, 0; 0, \phi_{bb}]$ (right).}
    \label{fig:self_externality_v0}
\end{figure}

Figure~\ref{fig:self_externality_v0} investigates the dependence of the collusive level on the within-side externalities in two controlled settings. In the first setting (left panel) $\Phi = [\phi_{bb}, 0; 0, 0]$, and in the second one (right panel) $\Phi = [\phi_{bb}, 0; 0, \phi_{bb}]$. In both cases $\phi_{bb}\in[-6,2]$. 

In the left panel, the collusive level forms a J-shape, where it decreases on $[-6,0]$ and increases sharply on $[0.5,2]$.
The minimum value of the collusive level is achieved when $\phi_{bb}\approx 0.5$ and it is slightly below the baseline collusive level. 
Note that this subfigure indicates a similar behavior of the collusive level to its univariate effect shown in the top left panel of Figure~\ref{fig:non_linear_main_effects}. Indeed, in this case, the collusive level depends on the single variable $\phi_{bb}$, so the other univariate and bivariate functions are irrelevant. However, the minimal value of the univariate effect in the top left panel of Figure~\ref{fig:non_linear_main_effects} is around zero, since it is separate from the baseline collusive level $\Delta_0$. By adding $\Delta_0$ to this univariate effect, we obtain a function similar to the collusive level described in the left panel of Figure~\ref{fig:self_externality_v0}. We remark that in the experiments of Section \ref{subsection:phi_relationship_Delta}, our domain was restricted by the underlying Gaussian model and thus the domain in Figure~\ref{fig:non_linear_main_effects} is narrower than that of Figure~\ref{fig:self_externality_v0}. 

In the right panel, the collusive level sharply increases when $\phi_{bb}$ exceeds $1$.  This behavior can be explained using  our previous findings. Indeed, as shown in Figures~\ref{fig:non_linear_main_effects} and~\ref{fig:non_linear_bbss}, when the within-side externalities, $\phi_{bb}$ and $\phi_{ss}$, are both large, the univariate effect is more significant than the negative bivariate effect, resulting in a significant increase.
We also notice that the collusive level remains flat and around  $\Delta_0$ when $\phi_{bb}=\phi_{ss}$ falls below 1. This observation also confirms our findings in the previous section.  Indeed, Figures~\ref{fig:non_linear_main_effects} and~\ref{fig:non_linear_bbss} indicate that when $\phi_{bb}$ and $\phi_{ss}$ are both sufficiently negative, the (positive) univariate and (negative) bivariate effects cancel each other.

\begin{figure}[H]
    \centering
    \includegraphics[width=0.45\linewidth]{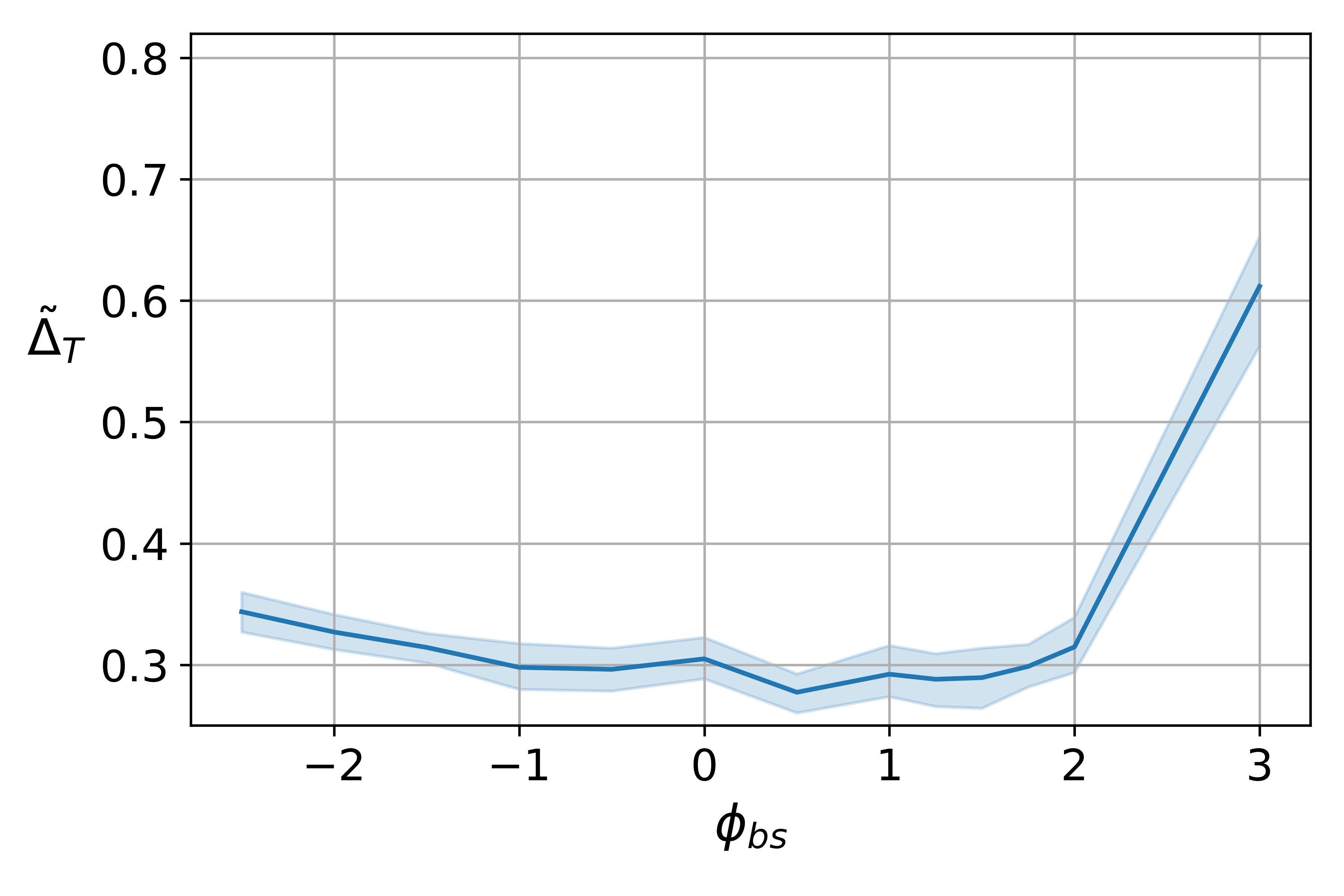}\includegraphics[width=0.45\linewidth]{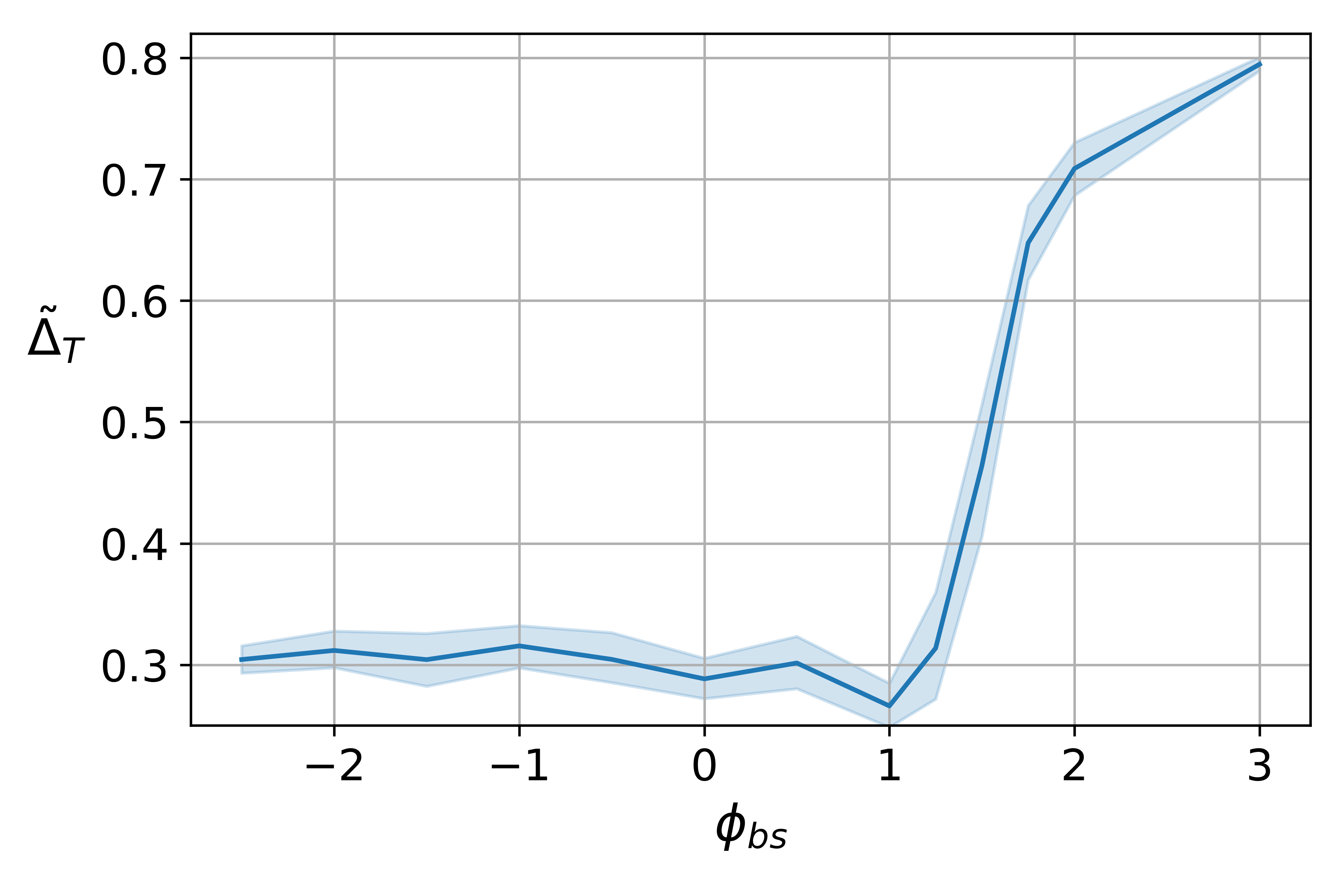}
    \caption{Collusive level with varying $\phi_{bs}$: $\Phi = [0, \phi_{bs}; 0, 0]$ (left) and $\Phi = [
        0 , \phi_{bs} ;
        \phi_{bs} , 0]$ (right).}
    \label{fig:cross_externality_v0}
\end{figure}

Figure~\ref{fig:cross_externality_v0} investigates the dependence of the collusive level on the cross-side externalities in two controlled settings. In the first setting (left panel) $\Phi = [0, \phi_{bs}; 0, 0]$, and in the second one (right panel) $\Phi = [0, \phi_{bs}; \phi_{bs}, 0]$. In both cases $\phi_{bs}\in[-2.5, 3]$.

In the left panel, the collusive level increases when the cross-side externality exceeds $2$, and it slightly decreases when the same externality falls below $-1$. 
This observation aligns with our findings in the previous section. Indeed, as shown by the bottom panels in Figure~\ref{fig:non_linear_main_effects}, the collusive level increases as $\phi_{bs}$ increases in absolute value with values above $1$.  

In the right panel, the collusive level increases when the cross-side externalities exceed $1$. It has a sharper increase than the one in the left panel. These observations agree with the findings of the previous section. Indeed, Figures~\ref{fig:non_linear_main_effects} and~\ref{fig:non_linear_bbss} show that the univariate and bivariate effects of $\phi_{bs}$ and $\phi_{sb}$ are both positive when $\phi_{bs}=\phi_{sb}>1$. Furthermore, Figure~\ref{fig:non_linear_bbss} shows the positive bivariate effect between $\phi_{bs}$ and $\phi_{sb}$, which explains the sharper increase in the right panel.  
The dependence of the collusive level in the right panel on smaller values of $\phi_{bs}$, which are not shown in this figure, is rather unique and thus deferred to Section~\ref{subsec:assymetric_collusion}.

\begin{figure}[H]
    \centering
    \includegraphics[width=0.45\linewidth]{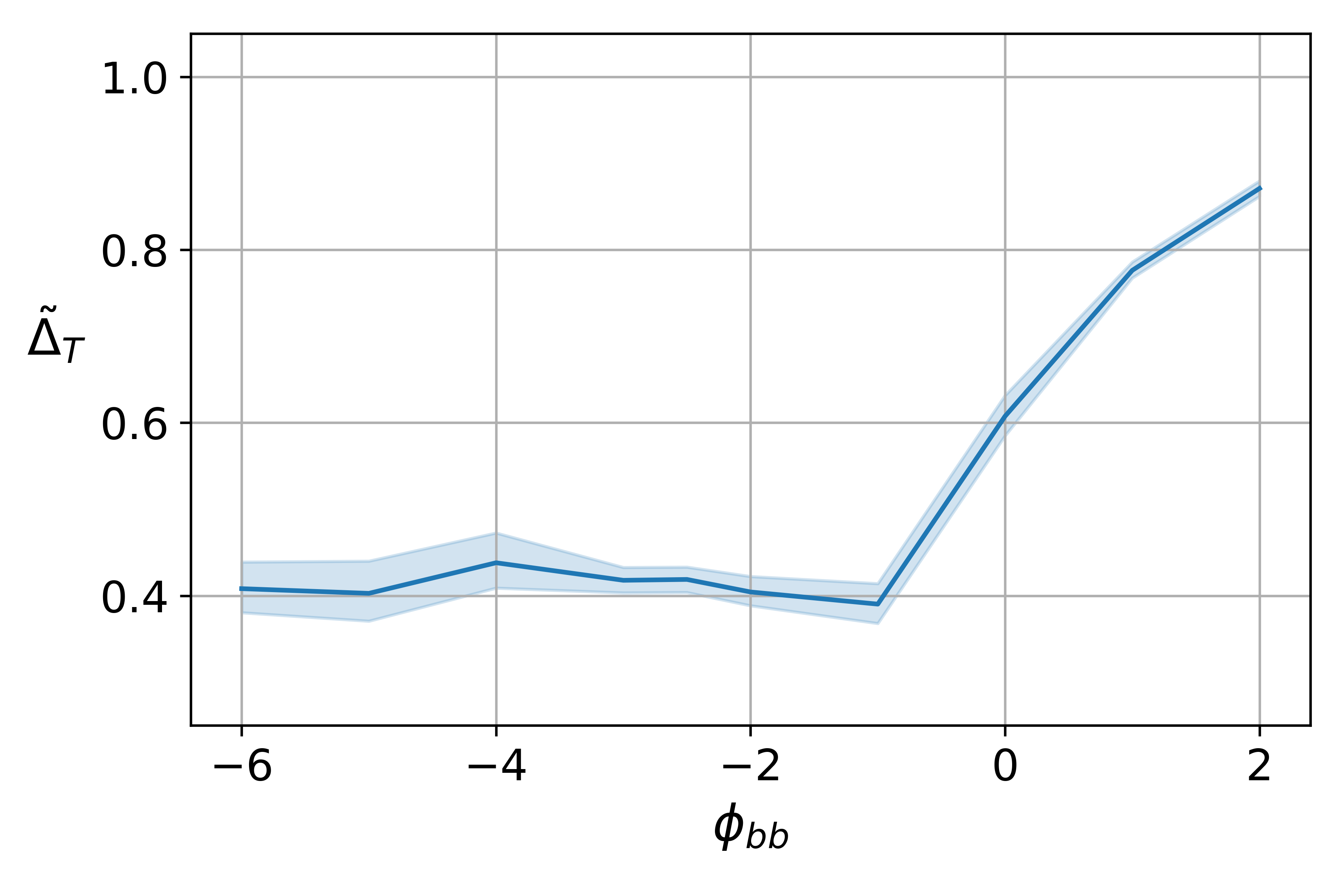}
    \includegraphics[width=0.45\linewidth]{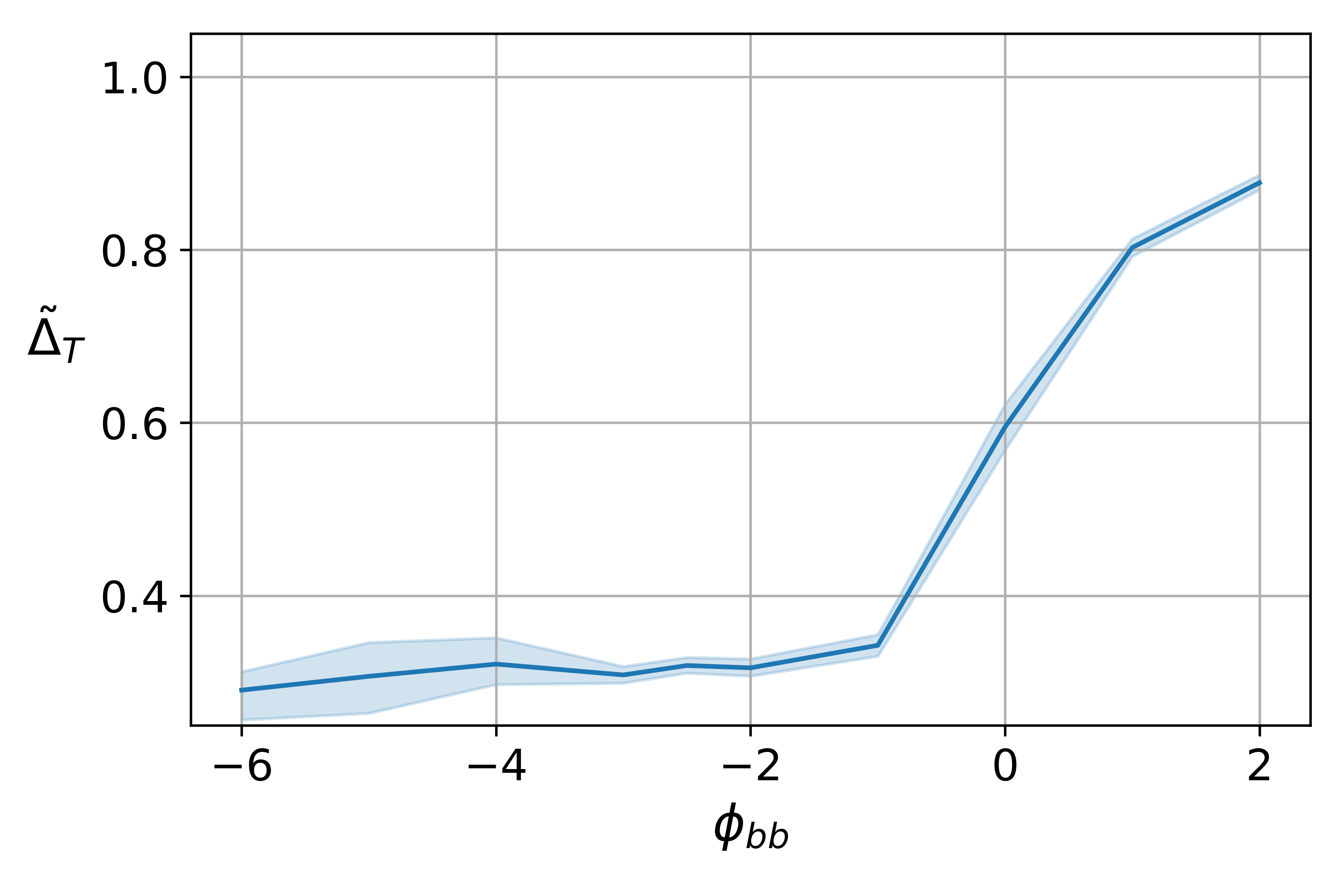}
    \caption{Collusive level with varying $\phi_{bb}$: $\Phi = [\phi_{bb}, 3; 0, 0]$ (left) 
    and $\Phi = [\phi_{bb}, 3 ;
        0 , \phi_{bb}]$ (right).}
    \label{fig:bivariate_cases}
\end{figure}

Figure~\ref{fig:bivariate_cases} investigates the dependence of the collusive level on the bivariate effects between the within- and cross-side externalities in two controlled settings. For simplicity, we fix the cross-side externality and vary the within-side externality. 
In the first setting (left panel) $\Phi=[\phi_{bb}, 3; 0, 0]$, and in the second one  
(right panel) $\Phi = [\phi_{bb}, 3 ;0 , \phi_{bb}]$. In both cases $\phi_{bb}\in[-6,2]$.

In the left panel, the collusive level is flat when the within-side externality falls below $-1$ and increases sharply as it exceeds $-1$. 
This increase can be explained when $\phi_{bb}$ exceeds $0$ by the univariate effect for $\phi_{bb}$ shown in Figure~\ref{fig:non_linear_main_effects}. 
On the other hand, the increase in $[-1, 0]$ can be explained by the bivariate effect between $\phi_{bb}$ and $\phi_{bs}$, shown in Figure~\ref{fig:non_linear_bbbs_sssb}. Indeed this bivariate effect increases with respect to $\phi_{bb}$ in $[-1, 0]$ when $\phi_{bs}=3.0$. 

In the right panel, the collusive level follows a similar pattern as in the left one. This follows from a similar explanation as above, where one should also note that when $\phi_{bs}=3$, the bivariate effect between $\phi_{ss}$ 
and $\phi_{bs}$ increases with respect to $\phi_{ss}$ in $[-1, 0]$, as shown in Figure~\ref{fig:non_linear_bbsb_ssbs}. 
Additionally, when $\phi_{bb} < 0$, the collusive level in the right panel is lower than that in the left panel (it is easiest to see this for $\phi_{bb} < -1$). 
We clarify this observation in view of the findings of Section \ref{subsection:phi_relationship_Delta} as follows. We note that according to the right panel of Figure~\ref{fig:non_linear_bbsb_ssbs}
, when $\phi_{bs}=3$, the bivariate effect between $\phi_{ss}$ and  $\phi_{bs}$ is negative when $\phi_{ss} < 0$, therefore the collusive level in the right panel is expected to be lower than the collusive level in the left panel when $\phi_{bb}=\phi_{ss} < 0$.

We make some additional remarks comparing Figures~\ref{fig:self_externality_v0} and~\ref{fig:bivariate_cases}. The left panel in Figure~\ref{fig:self_externality_v0} shows that the collusive level decreases with respect to $\phi_{bb}$, when $\phi_{bb}<0$ and $\phi_{bs}=0$. 
On the other hand, in the left panel of Figure~\ref{fig:bivariate_cases}, the collusive level is flat or increases with respect to $\phi_{bb}$ when $\phi_{bb}<0$ and $\phi_{bs} = 3$. This behavior can be explained by the contribution from the bivariate effect between $\phi_{bb}$ and $\phi_{bs}$ when $\phi_{bb}$ is negative. Indeed, this bivariate effect is almost flat with respect to $\phi_{bb}$ when $\phi_{bs}=0$, but is increasing with respect to $\phi_{bb}$ when $\phi_{bs}=3.0$ (see left panel of Figure~\ref{fig:non_linear_bbbs_sssb}). A similar comparison can be made for the right panels in Figures~\ref{fig:self_externality_v0} and~\ref{fig:bivariate_cases}, and the explanation similarly follows from the right panel of Figure~\ref{fig:non_linear_bbsb_ssbs} and the left panel of Figure~\ref{fig:non_linear_bbbs_sssb}.

\subsection{A Study of the Collusive Level under Special Market Parameters}\label{subsect:colllusive_idiosyncratic} 
We explore the dependence of the collusive level on the market parameters $\beta_k$, $u^{(0)}_k$ and $\delta$. In each experiment, we fix two of the latter parameters, using the setup described in Section \ref{subsec:problem}, and the matrix 
$\Phi$, where its choices change with the experiments, and vary the remaining parameter. 
For each experiment, we ran 100 simulations, averaged the collusive levels among the 100 runs and computed the uncertainty levels using bootstrapping with a 99\% confidence interval. 
Our figures present the averaged collusive level as a function of one parameter, where the shaded areas represent the uncertainty level.

\begin{figure}[H]
    \centering
    \includegraphics[width=0.45\linewidth]{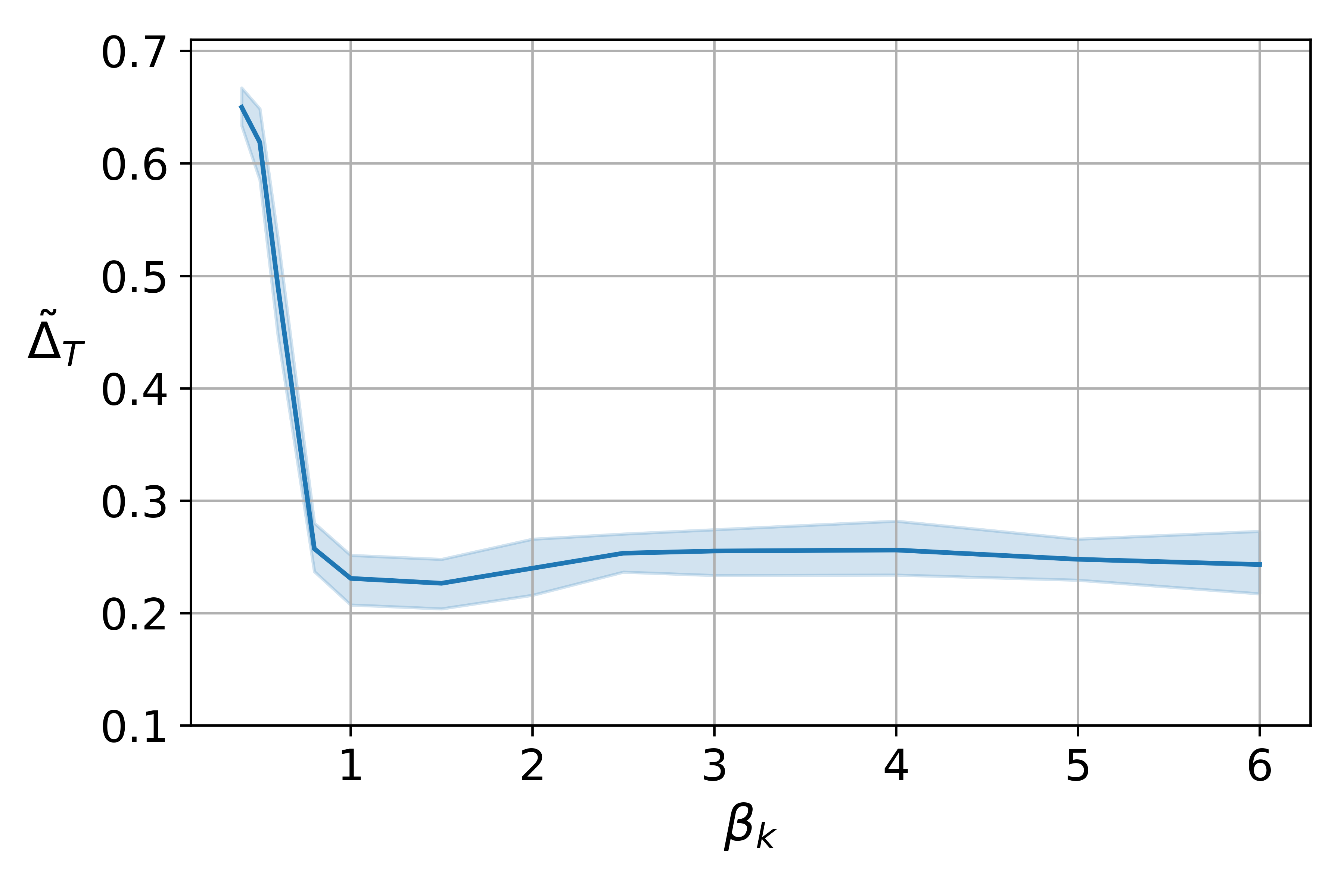}
\includegraphics[width=0.45\linewidth]{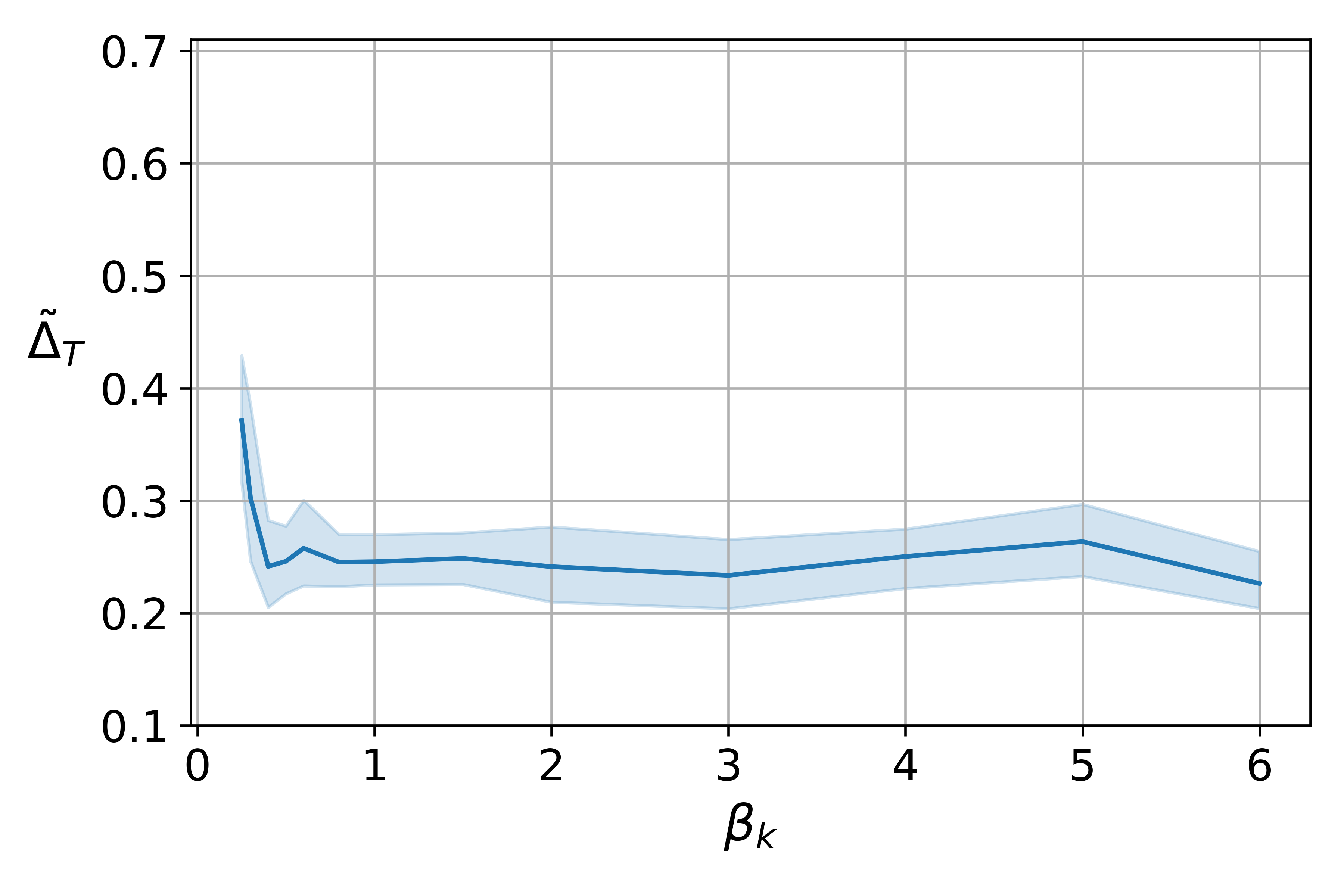}
    \caption{Collusive level with varying $\beta_b=\beta_s\in[0.2, 6]$ and different matrices $\Phi$, where $\Phi = [1, 0; 0, 1]$ (left) and $\Phi = [0, 1; -1, 0]$ (right).}
    \label{fig:beta_v}
\end{figure}

Figure~\ref{fig:beta_v} investigates the dependence of the collusive level on the idiosyncratic preference parameter $\beta_k$, while considering two different choices of the externality matrix $\Phi$: A symmetric one, where $\Phi = [1, 0; 0, 1]$ (left panel) and an asymmetric one, where $\Phi=[0, 1; -1, 0]$ (right panel). In both cases, we vary the idiosyncratic preference parameters and let $\beta_k \in[0.2, 6]$. 
In both panels, the collusive level sharply decreases when $\beta_k$ is sufficiently small. In particular, the collusive level is high only when the degree of heterogeneity in users' tastes is sufficiently small. Section \ref{sect:Discussion} interprets this behavior. 
We note that in the left panel the sharp decrease stops when $\beta_k$ exceeds $0.9$, compared to the right panel where this happens ses once $\beta_k$ exceeds $0.5$. 
Furthermore, we note that in both subfigures, the collusive levels remains almost flat, at a value slightly below the collusive level $\Delta_0=0.3$, as the degree of heterogeneity in users' tastes exceeds $1$. 
More experiments varying $\beta_k$ with different choices of $\Phi$ are presented in Appendix~\ref{append:other_simulation}.

\begin{figure}[H]
    \centering
\includegraphics[width=0.45\linewidth]{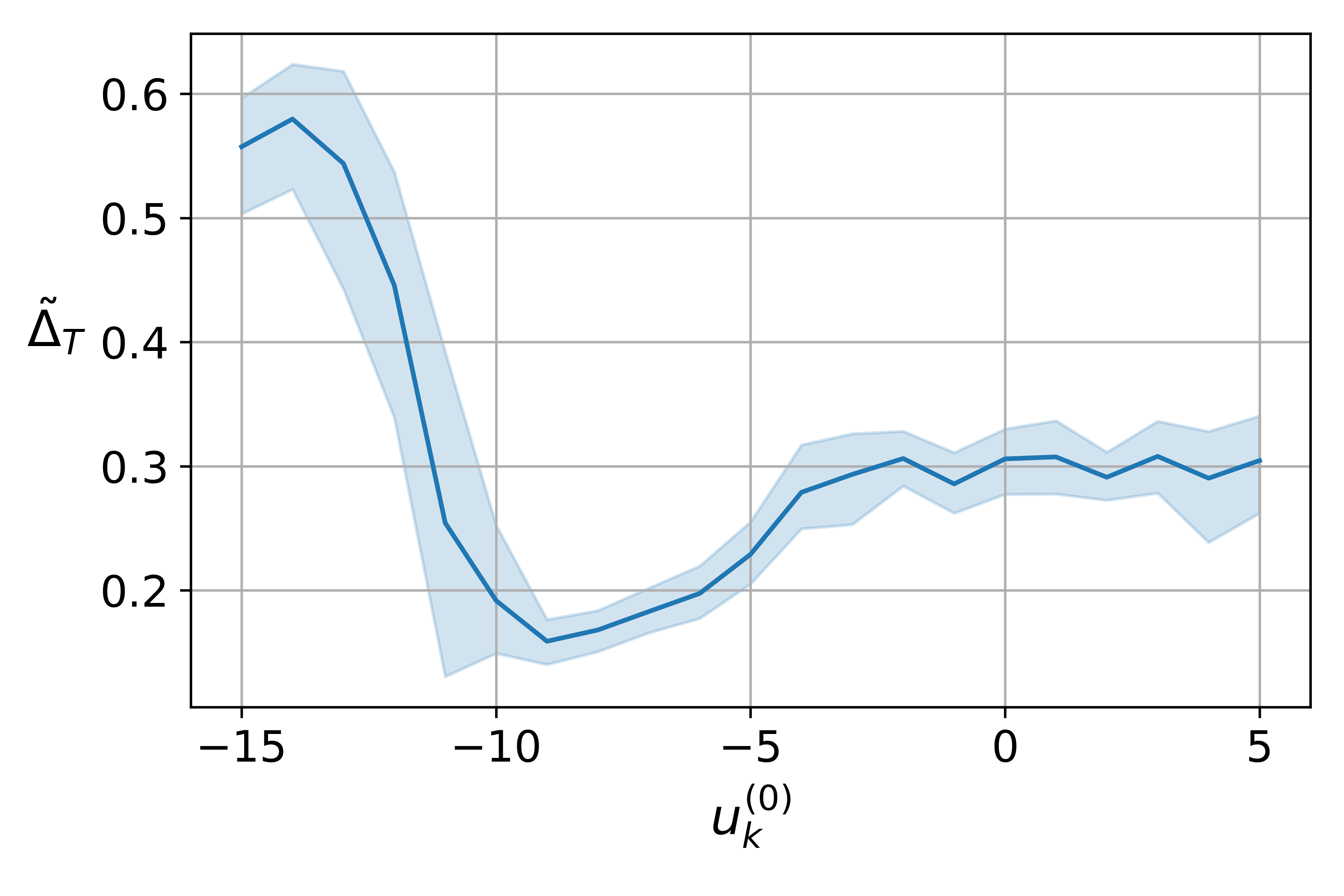}
\includegraphics[width=0.45\linewidth]{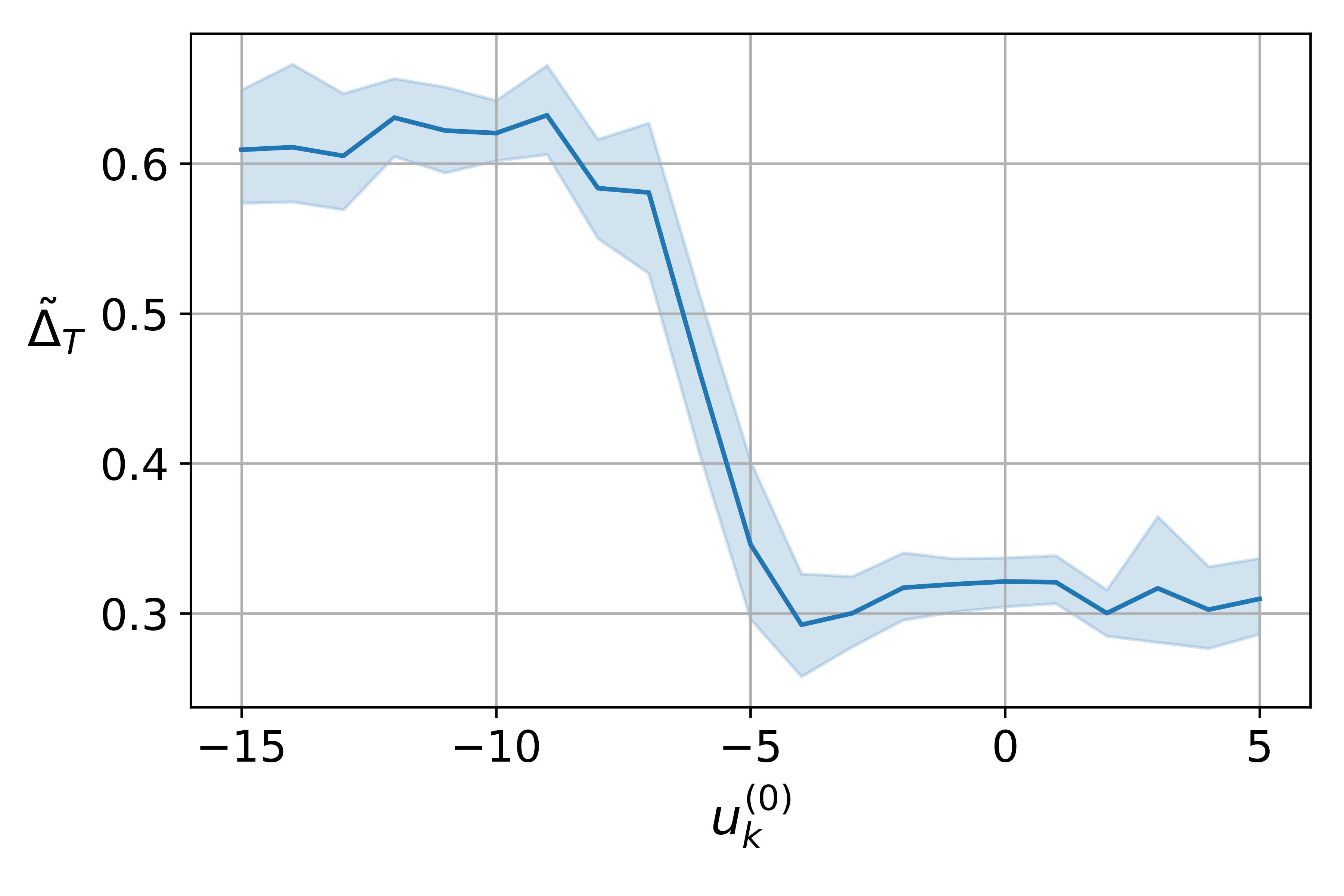}
\includegraphics[width=0.45\linewidth]{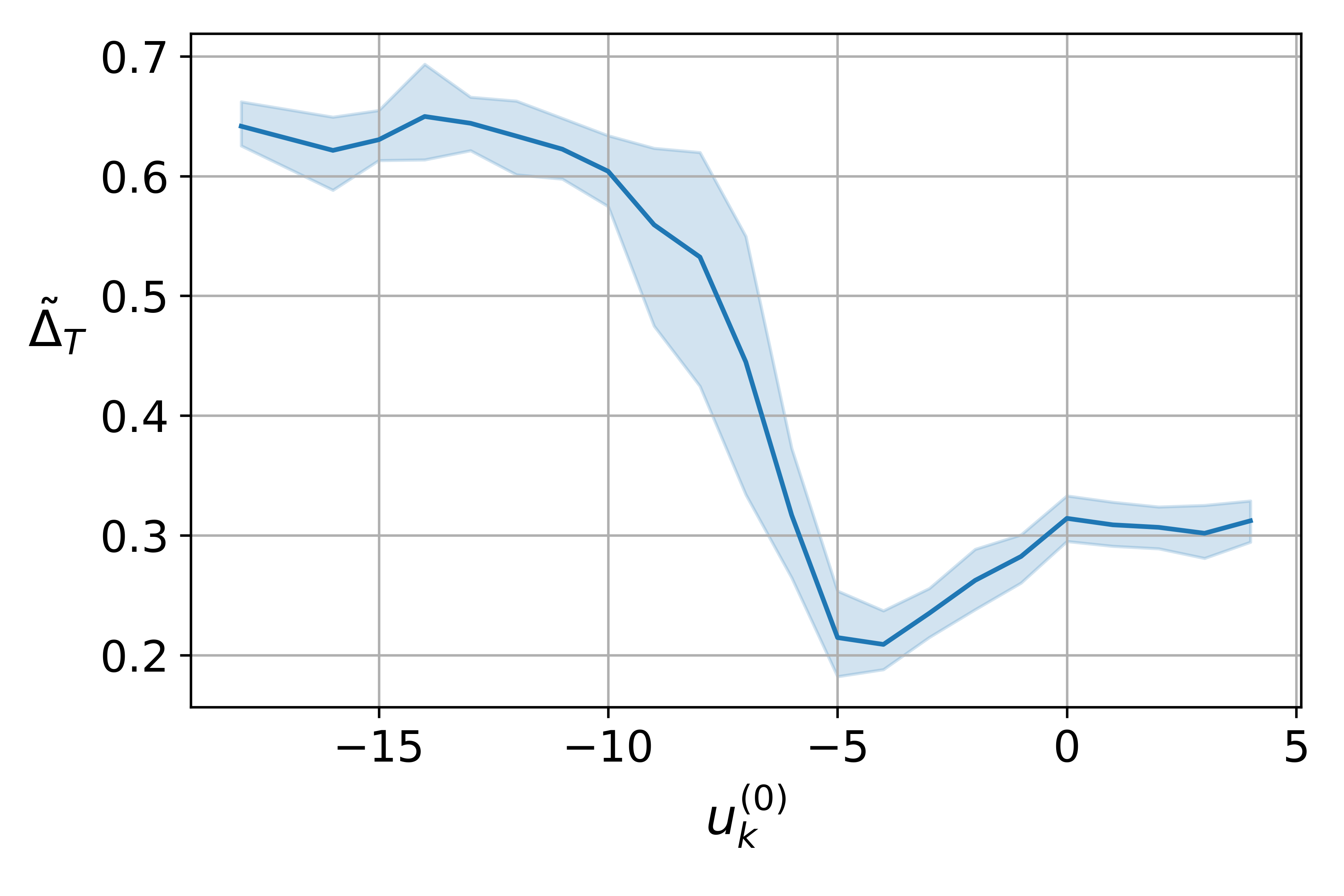}
\includegraphics[width=0.45\linewidth]{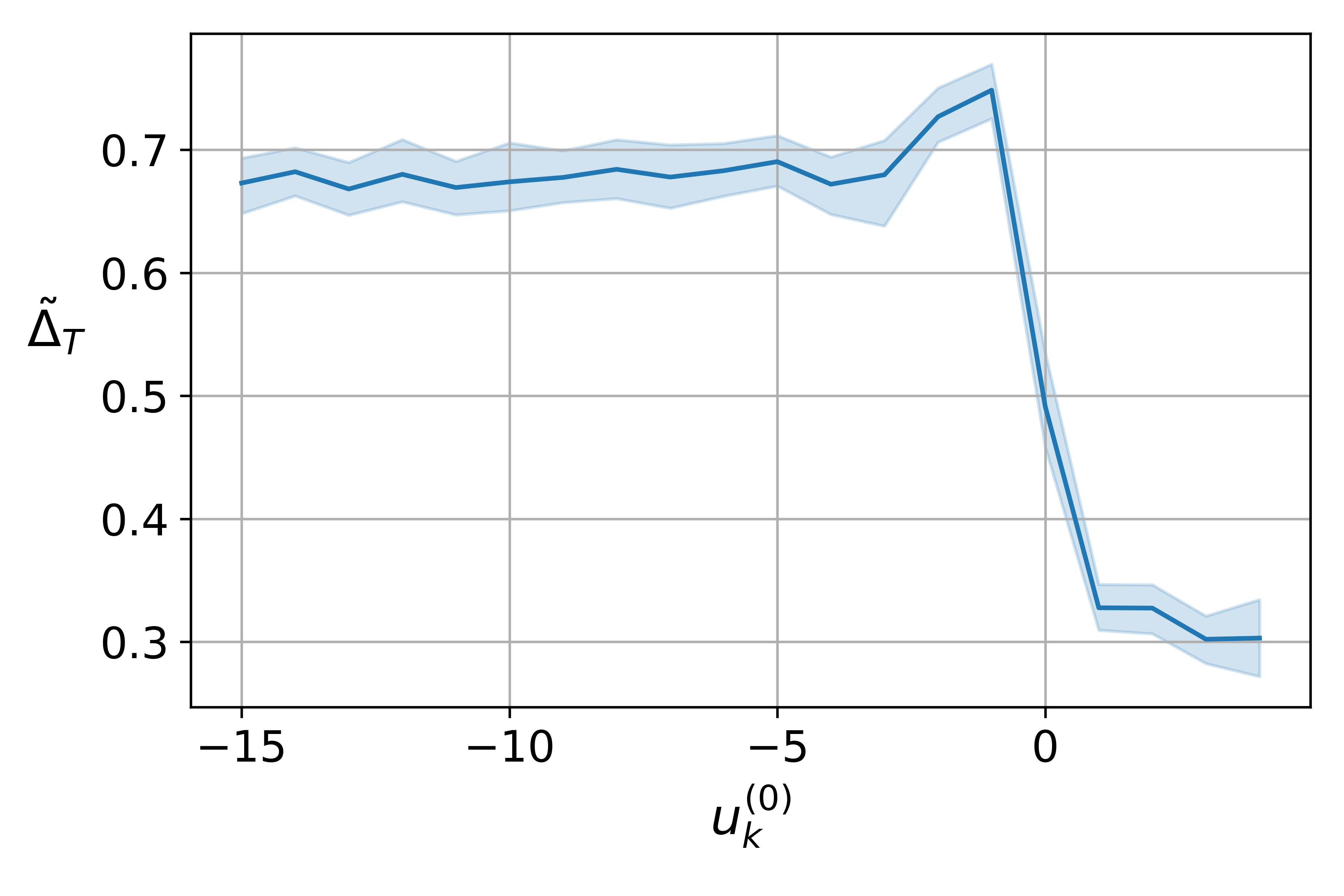}
    \caption{Collusive level with varying $u^{(0)}_k$,  $k\in\{b,s\}$, with $\Phi = [0, 0; 0, 0]$ (top left), $\Phi=[1, 0; 0, 1]$ (top right), $\Phi=[0, 1; 1, 0]$ (bottom left) and $\Phi=[1, 1; 1, 1]$ (bottom right).}
    \label{fig:u0_change_main}
\end{figure}

Figure~\ref{fig:u0_change_main} demonstrates the dependence of the collusive level on the outside option utility $u_k^{(0)}$ with the following choices for the externality matrix: $\Phi = [0, 0; 0, 0]$ (top left panel), $\Phi=[1, 0; 0, 1]$ (top right panel), $\Phi=[0, 1; 1, 0]$ (bottom left panel), and $\Phi=[1, 1; 1, 1]$ (bottom right panel). In all four panels, we observe a main trend of decrease of the collusive level as a function of the outside option utility over a sufficiently large domain. In all of these examples when $u_k^{(0)}$ is sufficiently small, the collusive level is at least $0.55$, and when $u_k^{(0)}$ is sufficiently large, the collusive level is around $0.3$, which is near the baseline level. The most significant reduction of the collusive level happens in a narrow range and the location of this significant decrease appears to be determined by the externalities as follows. It tends to move to the left when the network externalities are small and to the right when they are large. Additional examples in Figure~\ref{fig:u0_change_ones} support this conclusion. 
In addition, we observe that both left subfigures exhibit another small region of increase to the baseline level after the region of sharp decrease. This is not the case for the right subfigures. We also note a similar phenomenon in Figure~\ref{fig:u0_change_ones}. It seems that an increase to the baseline level after a sharp decrease occurs in cases of sufficiently small externalities, where the threshold on externalities required to guarantee such a short increase is smaller for within-side externalities than for cross-side externalities.  
For example, considering cases of both Figures~\ref{fig:u0_change_main} and \ref{fig:u0_change_ones}, this short increase is observed at $\Phi=[0, 1; 1, 0]$ but not at $\Phi=[0, 2; 2, 0]$, and is observed at $\Phi=[1, 0; 0, 0]$ but not at $\Phi=[1, 0; 0, 1]$. 
Lastly, it follows from the bottom right subfigure of Figure~\ref{fig:u0_change_main} and cases of Figure~\ref{fig:u0_change_ones} that for sufficiently large externalities, the collusive level may slightly increase before the sharp decrease.

\begin{figure}[H]
    \centering
\includegraphics[width=0.45\linewidth]{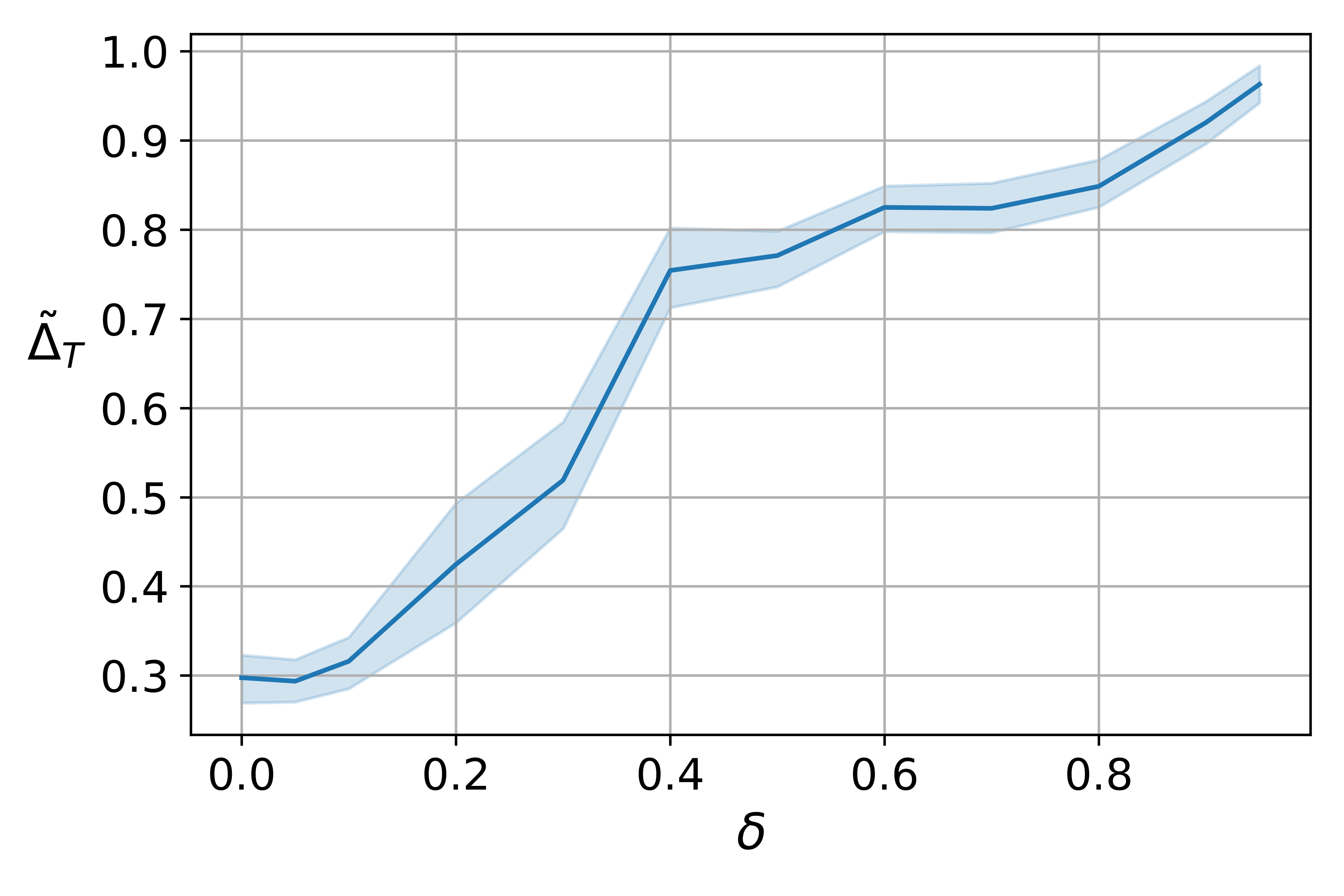}
\includegraphics[width=0.45\linewidth]{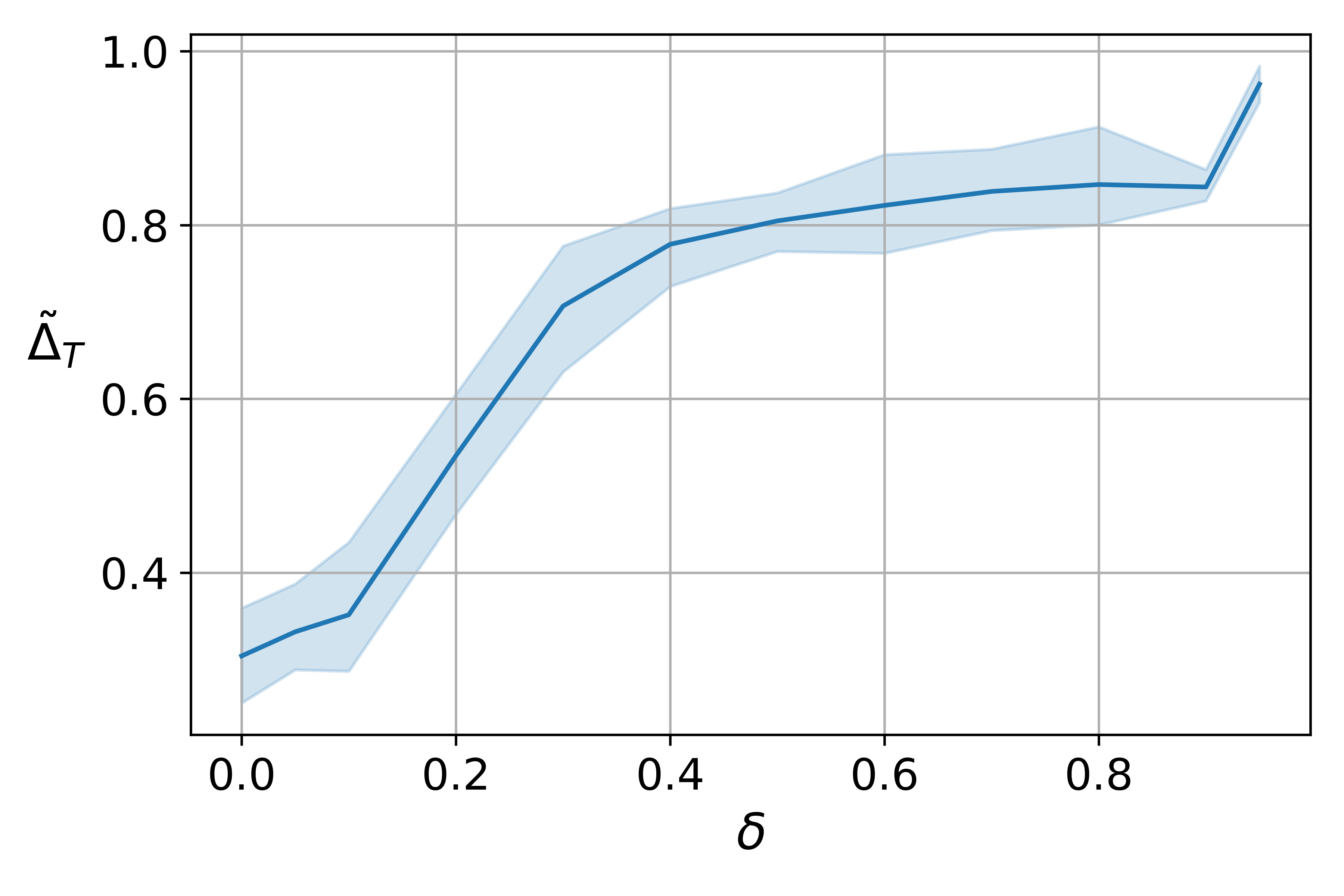}
    \caption{Collusive level with varying $\delta$, where $\Phi = [0,0;0,0]$ (left) and $\Phi = [1, 0; 0, 1]$ (right); in both cases $\delta\in[0.01,0.99]$.}
    \label{fig:delta_change}
\end{figure}

Figure~\ref{fig:delta_change} investigates the dependence of the collusive level on the discount rate $\delta$, where $\delta\in[0.01,0.99]$, with $\Phi = [0,0;0,0]$ (left) and  $\Phi = [1,0;0,1]$ (right). 
In both panels, the collusive level increases with the discount rate. Thus, the more patient platforms are about the future, the higher the collusive level becomes.
We note that the collusive level in the right panel is approximately a horizontal shift of that in the left panel. Thus, with a non-zero externality matrix the collusive level increases earlier than with the zero externality matrix. 
The positive relationship between the collusive level and the discount rate was first observed by \cite{calvano2020artificial} for Bertrand model and our experiments verify the same relationship for a multi-sided market.

\subsection{Discussion of Exceptional Cases}
\label{subsec:assymetric_collusion}

We numerically demonstrate two exceptional and uncommon scenarios: asymmetric optimal prices and competition prices larger than collusion prices. 

\textbf{Asymmetric collusion.} Our metric for the collusive level compares the platform's rewards at time $t$ with the symmetric equilibrium quantities $\pi^*$ and $\pi^C$, following previous simulations of collusion (see, e.g., \cite{calvano2020artificial} and \cite{klein2021autonomous}). 
However, our specific model may give rise to asymmetric equilibria and we thus study their possibility more carefully. 

To allow asymmetric equilibria, we modify the definitions of the maximum values of the total profits and the collusive level as follows. The total profit, $\Pi_a$, is 
\begin{equation}
\Pi_{a}(p_b^{(1)},p_s^{(1)},p_b^{(2)},p_s^{(2)}) 
 := \sum_{i=1}^2 (x^{(i)}_b p_b^{(i)}+x^{(i)}_s p_s^{(i)} )\equiv \sum_{i=1}^2\pi_t^{(i)},
 \label{pim_asymmetric}
\end{equation}
where unlike \eqref{pim} it does not assume symmetric prices (that is, it does not assume that $\pi_t^{(1)}=\pi_t^{(2)}$) and its subscript $a$ indicates asymmetry.
Similarly, the collusive level $\Delta_t$ is averaged among the two firms as follows 
$$\Delta_t:=(\Delta_t^{(1)}+\Delta_t^{(2)})/2 \equiv \frac{\Pi_a - 2\pi^*}{2(\pi^\text{C} -\pi^*)},$$ 
where $\Delta_t^{(i)}$, $i\in\{1,2\}$, was defined in \eqref{eqn:def_Delta}. 

\begin{figure}[H]
    \centering
\includegraphics[width=0.45\linewidth]{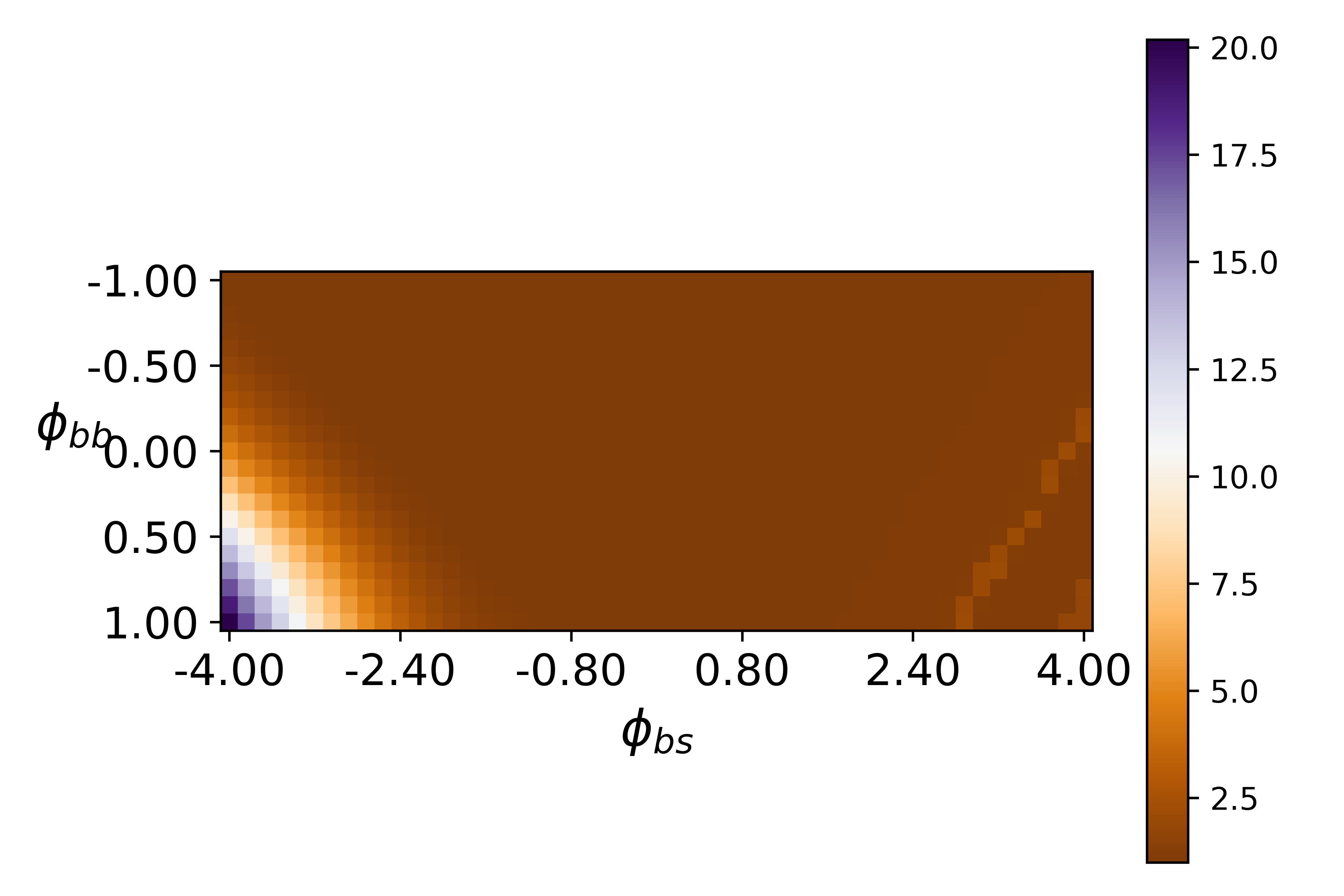}
\includegraphics[width=0.45\linewidth]{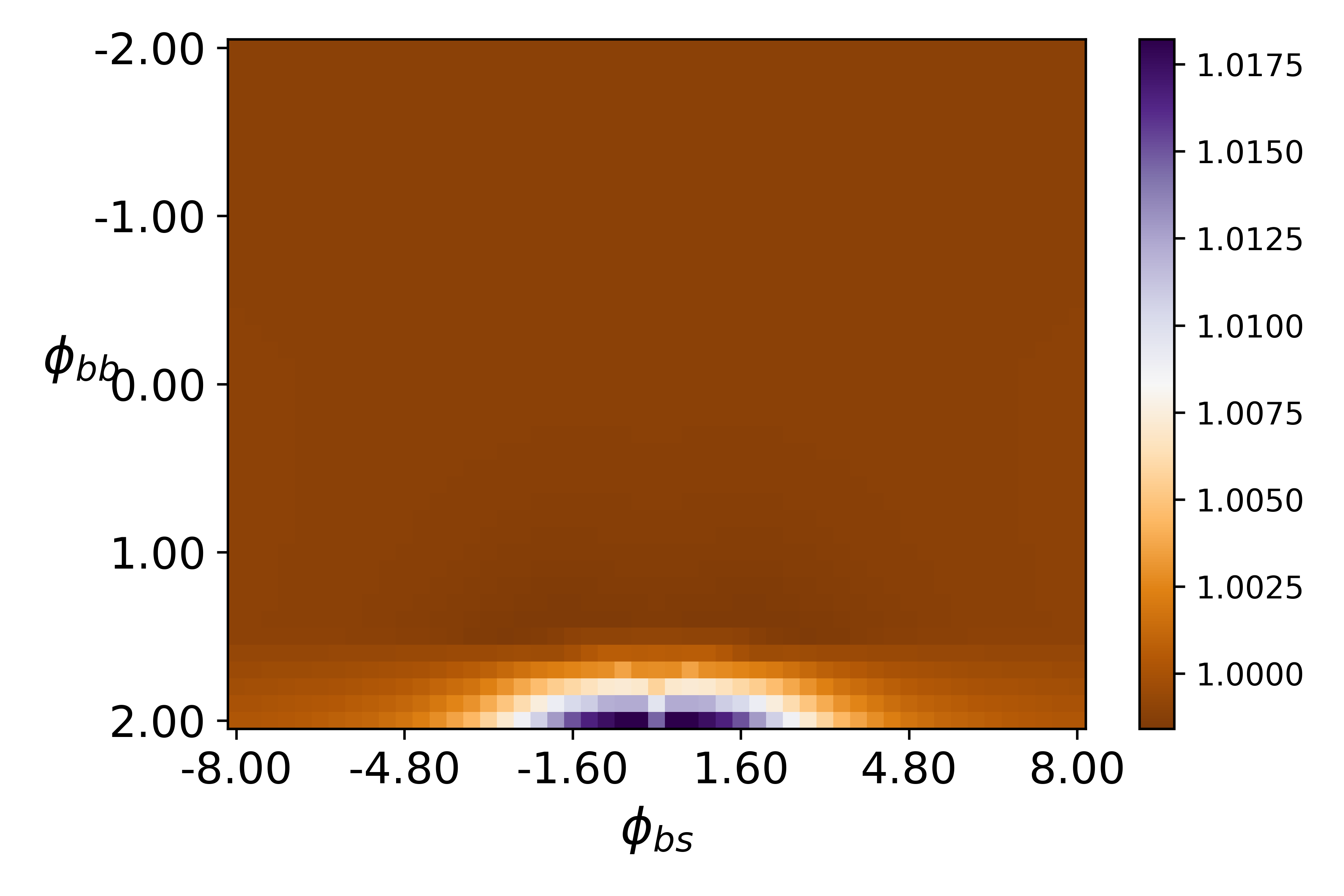}
    \caption{Maximal average collusive level
    with varying $\phi_{bb}$ and $\phi_{bs}$, where $\Phi = [\phi_{bb}, \phi_{bs};\phi_{bs},\phi_{bb}]$ (left) and  $\Phi = [\phi_{bb}, \phi_{bs};-\phi_{bs},\phi_{bb}]$ (right).}
    \label{fig:maximal_Delta}
\end{figure}

Figure~\ref{fig:maximal_Delta} demonstrates the maximum value that $\Delta_t$ can achieve in two controlled settings. In the first setting (left panel) $\Phi = [\phi_{bb}, \phi_{bs};\phi_{bs},\phi_{bb}]$ with $\phi_{bb}\in [-1,1]$ and $\phi_{bs}\in[-4,4]$, and in the second one (right panel) $\Phi = [\phi_{bb}, \phi_{bs};-\phi_{bs},\phi_{bb}]$ with $\phi_{bb}\in [-2,2]$ and $\phi_{bs}\in[-8,8]$. More precisely, we maximize $\Delta_t$ over $\vp^{(1)}_t\in\gA$ and $\vp^{(2)}_t\in\gA$, where $\gA$ was defined in \eqref{eqn:def:gA_gS}, and present the maximal values using a heatmap, whose values vary from purple ($\Delta_t$ greater than $1$) to orange ($\Delta_t$ less than $1$). In both panels, $\beta_k=1.0$ and $u_k^{(0)}=-2.0$, $k\in\{b,s\}$.

In the left panel, the maximal $\Delta_t$ exceeds $1$ when the within-side externality $\phi_{bb}$ is positive and close to $1$ and the cross-side externality $\phi_{bs}$ is sufficiently negative. We note that it is much larger than 1. In the right panel, the maximal $\Delta_t$ is larger than 1 when $\phi_{bb}$ is sufficiently large and $\phi_{bs}$ is close to zero. We note though that it never exceeds the value of $1.0175$. 
In both panels, for all other corresponding values of $\phi_{bb}$ and $\phi_{bs}$, the maximum value of $\Delta_t$ is either achieved at a symmetric vector price or at an asymmetric vector price with corresponding maximum value close to $1$. In the latter case, of the corresponding maximum value close to $1$, our measure of collusive level $\tilde{\Delta}$ using the symmetric assumption can still be used to quantify the level of collusion. That is, one may often follow up our analysis with the symmetric quantities, except for special cases where the estimated collusive level exceeds one, where one needs to use the asymmetric quantities.

\begin{figure}[H]
    \centering
    \includegraphics[width=0.45\linewidth]{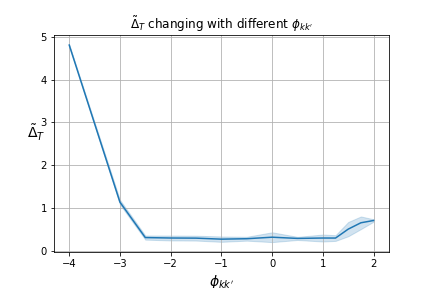}
    \caption{Collusive level with varying $\phi_{sb},$ where $\Phi = [0,\phi_{sb};\phi_{sb},0]$.}
    \label{fig:cross_externality_extend_neg}
\end{figure}

We remark that for almost all choices of $\beta_k$, $u_k^{(0)}$, $\delta$ and $\Phi$ in this paper, the asymmetric maximum value of $\Delta_t$ does not exceed the symmetric maximum value. Indeed, our simulations from Sections \ref{subsection:phi_relationship_Delta} through \ref{subsect:colllusive_idiosyncratic} show that $\tilde{\Delta}$ is smaller than or equal to $1$. Nevertheless, when extending the right panel of Figure~\ref{fig:cross_externality_v0} to more negative values a collusive level higher than 1 is noticed, which we depict in Figure~\ref{fig:cross_externality_extend_neg}.  In this figure, $\Phi = [0,\phi_{sb};\phi_{sb},0]$ with $\phi_{sb}\in [-4,2]$ and the asymmetric maximum value exceeds the symmetric maximum value when $\phi_{sb} < -2.5$. We notice that the collusive level reaches values close to $5$ when the within-side externalities are zero and the cross-side externalities are equal in magnitude and sufficiently negative. 
Note that this scenario agrees with the one depicted in the left panel of Figure~\ref{fig:maximal_Delta}, where one can notice a similar value of 5 when $\phi_{bb}=\phi_{ss}=0$ and $\phi_{bs}=\phi_{sb}=-4$. 

\begin{figure}[H]
    \centering
\includegraphics[width=0.95\linewidth]{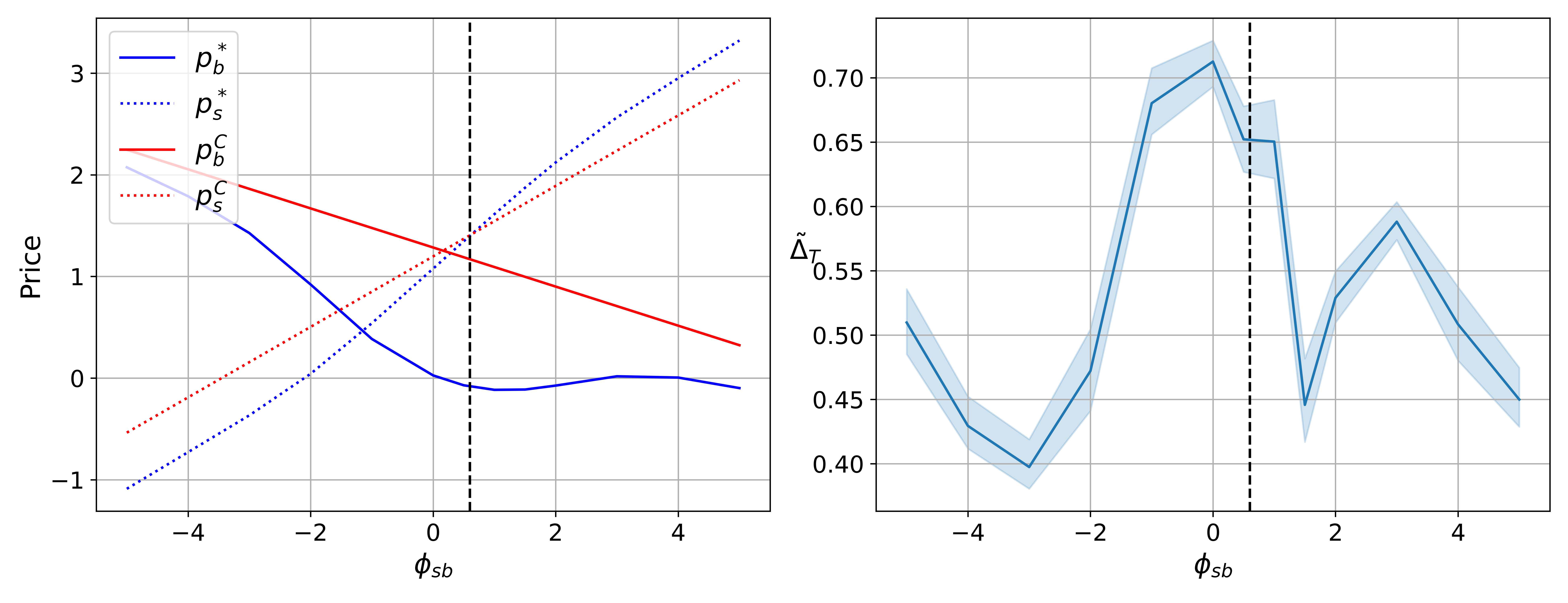}
    \caption{Demonstration of an uncommon scenario where the collusion prices can be lower than competition prices. 
    Here, $\Phi = [1,-\phi_{sb};\phi_{sb},-2]$, where $\phi_{sb} \in [-5,5]$.  
    The left panel demonstrates collusion and competition prices on both sides of the market. The right panel demonstrates the collusive level. 
    The black vertical dotted lines separate the two regimes:  $p^*_s < p_s^{\text{C}}$ on the left side of the black vertical dotted line and $p^*_s > p_s^{\text{C}}$ on the right side of the black vertical dotted line.}
    \label{fig:phibs_cross}
\end{figure}

\textbf{Competition prices larger than collusion prices.} We recall that both the simulated action space presented in \eqref{def:Pkset} and the formulated baseline model in Section \ref{sec:model1} allow for the competition prices ($p_k^*$) to be either smaller or larger than the collusion prices ($p_k^{\text{C}}$). 
We demonstrate here an uncommon situation where the collusion price can be smaller than the competition price for one side of the market. In this example $\Phi = [1,-\phi_{sb};\phi_{sb},-2]$, $\beta_k = 0.5$ and $u^{(0)}_k = -1.0$, $k\in\{b,s\}$. If $\phi_{sb}\in [-5,0.5)$, then $p_s^*<p_s^{\text{C}}$. If $\phi_{sb}\in (0.5,5]$, then $p_s^*>p_s^{\text{C}}$. 
The left panel of Figure~\ref{fig:phibs_cross} demonstrates the competition and collusion prices of both sides of the market. For side $b$, the collusion price is always higher than the competition price. However, for side $s$, there are two different regimes separated by 
$\tilde{\phi}_{sb} \approx 0.5$. When 
$\phi_{sb} < \tilde{\phi}_{sb}$, the collusion price is higher than the competition price on the seller side, and when $\phi_{sb} > \tilde{\phi}_{sb}$ it is lower. 

The right panel of 
Figure~\ref{fig:phibs_cross} shows the collusive level for this example. The black dotted line is drawn at $\tilde{\phi}_{sb}$ to separate the two regimes. 
In the left regime, where $p_s^*<p_s^{\text{C}}$, the collusive level decreases on $[-5,-3]$ and increases on $[-3,0]$. This behavior is somewhat similar to the one described in the left panel of Figure~\ref{fig:cross_externality_v0}. 
During the transition from $p_s^*<p_s^{\text{C}}$ to $p_s^*>p_s^{\text{C}}$, that is, in a small interval around $\tilde{\phi}_{sb}$, the collusive level decreases. Next, the collusive level increases on $[1.5, 3]$ and  decreases on $[3,5]$, which is the opposite behavior (in terms of decreasing and increasing) to the one demonstrated in the other regime.

%% file: discussion.tex
When $\Phi = 0$, our baseline platform competition game reduces to Bertrand competition games on each side of the market. For $\delta = 0.05$, $\beta_k = 1$ and $u_k^{(0)} = -2$, $k\in\{b,s\}$, our simulations show that $\Delta_0$ in \eqref{eqn:fitting_Delta} is approximately $0.3$. This means that when $\Phi = 0$, the profit gain relative to competition profits is about 30\%. It is interesting to note that \cite{calvano2020artificial} reported a different gain of approximately $20\%$ for the same value of $\delta$. To understand the difference, we note that in the latter paper, firms serve only one market and Bertrand competition is the baseline game. For that reason, the action space used in their simulations is smaller than the action space used in ours.  
The larger action space allows the platforms to algorithmically communicate more information, increasing the chance of achieving a higher collusive level. 

Next, we discuss the impact of $\Phi\neq 0$ on the collusive level  compared to Bertrand competition, where $\Phi=0$. When $\Phi$ has only one non-zero entry, the dependence of the collusive level on each possible entry is depicted in Figures~\ref{fig:non_linear_main_effects} and the left panel of \ref{fig:self_externality_v0}. We note that if the nonzero externality is either positive or sufficiently negative, then the collusion is higher in platform competition than in traditional Bertrand competition. 
Furthermore, when $\Phi$ is a diagonal matrix, the dependence of the collusive level on these entries is depicted in Figures~\ref{fig:non_linear_main_effects} and the left panel of \ref{fig:non_linear_bbss}. We note that when these entries are both positive, collusion is higher in platform competition than in single-sided Bertrand competition. These findings suggest that in markets such as online (or cloud) gaming, such as in \textit{Xbox}, mobile and computer games, where positive within-side externalities are significant, algorithmic pricing will increase collusion levels  above those observed in baseline Bertrand competition.  Finally, when  $\Phi$ has only off diagonal non-zero entries, the dependence on the collusive level on these entries is depicted in  Figures~\ref{fig:non_linear_main_effects} and the right panel of  \ref{fig:non_linear_bbss}. We note that when these entries are both positive, collusion sharply increases above the baseline Bertrand level. These findings suggest that in markets such as video streaming (e.g., \textit{Netflix}, \textit{Hulu}, and \textit{Amazon}) and social media markets (e.g., \textit{Instagram} and \textit{TikTok}), where positive cross-side externalities are significant, high levels of collusion can be expected if platforms use algorithmic pricing.  

Traditionally, platforms use positive network externalities to enhance demand and profits  by subsidizing one side of the market in order to attract population on the other sides (see, e.g., \cite{armstrong2007two}, \cite{tan2021effects}, and \cite{chica2021exclusive}). Our findings suggest that algorithmic driven platforms may also learn to use positive network externalities to significantly increase the profit. 

Next we discuss the other scenario, where network externalities result in relatively small levels of collusion. 
First, we note that the right panels of both Figures~\ref{fig:self_externality_v0} and~\ref{fig:cross_externality_v0} 
indicate examples where either the within-side externalities (Figure~\ref{fig:self_externality_v0}) or the cross-side externalities (Figure~\ref{fig:cross_externality_v0}) are both negative with the same magnitude and in these cases the collusive level remains flat at the baseline competition level $\Delta_0$.  
On the other hand, Figure~\ref{fig:bivariate_cases} shows a case where the cross-side externality ($\phi_{bs}$) is large and positive, and the within-side externality is sufficiently small and negative. In this scenario, the collusive level remains flat at a value slightly above $\Delta_0$. 
The latter example is relevant to ride-sharing markets, where drivers compete with each other for riders, while riders benefit from faster pickup times. In this case, when using algorithmic pricing, our experiments indicate that collusive levels are close to the baseline level $\Delta_0$.

Our findings reveal some interesting patterns in the dependence of the collusive level on three different market parameters: the degree of heterogeneity in users' tastes $\beta_k$, the constant term of the outside option utility $u^{(0)}_k$, and the discount rate $\delta$. As shown by Figure~\ref{fig:beta_v}, the collusive level sharply decreases as the degree of heterogeneity in users' tastes increases from $0.2$ to $1$. Afterwards, it remains flat around values lower than the baseline collusive level $\Delta_0$. These findings can also be observed in Appendix \ref{append:other_simulation} for multiple choices of the externality matrix $\Phi$. 
We are not aware of any previous result like this. A different result states that higher degree of heterogeneity in users' tastes leads to inelastic demand and higher price, which in turn leads to higher individual profits (see, e.g., \cite{perloff1985equilibrium} and \cite{anderson1992logit}). However, the collusion level generally does not correlate with individual profit values.   

Next, Figure~\ref{fig:u0_change_main} indicates a main trend of decrease of the collusive level as a function of the outside option utility. This behavior coincides with the 
observation of \cite{cristian2023competition} (see Proposition 4.6) that as the value of the outside option increases, market power held by the platforms decreases. We further noticed some very local trends, but they depend on specific choices of the externalities. This behavior has not been observed before and needs to be further explored.  
The main trend of the dependence of the collusive level on $u_k^{(0)}$ has relevant market implications. For instance, we note that for ride-sharing platforms using AI pricing, our analysis shows that an emphasis on increasing the outside option utility would decrease price and collusion levels. Such outside option utility can be increased by enhancing the public transportation system, and increasing mobility options such as e-bikes. Instead of investing further in such options and, in particular, in the safety of public transportation, the city of Minneapolis chose to decrease prices by passing an ordinance\footnote{The ordinance can be found at \href{https://lims.minneapolismn.gov/Download/FileV2/32072/Transportation-Ride-Share-Worker-Protection-Ordinance.pdf}{https://lims.minneapolismn.gov/Download/FileV2/32072/Transportation-Ride-Share-Worker-Protection-Ordinance.pdf}} that would force the two major ride-sharing platforms in the city, \textit{Uber} and \textit{Lyft}, to pay drivers the city's minimum hour wage. As a result, \textit{Uber} and \textit{Lyft} announced plans to leave the market. Our analysis indicates that there are other strategies to mitigate the problem. 

Finally, as shown in Figure \ref{fig:delta_change}, the collusive level increases as the discount rate increases. This result coincides with earlier results by \cite{calvano2020artificial}. However, the rate at which the collusive level increases w.r.t.~$\delta$ seems to be larger for the case of platform competition compared to Bertrand competition. In fact, most of our experiments in section \ref{sect:experiments} use a value of $\delta = 0.05$, which would be considered a low discount rate. Our sensitivity analysis in Appendix \ref{appendix} indicates that this is likely tacit collusion and not just high prices. As mentioned earlier, for a more specific case, we believe that the main reason for the higher rate of increase of the collusive level in the platform competition model is due to a larger action space.

%% file: policy.tex
U.S.~senator Amy Klobuchar introduced the \textit{S.3686 - Preventing Algorithmic Collusion Act of 2024} in January 2024,\footnote{The act can be consulted at  \href{https://www.congress.gov/bill/118th-congress/senate-bill/3686/text}{https://www.congress.gov/bill/118th-congress/senate-bill/3686/text},} whose abstract is as follows:
\begin{center}
``\textit{A bill to prevent anticompetitive conduct through the use of pricing algorithms by prohibiting the use of pricing algorithms that can facilitate collusion through the use of nonpublic competitor data, creating an antitrust law enforcement audit tool, increasing transparency, and enforcing violations through the Sherman Act and Federal Trade Commission Act, and for other purposes.}'' 
\end{center}

This bill reflexes the increased concern by congress members and governmental institutions for the use of algorithmic price collusion. If such legislation succeeds, it will constitute a major advancement for consumer safety against the potential threats of AI. However, the above act is rather limited. 
Indeed, it only targets AI algorithms trained with nonpublic competitor data and characterizes them as unlawful:

\begin{center}
``\textit{
SEC. 4. PREVENTING COLLUSIVE ACTIVITY IN PRICING ALGORITHMS. (a) In General.—It shall be unlawful for a person to use or distribute any pricing algorithm that uses, incorporates, or was trained with nonpublic competitor data.
}'' 
\end{center}
Potentially collusive AI algorithms, such as the ones presented in this work using $Q$-learning, are left out of the scope of this act. 
We showed that these algorithms can learn to sustain high levels of collusion using only publicly available data, even in cases where agents have limited memory capacity. While this general observation has been established in previous works (\cite{waltman2008q}, \cite{calvano2020artificial}, \cite{klein2021autonomous} and \cite{clark2023algorithmic}), this work shows that in the presence of positive network externalities and two-sided markets, algorithmic driven platforms achieve collusive levels higher than those shown in previous works. 

The question of why these algorithms can achieve high levels of collusion and whether there exists simple conditions to avoid it remains open. 
One way of decreasing collusion levels is obtained by increasing the value of the outside option utility, which has to be done in different ways for different markets (see e.g., the discussion in Section \ref{sect:Discussion} for the ride-sharing platforms). Nevertheless, we next suggest a preliminary  
policy recommendation that can help avoid the risk of collusion by algorithmic driven AI agents in multiple markets. It is based on penalizing the $Q$-learning rewards.

\vspace{5pt}
\noindent \textbf{Policy Recommendation (Q-learning with penalty term).}
We describe a very basic approach for reducing the collusive level by a potential intervention method. 
It can only be effective if it is enforced by regulators. The method introduces a penalty coefficient $\rho \geq 0$ that regularizes the $Q$-learning update formula as follows: 
\begin{align*}
    Q^{(i)}_{t+1}(\vs_t^{(i)}, \va_t^{(i)}) &:= (1-\alpha) Q^{(i)}_t(\vs_t^{(i)}, \va_t^{(i)}) \notag \\
    &+\alpha \left({\pi}_t^{(i)} + \delta \max_{\va} Q^{(i)}_t(\vs_t^{(i)}, \va) - \rho  \left((p^{(i)}_b - \bar{p}_b)_+ + (p_s^{(i)} - \bar{p}_s)_+\right)
\right), \notag
\end{align*}
where $(p_k - \bar{p}_k)_+ = \max\{p_k - \bar{p}_k, 0\}$ and $\bar{p}_k = \frac{1}{N}\sum_{i=1}^N p^{(i)}_k$, for $k\in\{b, s\}$. 
Notice that the penalty term, $\rho  \left((p^{(i)}_b - \bar{p}_b)_+ + (p_s^{(i)} - \bar{p}_s)_+\right)$, only becomes active, if either of the current prices is larger than the average price charged by all the platforms in the market at time $t$. 

One should fix $\rho$ to ensure a tolerable collusive level. For example, the strongest requirement of having no collusion fixes $\rho$ such that $\tilde{\Delta}_T = 0$. 
If on the other hand, one accepts the collusion level with no externalities, i.e., when $\Phi=0$, then one may fix $\rho$ 
such that $\tilde{\Delta}_T = \Delta_0$. We note that any such chosen value of $\rho$ is a function of the parameters of the model $\{\beta_k,\delta,\Phi,u_k^{(0)}\}$.  
Indeed, our findings suggest that any policy recommendation aimed to reduce the risk of algorithmic price collusion needs to be dependent on market parameters.

\begin{figure}[H]
    \centering
    \includegraphics[width=0.45\linewidth]{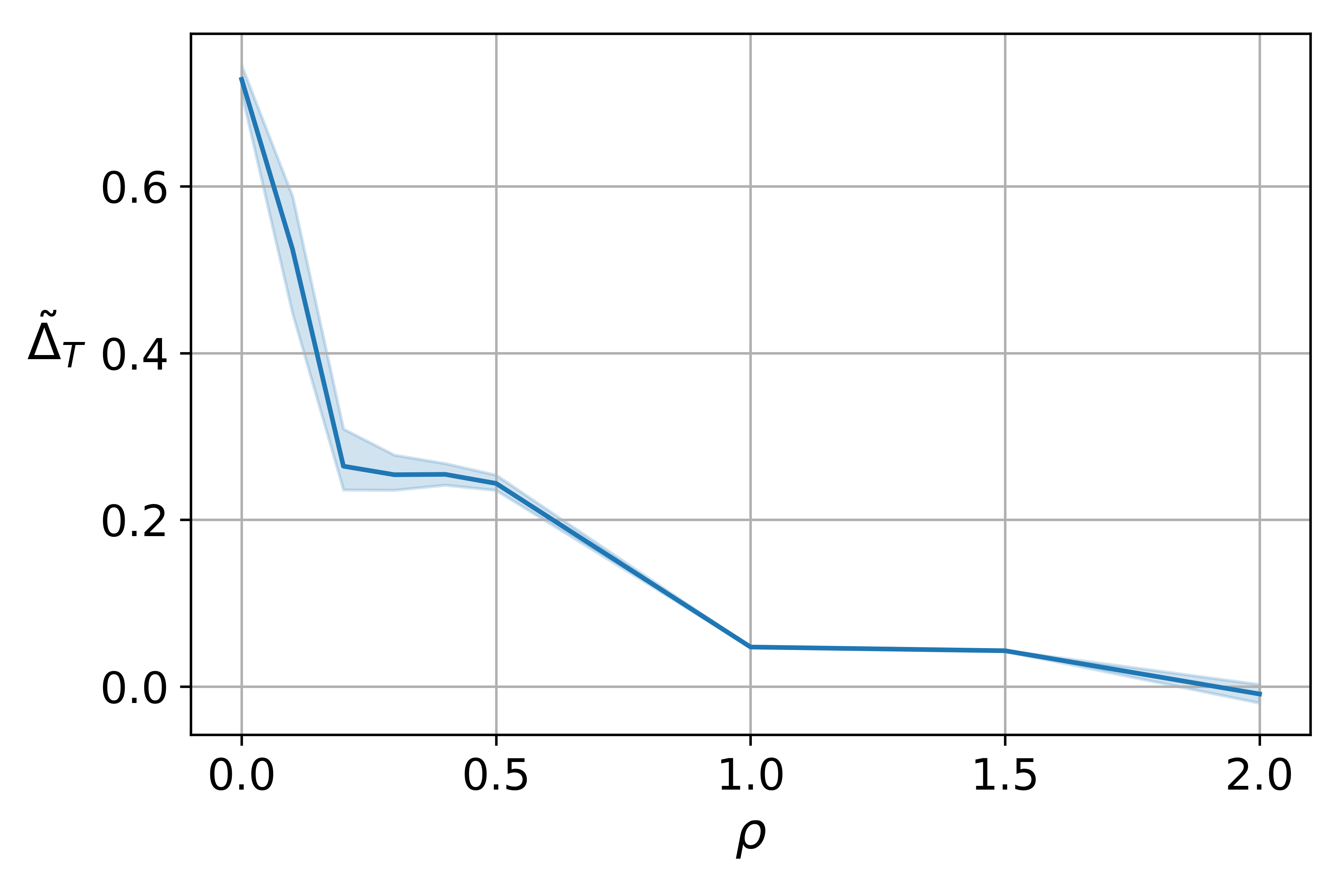}
    \includegraphics[width=0.45\linewidth]{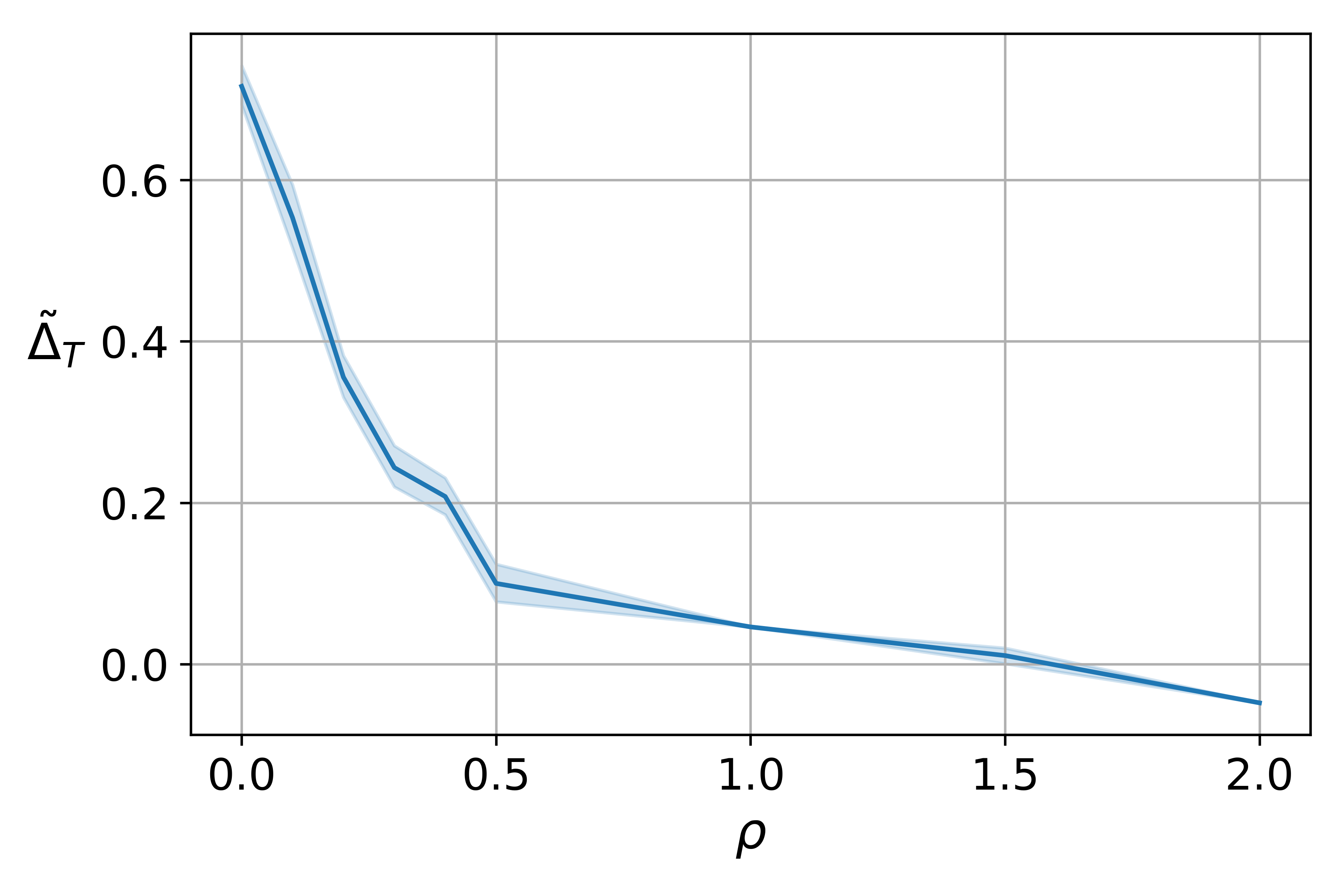}
    \caption{Collusive level with varying $\rho\in[0,2]$: $\Phi = [2, 0; 0, 2]$ (left) and $\Phi = [0, 2; 2, 0]$ (right).}
    \label{fig:penalty}
\end{figure}

Figure~\ref{fig:penalty} investigates the dependence of the collusive level on the penalty coefficient $\rho$ for a setting with $\Phi = [2, 0; 0, 2]$ (left panel) and $\Phi = [0, 2; 2, 0]$ (right panel). 
Without penalty, these two cases have shown significantly high collusive levels (above 0.7) as shown in the right panel of Figure~\ref{fig:self_externality_v0} and the right panel of Figure~\ref{fig:cross_externality_v0}.
In both cases, the collusive level reduces sharply as $\rho$ increases and it reaches $\Delta_0$ when $\rho$ is approximately $0.2$ (left panel) and $0.3$ (right panel). For completeness, we also show extreme cases where the collusive level can be negative, which are different from our above recommendations for choosing $\rho$.

%% file: concludingRemarks.tex
We provided a framework for exploring algorithmic price collusion in a model of platform competition and numerically studied how the collusive level depends on the externality matrix. 

For the case of zero network externalities, the profit gain relative to the competition profits is about 30\%, higher than the value reported by \cite{calvano2020artificial} for the same value of $\delta$. We attribute this difference to the larger action space, which allows the platforms to algorithmically communicate more information, increasing the chances of achieving a high collusive level. 

In common economic scenarios with positive network externalities, particularly when the positive elements are either single entries, diagonal elements, or off-diagonal elements of $\Phi$, the collusive level in platform competition is significantly higher than in traditional Bertrand competition. This suggests that in markets such as online gaming, video streaming, and social media, AI-driven platforms may exploit positive network externalities to significantly increase profits through algorithmic pricing, indicating potential for high levels of collusion.

Our findings also reveal patterns in how the collusive level depends on user heterogeneity, discount rates, and outside option utility. Specifically, greater heterogeneity in user tastes leads to lower collusive levels. Additionally, collusion increases with higher discount rates, consistent with findings by  \cite{calvano2020artificial} for Bertrand competition, though the rate of increase is higher in platform competition. We provide evidence in the appendix that this collusion is tacit. Furthermore, higher outside option utility generally decreases collusion levels within a sufficiently large domain. This suggests that market regulators can reduce collusion by enhancing the value of outside options, such as improving public transportation in ride-sharing markets.

We also proposed a version of $Q$-learning with a penalty term to reduce the risk of algorithmic price collusion. It can be used for policy recommendations by considering market parameters.

There are several open future research directions. First, while the platform competition model in \cite{cristian2023competition} allows for an unlimited number of platforms, we only conducted numerical experiments with two platforms due to the exponential growth of the state space and computational resource requirements with more platforms. Future research could explore alternative methods to estimate the $Q$-function to mitigate high computational demands. Second, replicating our experiments using a model that includes nonlinear network externality effects and allows for multi-homing, where users join multiple platforms, would be valuable.

%% file: appendix.tex
Section \ref{appd:sensitive} presents sensitivity analysis 
and exemplifies convergence paths. Section \ref{appendix:fitting_method}  provides extra details on how to fit the univariate and bivariate functions of equation \eqref{eqn:fitting_Delta}. Section \ref{append:other_simulation} provides additional numerical experiments on the dependence of the collusive level on $\beta_k$ and $u_k^{(0)}$.

\subsection{Sensitivity Analysis and Examples of Convergence Paths}
\label{appd:sensitive}

We assume special parameterizations of the network externality matrix, $\Phi$, and perform a sensitivity analysis for our simulation framework. For each specific $\Phi$, we ran 500 simulations and classify the behavior in the last 5000 time steps of each run into one of five categories: 
\begin{enumerate}
    \item Symmetric 1-cycle (or 1-Sym), where platforms $1$ and $2$ repeatedly choose actions $p_1$ and $p_2$, respectively, and $p_1=p_2$.  
    \item Asymmetric 1-cycle (or 1-Asym), where platforms $1$ and $2$ repeatedly choose actions $p_1$ and $p_2$, respectively, and $p_1\neq p_2$. 
    \item Cycle of length 2-4 (or C2-4), where platforms $1$ and $2$ repeatedly choose actions that follow a pattern of length two to four. For example, the actions in a cycle of length 3 are $\{(p_{1},p_2),(q_1,q_2),(r_1, r_2)\}$. 
    \item Cycle of length 5-8 (or C5-8), where the two  platforms repeatedly choose actions that follow a pattern of length five to eight. 
    \item Cycle of length at least 9 (or C9), where the two  platforms repeatedly choose actions that follow a pattern of length at least nine.  
\end{enumerate}  
Let $k\in\{1,\dots,5\}$ index the five categories described above, and $C_k$ denote the set of simulations within category $k$. For each $k\in \{1,\dots,5\}$ and a chosen $\Phi$, we report the following measures: 
\begin{itemize}
\item Frequency (Freq.) of the category among the 500 simulations.

\item Average (Avg.) and standard deviation (S.D.) of the collusive level $\tilde{\Delta}$ within the category. 

\item Frequency of equilibrium, which quantifies the proportion of simulations in which the AI driven platforms learn the action that achieves maximal total future reward given that the opponent plays with the final $Q$-function. We compute this frequency, denoted by $Y_k\in [0,1]$ for $k \in \{1, \ldots, 5\}$, according to the following three steps: 
    \begin{enumerate}
    \item  
Let $Q^{*}_i$ be the final $Q$-function of platform $i\in\{1,2\}$ in a given simulation $s\in C_k$. 
It is the numerical approximation to $Q^{*}_i$ defined in \eqref{eqn:bellman}, as explained in Section    \ref{subsec:problem}. We use this notation in defining other frequencies below. 
We estimate, without loss of generality, platform's $1$ best response to $Q^{*}_2$ using backward induction. For $\hat{T}=10$, initiate $$Q^{(\hat{T})}_1(p, x) = \pi^{(1)} (p, \argmax_{p'} Q^*_2(p', x) ),$$ 
    where $p$ is a vector of prices and $x$ is the state variable.
    
    For $t\in\{1,\dots, \hat{T}\}$, compute via backward induction 
    $$
    Q^{(t-1)}_1(p, x) = \pi^{(1)} (p, \argmax_{p'} Q^*_2(p', x)) + \delta \max_{p''} Q^{(t)}_1(p'', x'), 
    $$
    where $x' = (p, \argmax_{p'} Q^\ast_2(p', x))$. The matrix $Q^{(0)}_1$ is used as an approximation to platform's $1$ best response to $Q^{*}_2$ in the given simulation $s\in C_k$.
    
    \item Let $P_s$ denote the set of states used in the convergence path of $Q_i^{*}$ in the given simulation $s\in C_k$. We use this definition in defining the other frequencies below. For each state $(p_1, p_2)\in P_s$, we form an indicator variable $X_s(p_1,p_2)$ that checks if the action that platform $1$ takes using $Q^{*}_1$ approximates the action that it would take using $Q^{(0)}_1$. Thus, for each $(p_1, p_2)\in P_s$, let $p^{0}:= \argmax_{p'} Q^{(0)}_1(p', (p_1,p_2) )$ and $p^*:= \argmax_{p'} Q^{*}_1(p', (p_1,p_2) )$, then $X_s(p_1,p_2) = 1$ if and only if either $p^{0}=p^{*}$ or $p^{0}$ and $p^{*}$ are neighbors with distance at most $1$ in the set of indices of the $Q$-matrix function. 
    \item Let $Y_k =\frac{1}{|C_k|}\sum_{s\in C_k}\frac{1}{|P_s|}\sum_{(p_1,p_2)\in P_s}X_s(p_1,p_2)$. 
    \end{enumerate}

\item Frequency of one-step equilibrium, which quantifies the proportion of simulations in which the AI driven platforms learn the action that achieves maximal one-step reward given that the opponent plays with the final $Q$-function. Note that the underlying learning involves no memory. 
We compute this frequency, denoted by $Y_k^{(\textnormal{one})}\in[0,1]$ for $k \in \{1, \ldots, 5\}$, according to the following three steps: 
    \begin{enumerate}
    \item  Let $Q^{(\textnormal{one})}_i(p,x) = \pi^{(1)} (p, \argmax_{p'} Q^*_2(p', x) ).$  

\item Let $p^{(\textnormal{one})}:= \argmax_{p'} Q^{(\textnormal{one})}_1(p', (p_1,p_2) )$. We define $X_s^{(\textnormal{one})}(p_1,p_2) = 1$ if and only if either $p^{(\textnormal{one})}=p^{*}$ or $p^{(\textnormal{one})}$ and $p^{*}$ are neighbors with distance at most $1$ in the set of indices of the $Q$-matrix function.

\item Let $Y_k^{(\textnormal{one})} =\frac{1}{|C_k|}\sum_{s\in C_k}\frac{1}{|P_s|}\sum_{(p_1,p_2)\in P_s}X_s^{(\textnormal{one})}(p_1,p_2)$. 

\end{enumerate}

\item Frequency of converging back to the limiting action, which quantifies the proportion of simulations in which after one platform unilaterally deviates to the one-stage Nash equilibrium price, platforms return back to the limiting action. We compute this frequency, $Y_k^{(\textnormal{b})}\in[0,1]$ for $k \in \{1, \ldots, 5\}$, according to the following three steps: 
\begin{enumerate}
    \item At time $T+1$, set $p_{T+1}^{(1)} = (p_b^*,p_s^*)$, while $p_{T+1}^{(2)}\in \argmax_{p} Q_2^{*}(p,(p_{T}^{(1)},{p_{T}^{(2)}}))$. For $\tau \in\{2,\ldots,101\}$ and $i \in \{1,2\}$, let $p_{T+\tau}^{(i)}\in \argmax_{p} Q_i^{*}(p,(p_{T+\tau-1}^{(1)},{p_{T+\tau-1}^{(2)}}))$.
    \item Let $p^{(\textnormal{b}),(i)}:=p_{T+101}^{(i)}$. We define $X_s^{(\textnormal{b})}(p^{(\textnormal{b}),(1)},p^{(\textnormal{b}),(2)}) = 1$ if and only if $(p^{(\textnormal{b}),(1)},p^{(\textnormal{b}),(2)})\in P_s$.

\item Let $Y_k^{(\textnormal{b})} =\frac{1}{|C_k|}\sum_{s\in C_k}X_s^{(\textnormal{b})}(p^{(\textnormal{b}),(1)},p^{(\textnormal{b}),(2)}) $. 
\end{enumerate}

\item  Frequency of converging back to the limiting action for both platforms, which quantifies the proportion of simulations in which after both platforms deviate to the one-stage Nash equilibrium price, they return back to the limiting action. This frequency is computed similarly to the one above, where the only difference is that  $p_{T+1}^{(1)} = p_{T+1}^{(2)} = (p_b^*,p_s^*)$.

\item Average and standard deviation of the $Q$-loss w.r.t.~$P_s$. The $Q$-loss quantifies how close are the observed rewards on path $P_s$, $\sum_{\tau=0}^{100} \delta^\tau\pi_{t+\tau}^{(i)}$, to the optimal rewards given by the best response $Q_i^{(0)}$. We highlight that after convergence, the observed rewards should be pretty close in value to $\max_{p\in\gA}Q^{*}_i(p,x)$ for each state $x\in P$.  Let $T_0 = T - 5000$, we compute the $Q$-loss for a given simulation $s$ as 
$$
\frac{1}{|P_s|}\sum_{t = T_0}^{T_0+|P_s|}\left| \sum_{\tau=0}^{100} \delta^\tau\pi_{t+\tau}^{(i)} - \max_{p\in\gA} Q^{0}_i(p,x_t)\right|.
$$
The average and standard deviation are taken w.r.t.~all simulations $s \in C_K$ for any $k \in \{1, \ldots, 5\}$.

\item Average and standard deviation of the $Q$-loss w.r.t.~all states. The $Q$-loss w.r.t.~all states quantifies in average how close are the maximum values of $Q^{*}_i(\cdot,x)$ and $Q^{0}_i(\cdot, x)$, where the maximum is taken over all actions and the average over all states $x\in \gS$. We compute the $Q$-loss w.r.t.~all states for a given simulation $s$ as
$$\frac{1}{|\gS|}\sum_{x\in \gS}\left|\max_{p\in\gA}Q^{*}_i(p,x)-\max_{p\in\gA} Q^{0}_i(p,x)\right|.$$
The average and standard deviation are taken with respect to all simulations $s \in C_K$ for any $k \in \{1, \ldots, 5\}$. When the $Q$-loss w.r.t.~all states equals $0$, platforms' maximal rewards of the final $Q$-function yield the same maximal rewards of the best response $Q$-function. We interpret this as follows: the final $Q$-function exhibits behavior consistent with equilibrium off-path, also known as subgame perfection.
\end{itemize}

Tables~\ref{tab:sensitivity_zero}, \ref{tab:sensitivity_diag} and \ref{tab:sensitivity_asym} show the above measures for the following three respective choices of $\Phi$: $\Phi_1:=[0, 0; 0, 0]$, $\Phi_2:=[1, 0; 0, 1]$ and $\Phi_3:=[0, 1; 1, 0]$. The other parameters are set as in the default setting, i.e., $\delta=0.05$, $\beta_k = 1$, $u_k^{(0)} = -2$. 

In all tables, the frequency of symmetric 1-cycles is relatively small, where its average is 5.8\%. Even though this event could be considered rare, this category is important since within it we can find one-memory stationary equilibria, which are of great interest (see, e.g., \cite{barlo2016bounded} and \cite{chicaguolerman2024}). Note that the frequency of asymmetric 1-cycles is very small, where its average is 2.7\%. We will thus omit interpretations of results for this category. The average of both frequencies of cycles of length 2-4 and 5-8 is 34\% and the one for cycles of length at least 9 is 23.3\%. 

The average collusive level among all tables and categories, excluding 1-Asy, is approximately 29\%. This value is only slightly below the baseline collusive level $\Delta_0$. 

For 1-Sym, the frequencies of equilibrium are 46\%, 28\% and 52\% for $\Phi_1$, $\Phi_2$ and $\Phi_3$, respectively. Thus, approximately half of the simulations within 1-Sym result in equilibrium behavior for $\Phi_1$ and $\Phi_3$, with a smaller value of 28\% for $\Phi_2$. These numbers suggest that in cases of zero-externalities or non-zero cross-side externalities, equilibrium behavior is more likely than in cases of non-zero within-side externalities. Note that for the frequency of one-step equilibrium in 1-Sym, we still observe a larger proportion of one-step equilibrium for $\Phi_1$ and $\Phi_3$ compared to $\Phi_2$. 

For the categories C2-4, C5-8 and C9, the average frequencies of equilibrium are 28.8\%, 42.9\% and 49.4\% for $\Phi_1$, $\Phi_2$ and $\Phi_3$, respectively. Thus, on average 46\% of the simulations with cycle lengths greater than one result in equilibrium behavior for $\Phi_2$ and $\Phi_3$, with a smaller value of 28.8\% for $\Phi_1$. These numerical results suggest that in cases of non-zero externalities and for cycles of lengths greater than one, equilibrium behavior is more likely than in the zero-externality case. The above results for 1-Sym, C2-4, C5-8 and C9 suggest that in a sufficiently large percentage of cases, AI driven platforms exhibit behavior that is consistent with Nash equilibrium. 

The frequencies of converging back to the limiting action for one platform in the 1-Sym category are 84\%, 81\% and 76\% for $\Phi_1$, $\Phi_2$ and $\Phi_3$, respectively. Thus, in at least 76\% of the cases, platforms converge back to the limiting action after one unilateral price change to the Nash equilibrium price.  For the categories C2-4, C5-8 and C9, the frequencies of converging back to the limiting action for one platform are 90\%, 92\% and 95\% for $\Phi_1$, $\Phi_2$ and $\Phi_3$, respectively.  Note that the smallest frequency of converging back to the limiting action for one platform in the 1-Sym category is achieved at $\Phi_3$, which suggests that non-zero cross-side externalities make it harder to sustain collusion. A similar result is shown by \cite{ruhmer2010platform} for a case of platform competition without AI agents. Nonetheless, this result does not hold for cycles of lengths greater than one, where the smallest frequency is achieved at $\Phi_1$, i.e., at the zero-externalities case. As a general observation, the above results suggest that AI driven platforms can learn behavior that is consistent with tacit collusion. Note that similar behavior is found for the frequencies of converging back to the limiting action for both platforms.

The average $Q$-losses (on path) for all tables and categories, excluding 1-Asy, are all less than $0.019$. This observation and the definition of the $Q$-loss imply that the observed rewards on path are sufficiently close to the optimal rewards given by the best response matrix. These results suggest that the likelihood of the final action achieving maximal total future reward is very high. Similarly, the average $Q$-losses w.r.t.~all states for all categories, excluding 1-Asy, are less than $0.043$. Thus, platforms' maximal rewards of the final $Q$-function are relatively close to the maximal rewards of the best response $Q$-function. Furthermore, the $Q$-function exhibits behavior consistent with the equilibrium off-path or subgame perfection.

\begin{table}[H]
    \centering
\begin{tabular}{lllllll}
\toprule
Metric                                    &  1-Sym &  1-Asy &  C2-4 &  C5-8 &  C9 &    All \\
\midrule
Freq.                                   &   0.040 &  0.025 &      0.300 &      0.307 &     0.328 &  1.000 \\
Avg. Collusive Level                 &   0.273 &  0.326 &      0.294 &      0.301 &     0.295 &  0.297 \\
S.D. of Collusive Level   &   0.082 &  0.069 &      0.063 &      0.049 &     0.042 &  0.054 \\
Freq. of Eq.                    &   0.462 &    0.0 &      0.279 &      0.296 &      0.29 &  0.288 \\
Freq. of one-step Eq.      &   0.385 &  0.125 &      0.314 &      0.324 &     0.345 &  0.325 \\
Freq. Conv. back (one)  &   0.846 &  0.750 &      0.907 &      0.909 &     0.896 &  0.898 \\
Freq. Conv. back (both) &   0.846 &  0.750 &      0.856 &      0.879 &     0.925 &  0.882 \\
Avg. Q loss (on path)                &   0.003 &  0.006 &      0.005 &      0.005 &     0.005 &  0.005 \\
S.D. of Q loss (on path)  &   0.001 &  0.004 &      0.003 &      0.002 &     0.001 &  0.002 \\
Avg. Q loss (all)                    &   0.009 &  0.009 &      0.009 &      0.009 &     0.009 &  0.009 \\
S.D. of Q loss (all)      &   0.000 &  0.000 &      0.000 &      0.000 &     0.000 &  0.000 \\
\bottomrule
\end{tabular}

    \caption{Sensitivity analysis with $\Phi=\Phi_1:=[0, 0; 0, 0]$.}
    \label{tab:sensitivity_zero}
\end{table}

\begin{table}[H]
    \centering
\begin{tabular}{lllllll}
\toprule
Metric                                    &  1-Sym &  1-Asy &  C2-4 &  C5-8 &  C9 &    All \\
\midrule
Freq.                                   &   0.084 &  0.013 &      0.355 &      0.368 &     0.179 &  1.000 \\
Avg. Collusive Level                 &   0.297 &  0.433 &      0.298 &      0.301 &     0.322 &  0.305 \\
S.D. of Collusive Level   &   0.061 &  0.230 &      0.065 &      0.058 &     0.067 &  0.068 \\
Freq. of Eq.                    &   0.281 &    0.0 &       0.44 &      0.436 &     0.411 &  0.414 \\
Freq. of one-step Eq.     &   0.063 &  0.000 &      0.366 &      0.326 &     0.320 &  0.313 \\
Freq. Conv. back (one)  &   0.812 &  0.400 &      0.904 &      0.957 &     0.912 &  0.911 \\
Freq. Conv. back (both) &   0.812 &  0.400 &      0.889 &      0.943 &     0.897 &  0.897 \\
Avg. Q loss (on path)                &   0.011 &  0.042 &      0.015 &      0.017 &     0.019 &  0.016 \\
S.D. of Q loss (on path)  &   0.005 &  0.030 &      0.011 &      0.010 &     0.011 &  0.011 \\
Avg. Q loss (all)                    &   0.043 &  0.043 &      0.043 &      0.043 &     0.043 &  0.043 \\
S.D. of Q loss (all)      &   0.001 &  0.002 &      0.001 &      0.001 &     0.001 &  0.001 \\
\bottomrule
\end{tabular}

\caption{Sensitivity analysis with $\Phi=\Phi_2:=[1, 0; 0, 1]$.}
    \label{tab:sensitivity_diag}
\end{table}

\begin{table}[H]
    \centering
\begin{tabular}{lllllll}
\toprule
Metric                                    &  1-Sym &  1-Asy &  C2-4 &  C5-8 &  C9 &    All \\
\midrule
Freq.                                   &   0.049 &  0.042 &      0.364 &      0.353 &     0.193 &  1.000 \\
Avg. Collusive Level                 &   0.264 &  0.282 &      0.279 &      0.265 &     0.287 &  0.275 \\
S.D. of Collusive Level   &   0.057 &  0.079 &      0.075 &      0.053 &     0.061 &  0.065 \\
Freq. of Eq.                    &   0.524 &  0.336 &      0.533 &      0.499 &     0.451 &  0.497 \\
Freq. of one-step Eq.      &   0.245 &  0.342 &      0.369 &      0.365 &     0.321 &  0.351 \\
Freq. Conv. back (one)  &   0.762 &  0.500 &      0.930 &      0.967 &     0.976 &  0.926 \\
Freq. Conv. back (both) &   0.667 &  0.333 &      0.904 &      0.954 &     0.964 &  0.898 \\
Avg. Q loss (on path)                &   0.009 &  0.019 &      0.013 &      0.012 &     0.015 &  0.013 \\
S.D. of Q loss (on path)  &   0.004 &  0.010 &      0.009 &      0.006 &     0.009 &  0.008 \\
Avg. Q loss (all)                    &   0.038 &  0.038 &      0.038 &      0.038 &     0.038 &  0.038 \\
S.D. of Q loss (all)      &   0.001 &  0.001 &      0.001 &      0.001 &     0.001 &  0.001 \\
\bottomrule
\end{tabular}
    \caption{Sensitivity analysis with $\Phi=\Phi_3:=[0, 1; 1, 0]$.}
    \label{tab:sensitivity_asym}
\end{table}

\begin{figure}[H]
    \centering
\includegraphics[width=0.45\linewidth]{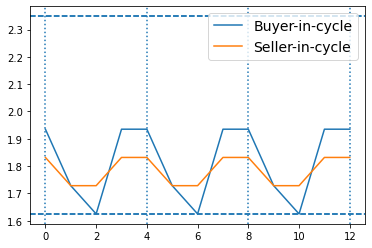} 
\includegraphics[width=0.45\linewidth]{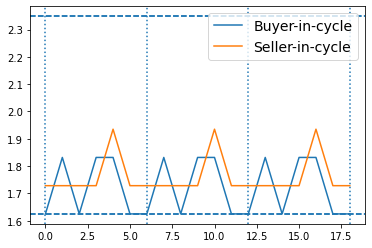}  
\includegraphics[width=0.45\linewidth]{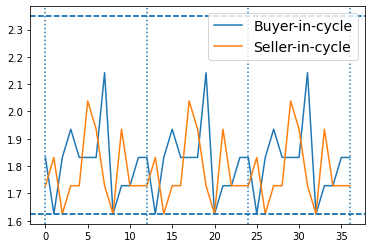} 
\includegraphics[width=0.45\linewidth]{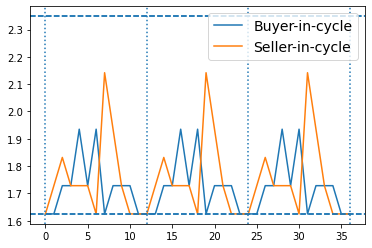} 
\caption{Examples of convergence paths from categories C2-4 (top left panel), C5-8 (top right panel) and C9 (bottom left and right panels), where $\Phi = [0, 1; 1, 0]$. The horizontal blue dashed lines at the bottom and top of each panel represent the lines at $p_b^*=p_s^*$ and $p_b^\text{C}=p_s^\text{C}$, respectively. The vertical blue dotted lines indicate where a cycle ends. The blue and orange curves represent the buyer and seller prices, respectively, chosen by platform 1.}
    \label{fig:converge_final_state}
\end{figure}

Figure~\ref{fig:converge_final_state} shows four examples of convergence paths, one from C2-4 (top left panel), one from C5-8 (top right panel) and two from C9 (bottom left and right panels), where $\Phi = [0, 1; 1, 0]$. The horizontal blue dashed lines at the bottom and top portion of each panel represent the Nash and Collusion equilibrium prices, respectively, i.e., the lines at $p_b^*=p_s^*$ and $p_b^\text{C}=p_s^\text{C}$, respectively. Note that buyer and seller prices are equal due to the symmetric choice of $\Phi$. The vertical blue dotted lines represent where a cycle ends, while the blue and orange curves represent the buyer and seller prices, respectively, chosen by platform 1 on the equilibrium path. The top left panel shows a convergence path of length four, in which the buyer's price starts above the Nash equilibrium price. Then, in two steps, it reaches the Nash equilibrium price, followed by a sudden increase to the starting price. This price pattern is similar to an Edgeworth cycle, where prices start well above the Nash equilibrium price, then slowly converge to the Nash equilibrium, followed by a sudden increase to the initial high price (see, .e.g., \cite{maskin1988theory}). Note that the seller's price oscillates between two levels above the Nash equilibrium price. The top right and bottom left and right subfigures show more intricate patterns. However, a common feature among them is that prices oscillate between the Nash equilibrium price and a level higher than the Nash price. This indicates that platforms learn behavior consistent with punishment and reward strategies.   
\color{black}

\begin{figure}[H]
    \centering
\includegraphics[width=0.45\linewidth]{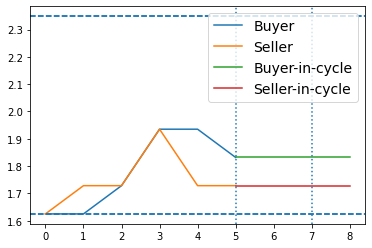} 
\includegraphics[width=0.45\linewidth]{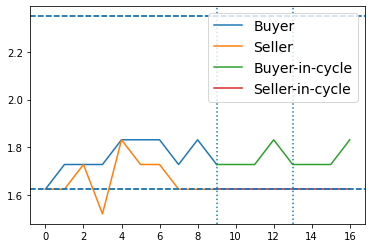}  
    \includegraphics[width=0.45\linewidth]{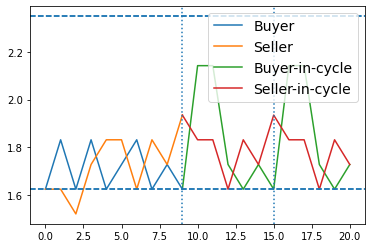} 
    \includegraphics[width=0.45\linewidth]{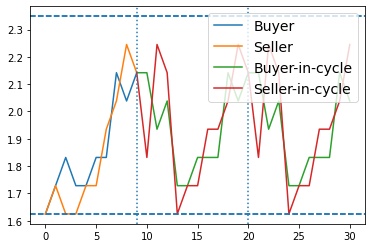} 
    \caption{Four examples of convergence paths after both platforms revert to the limiting action following a deviation to the one-stage Nash equilibrium price by both platforms. There are two examples for $\Phi_1$ (top left and right panels), one for $\Phi_2$ (bottom left), and one for $\Phi_3$ (bottom right). The blue and orange curves represent the buyer and seller prices, respectively, chosen by platform 1 immediately after the deviation to the Nash equilibrium price. The green and red curves represent platform 1's buyer and seller prices once they revert to the limiting action.}
    \label{fig:converge_final_state_deviation}
\end{figure}

Finally, Figure~\ref{fig:converge_final_state_deviation} illustrates four scenarios depicting the convergence path of platform 1 after both platforms revert to the limiting action following a deviation to the one-stage Nash equilibrium price by both platforms. The top left and right panels present an example for $\Phi_1$, while the bottom left and right panels provide examples for $\Phi_2$ and $\Phi_3$, respectively. In all four cases, the horizontal blue dashed lines at the bottom and top of each panel represent the lines at $p_b^*=p_s^*$ and $p_b^\text{C}=p_s^\text{C}$, respectively. The vertical blue dotted lines indicate where each cycle starts and ends. The blue and orange curves represent the buyer and seller prices, respectively, chosen by platform 1 immediately after the deviation to the Nash equilibrium price. The green and red curves represent platform 1's buyer and seller prices once they revert to the limiting action. In all four panels, we observe that in less than ten steps, platform 1's prices revert to the limiting action (the leftmost vertical dotted line appears before step 10). The top left panel shows that after one deviation to the Nash equilibrium price by both platforms, platform 1's prices slowly increase above the Nash equilibrium price, followed by a small decrease and then a convergence to a price above the Nash equilibrium price. Similar behavior is observed in the top right panel. The bottom left and right panels show more intricate patterns. However, as mentioned earlier, in both figures, prices converge back to the limiting action after 9 steps. The patterns they follow after reaching the limiting action indicate that firms oscillate between the Nash equilibrium price and a price higher than the Nash equilibrium price.
\color{black}

\subsection{Fitting a Sequence of Non-linear Functions}\label{appendix:fitting_method}
We provide the details of how to fit the univariate and bivariate functions in \eqref{eqn:fitting_Delta} in order to describe how the collusive level depends on the externality matrix $\Phi$ via an additive model. 
We first let $\Delta_0$ equal the sample mean of $\tilde{\Delta}$. In order to find the four univariate functions $f_{bb}(\phi_{bb}), f_{ss}(\phi_{ss}), f_{bs}(\phi_{bs})$ and $f_{sb}(\phi_{sb})$, we fit them sequentially using residuals. We consider all the twenty-four permutations of 4 elements, represented by $bb$, $ss$, $bs$, $sb$. Without loss of generality, we describe the fitting procedure for a given permutation $o := (bb, ss, bs, sb)$, where the same procedure is followed for all other permutations. 
We follow a regression setting with the predictor $\phi_{bb}$ and the response $$y_{bb}:=\tilde{\Delta} - \Delta_0,$$ and we apply XGBoost, which is a nonlinear regression method, to fit $\hat{f}^{(o)}_{bb}(\phi_{bb})$ that approximates $y_{bb}$. For the above $o$, we proceed from $bb$ to $ss$ as follows. We define $$y_{ss}:= y_{bb} - \hat{f}_{bb}^{(o)}(\phi_{bb})$$ and consider a regression setting with $y_{ss}$ as response and $\phi_{ss}$ as predictor. We apply XGBoost to approximate $y_{ss}$ by  $\hat{f}_{ss}^{(o)}(\phi_{ss})$. Similarly, we transition from $ss$ to $bs$ using the residuals $$y_{bs}:= y_{ss} - \hat{f}_{ss}^{(o)}(\phi_{ss})$$ 
and fitting $\hat{f}^{(o)}_{bs}(\phi_{bs})$ to approximate $y_{bs}$ by XGBoost, and we transition from $bs$ to $sb$ using 
the residual $$y_{sb}:= y_{bs} - \hat{f}^{(o)}_{bs}(\phi_{ss})$$ and fitting $\hat{f}^{(o)}_{sb}(\phi_{sb})$ to approximate $y_{sb}$ by XGBoost. We repeat this process for the rest of the twenty-three permutations and thus obtain twenty-four approximation for the functions $f_{bb}(\phi_{bb})$, $f_{ss}(\phi_{ss})$, $f_{bs}(\phi_{bs})$ and $f_{sb}(\phi_{sb})$, specified in \eqref{eqn:fitting_Delta}. We average over the twenty-four approximations to obtain the estimator $\hat{f}_{bb}(\phi_{bb}), \hat{f}_{ss}(\phi_{ss}), \hat{f}_{bs}(\phi_{bs})$ and $\hat{f}_{sb}(\phi_{sb})$.

Next, we apply the procedure described above for the six bivariate functions in \eqref{eqn:fitting_Delta}, such as $f_{bb, ss} (\phi_{bb}, \phi_{ss})$ and  $f_{bs, sb} (\phi_{bs}, \phi_{sb})$. 
For a given permutation of the six pairs of variables, e.g., $$o=((bb, ss), (bs, sb), (bb, sb) , (bb, bs), (ss, sb), (ss, bs)),$$ we follow the same procedure introduced above to iteratively fit the bivariate functions using XGBoost. 
In the first iteration, the response variable is 
$$
y_1 :=\tilde{\Delta} - \left(\Delta_0 + \hat{f}_{bb}(\phi_{bb})+ \hat{f}_{ss}(\phi_{ss})+ \hat{f}_{bs}(\phi_{bs})+\hat{f}_{sb}(\phi_{sb})\right)$$ 
and for the above permutation $o$ there are two predictors $\phi_{bb}$
 and $\phi_{ss}$. XGBoost then approximates $y_1$ by the function $\hat{f}^{(o)}(\phi_{bb}, \phi_{ss})$.  
In the following iterations, the response is the corresponding residual and the predictors are the two corresponding variables. 
For example, assuming the permutation $o$, in the second iteration the response variable is 
$$
y_1 - \hat{f}^{(o)}(\phi_{bb}, \phi_{ss})
$$
and the predictors are $\phi_{bs} $ and $\phi_{sb}$.
We average the obtained approximations over all 720 possible permutations of these six bivariate functions to approximate each bivariate function.

\subsection{Other Simulation Results}
\label{append:other_simulation}

\begin{figure}[H]
    \centering
\includegraphics[width=0.45\linewidth]{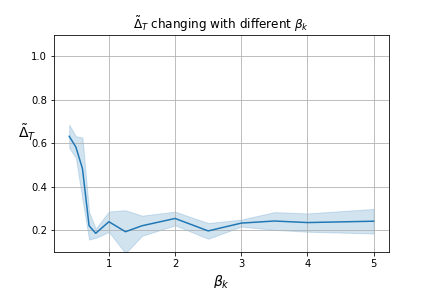} 
\includegraphics[width=0.45\linewidth]{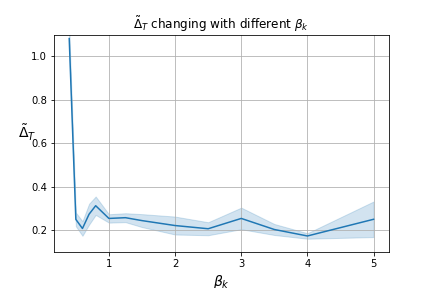}  
    \includegraphics[width=0.45\linewidth]{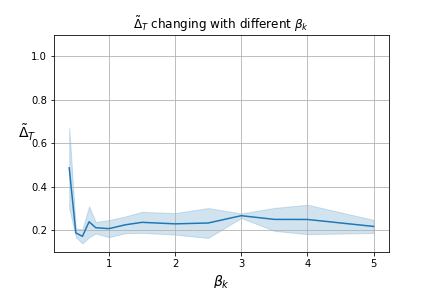} 
    \includegraphics[width=0.45\linewidth]{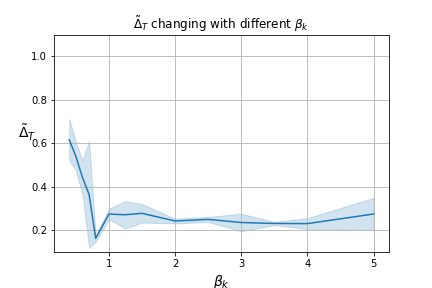} 
    \caption{The collusive level with varying $\beta_k\in [0.2, 5]$; Top left: $\Phi = [0, 1; 1, 0]$; Top right: $\Phi = [0, -1; -1, 0]$; Bottom left: $\Phi = [0, 1; 0, 0]$; Bottom right: $\Phi = [1, 0;0, -1]$.}
    \label{fig:beta_v_appendix}
\end{figure}

\begin{figure}[H]
    \centering
\includegraphics[width=0.45\linewidth]{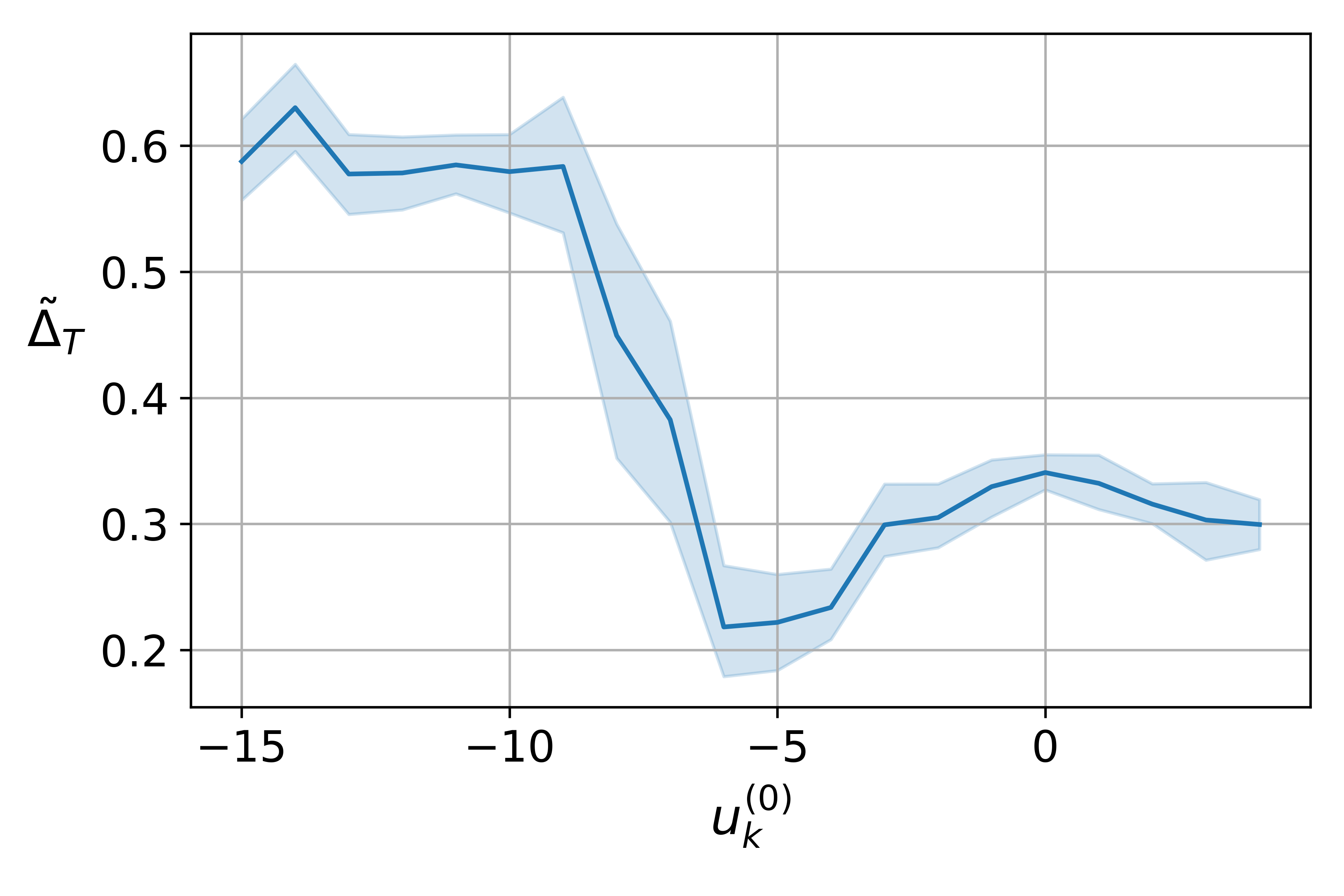}
\includegraphics[width=0.45\linewidth]{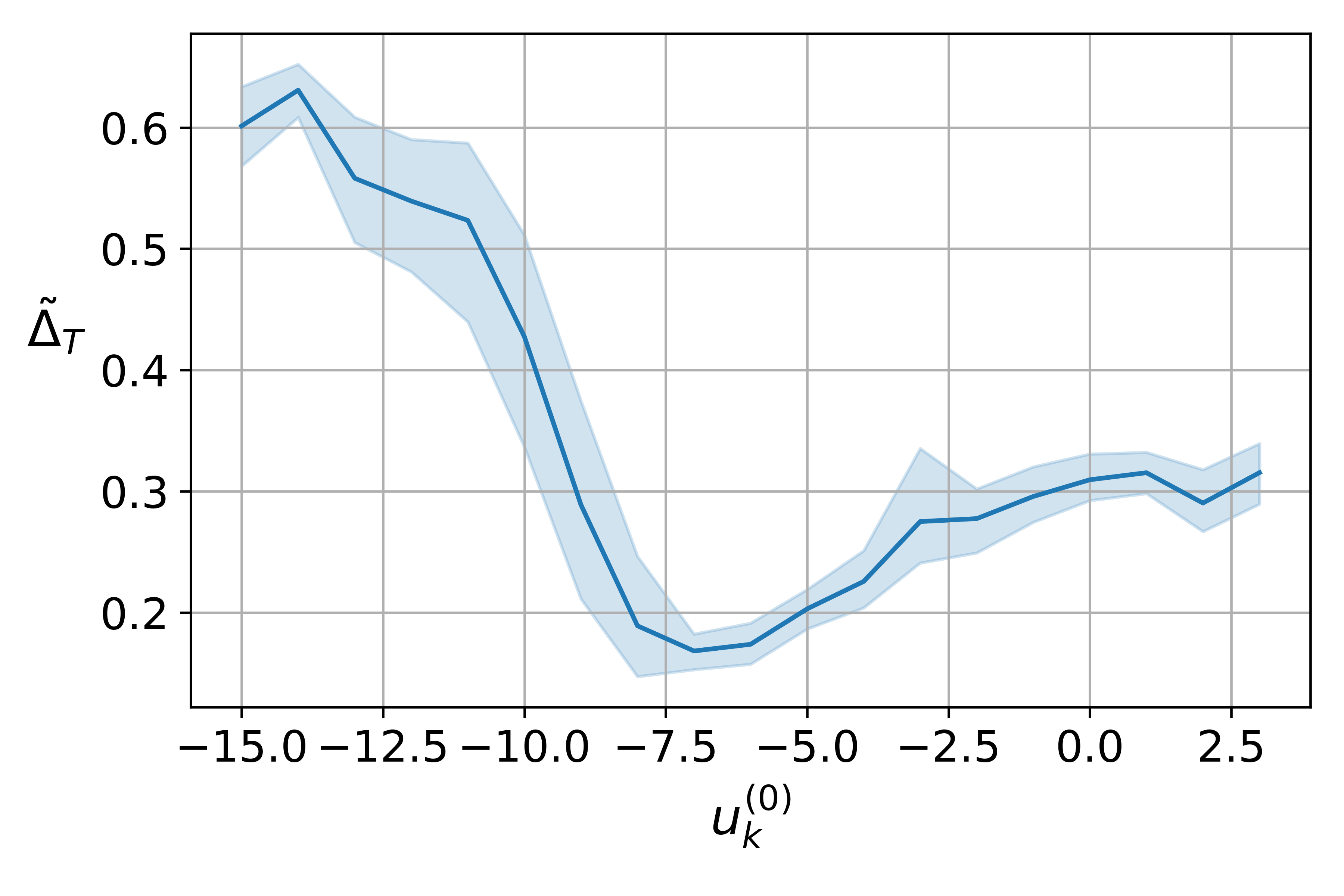}  
\includegraphics[width=0.45\linewidth]{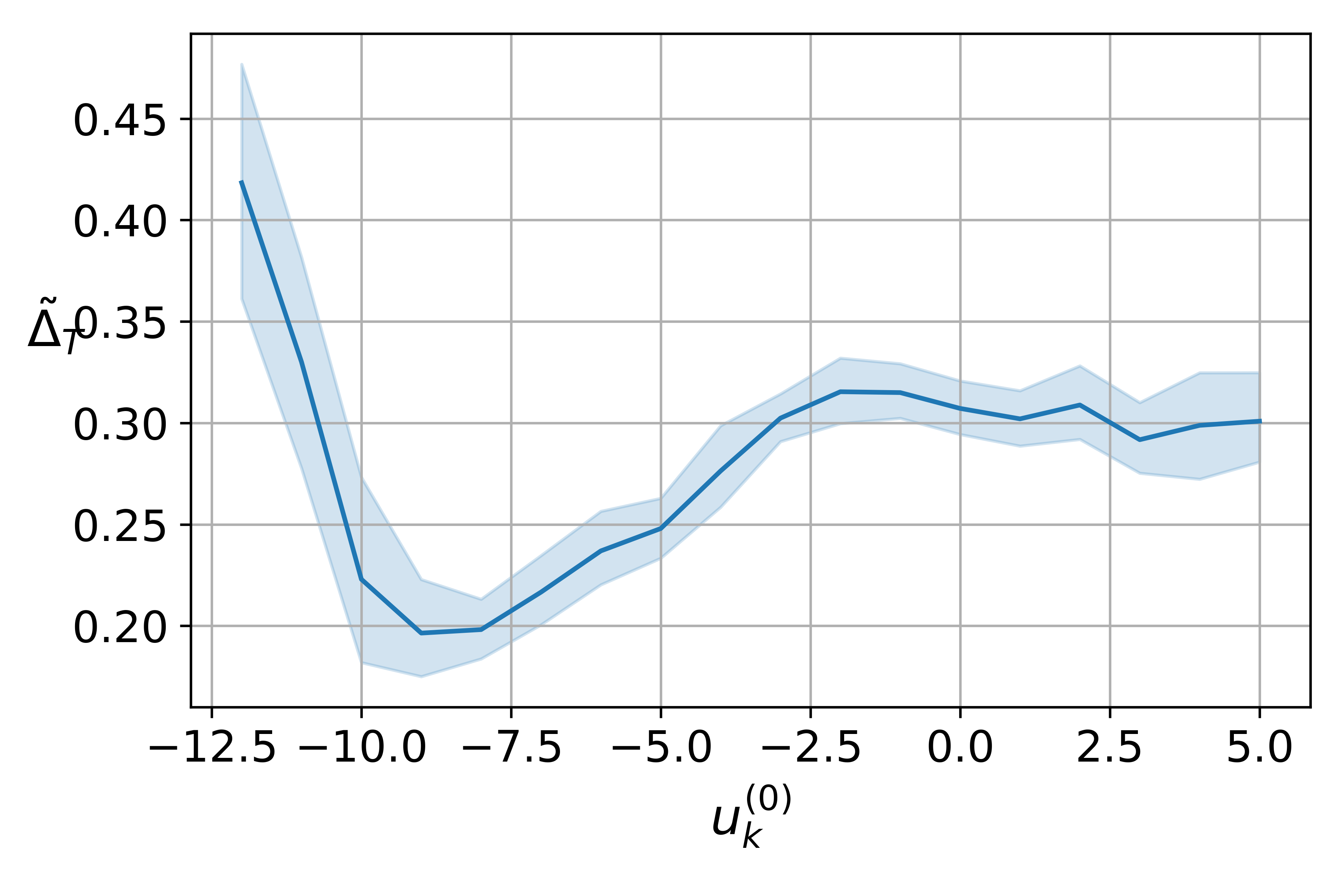} 
\includegraphics[width=0.45\linewidth]{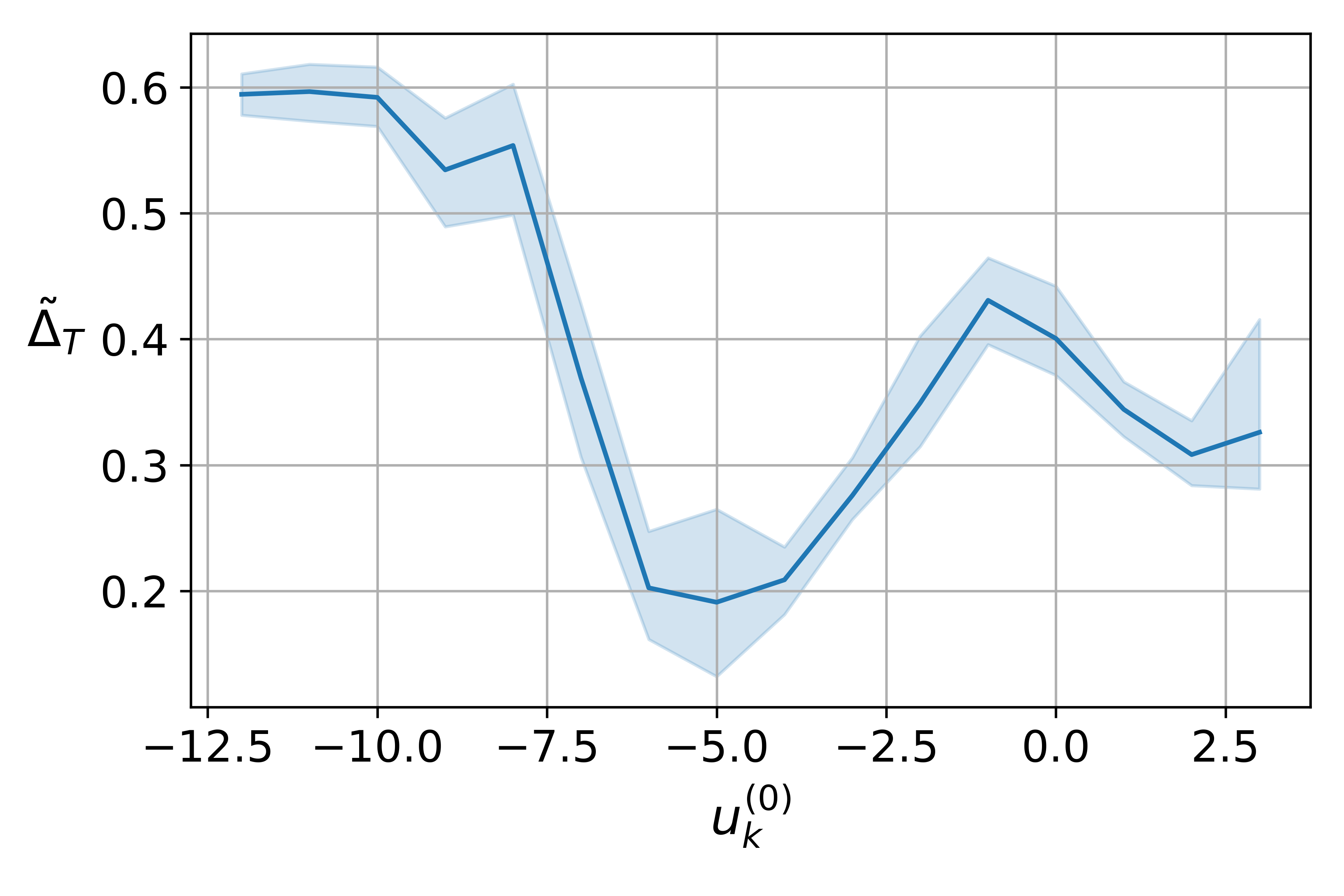} 
\includegraphics[width=0.45\linewidth]{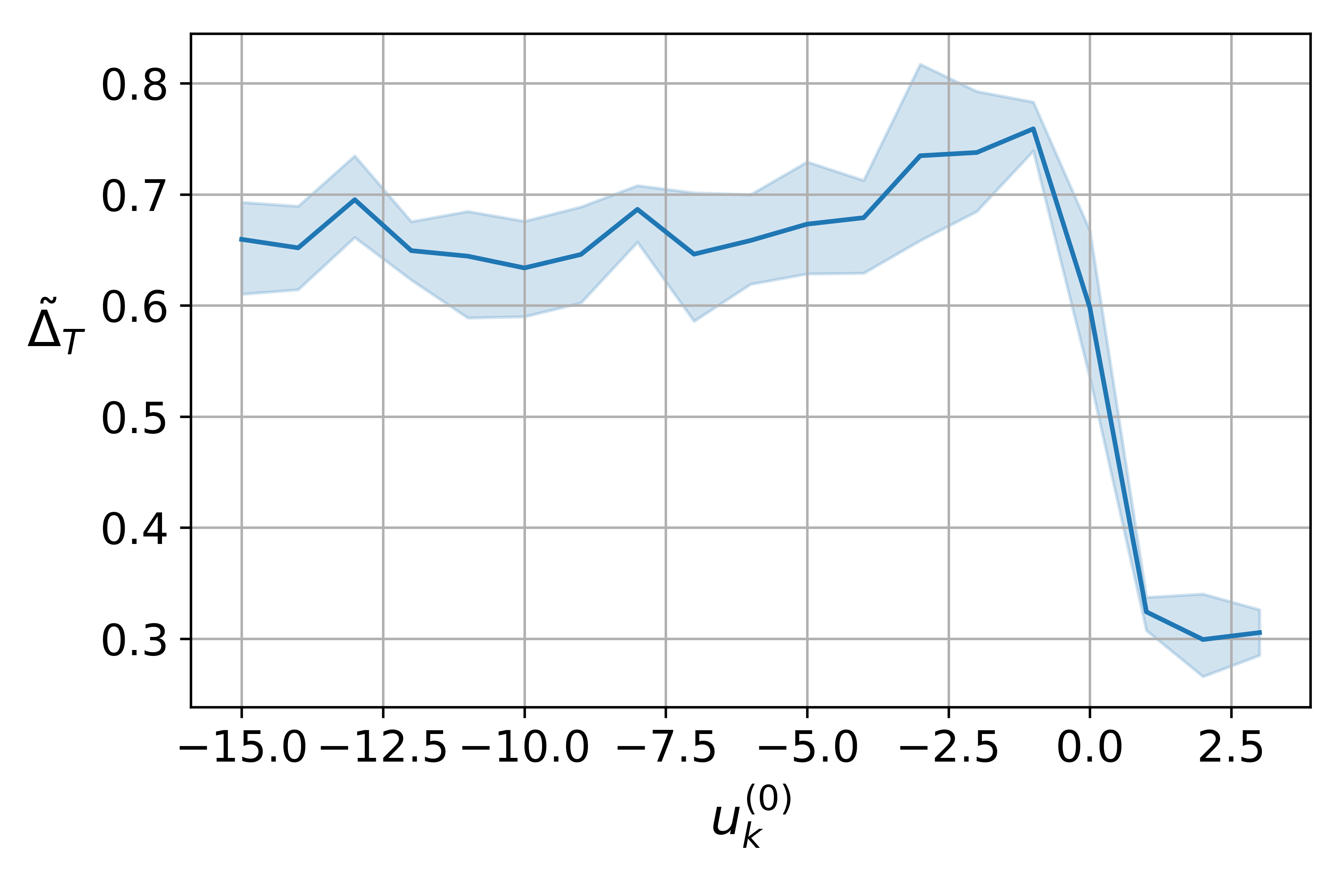}  
\includegraphics[width=0.45\linewidth]{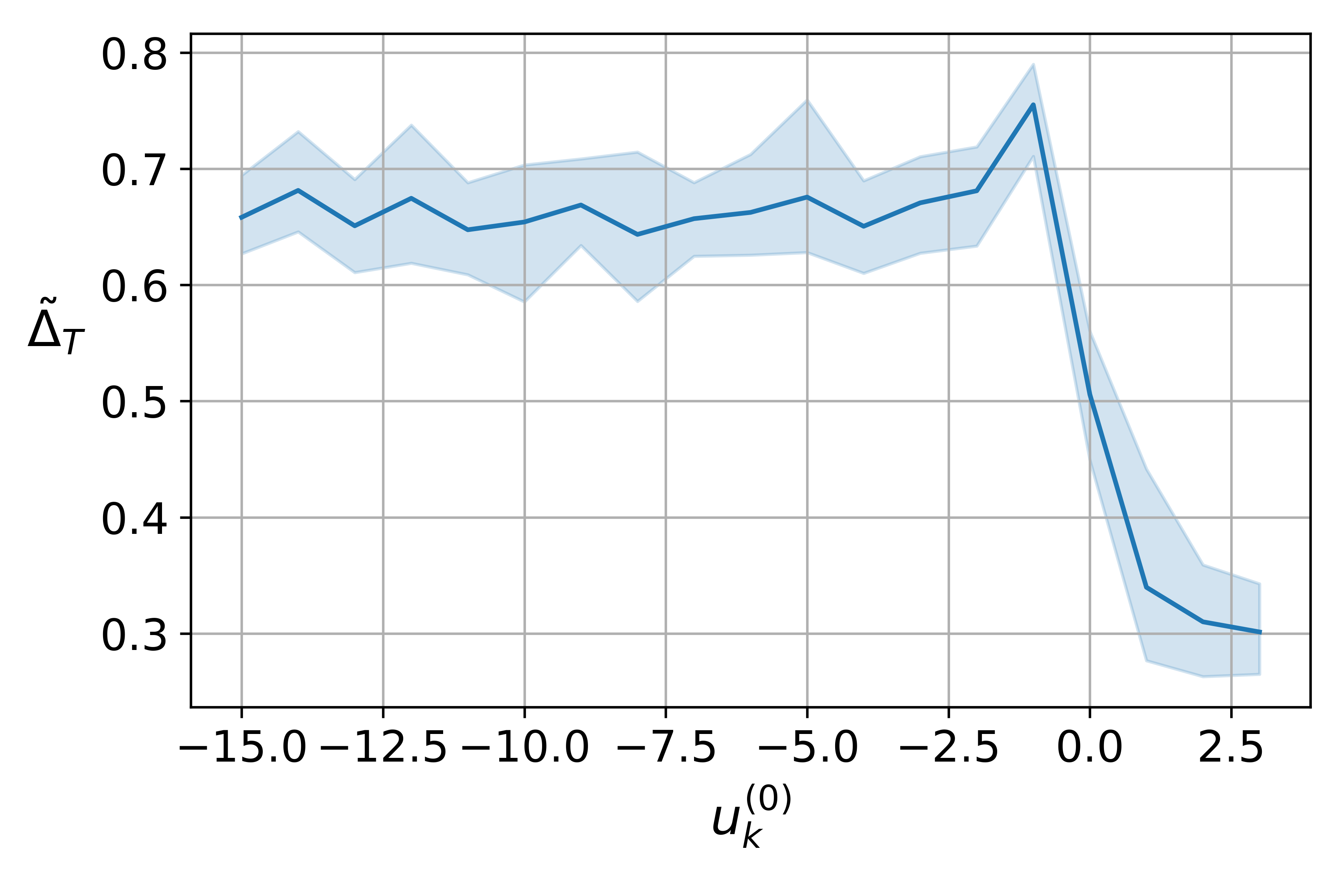}
    \caption{The collusive level with varying $u^{(0)}_k$, for $k\in\{b,s\}$, where $\Phi = [1, 0; 0, 0]$ (top left), $\Phi=[0, 1; 0, 0]$ (top right), $\Phi = [0, 1; -1, 0]$ (middle left), $\Phi = [1, 0; 0, -1]$ (middle right), $\Phi = [2, 0; 0, 2]$ (bottom left), and $\Phi=[0, 2; 2, 0]$ (bottom right).}
    \label{fig:u0_change_ones}
\end{figure}

We provide Figures~\ref{fig:beta_v_appendix} and~\ref{fig:u0_change_ones}, which 
complement Figures \ref{fig:beta_v} and \ref{fig:u0_change_main}, respectively, with additional choices of the externality matrix.

A common trend in both Figures~\ref{fig:beta_v} and~\ref{fig:beta_v_appendix} is that the collusive level decreases as $\beta_k$ increases.

Section~\ref{subsect:colllusive_idiosyncratic} summarizes the main common observation to both 
Figure~\ref{fig:u0_change_main} and Figure~\ref{fig:u0_change_ones}.  
